\documentclass[twocolappendix]{emulateapj}
\usepackage{times}
\usepackage{bm}% bold math
\usepackage{amsmath,amssymb}
\usepackage{color}
\usepackage{float}
\usepackage{xcolor}
\usepackage{hyperref}
\hypersetup{colorlinks=true,citecolor=olive,linkcolor=red}

\newcommand{\bolE}{{\bm  E}}

\newcommand{\bolB}{{\bm  B}}

\newcommand{\beq}{\begin{equation}}
\newcommand{\eeq}{\end{equation}}
\newcommand{\beqn}{\begin{eqnarray}}
\newcommand{\eeqn}{\end{eqnarray}}
\newcommand{\beqno}{\begin{equation*}}
\newcommand{\eeqno}{\end{equation*}}
\newcommand{\beqnno}{\begin{eqnarray*}}
\newcommand{\eeqnno}{\end{eqnarray*}}
\newcommand{\etal}{{et al. }}

\newcommand{\ASIAA}{Institute of Astronomy \& Astrophysics, Academia
Sinica, 11F of Astronomy-Mathematics Building, AS/NTU No. 1, Taipei
10617, Taiwan}
\newcommand{\NAOJM}{Mizusawa VLBI Observatory, National Astronomical
Observatory of Japan, Osawa, Mitaka, Tokyo 181-8588, Japan}
\newcommand{\PITP}{Perimeter Institute for Theoretical Physics, 31
Caroline Street North, Waterloo, ON, N2L 2Y5, Canada}
\newcommand{\UTULSA}{Department of Physics and Engineering Physics,
University of Tulsa, Tulsa, OK 74104, USA}
\newcommand{\GODDARD}{Gravitational Astrophysics Laboratory,
NASA Goddard Space Flight Center, Greenbelt, MD 20771, USA}
\newcommand{\NTU}{Department of Physics, National Taiwan University,
Taipei 10617, Taiwan}
\newcommand{\UTOHOKU}{Astronomical Institute, Tohoku University, Sendai
980-8578, Japan}
\newcommand{\KOGAKUINU}{Kogakuin University, Academic Support Center,
2665-1 Nakano, Hachioji, Tokyo 192-0015, Japan}
\newcommand{\NAOJ}{National Astronomical Observatory of Japan, Osawa,
Mitaka, Tokyo 181-8588, Japan}
\newcommand{\YITP}{Yukawa Institute for Theoretical Physics,
Kyoto University, Kyoto, 606-8502, Japan}
\newcommand{\SNU}{Department of Physics and Astronomy, Seoul National
University, 1 Gwanak-ro, Gwanak-gu, Seoul 08826, Korea}
\newcommand{\INAF}{INAF-Istituto di Radio Astronomia, via P. Gobetti
101, I-40129 Bologna, Italy}
\newcommand{\NRAO}{National Radio Astronomy Observatory, 520 Edgemont
Rd, Charlottesville, VA, 22903, USA}
\newcommand{\MIT}{Massachusetts Institute of Technology, Haystack
Observatory, 99 Millstone Road, Westford, MA 01886, USA}
\newcommand{\ISAS}{Institute of Space and Astronautical Science, Japan
Aerospace Exploration Agency, 3-1-1 Yoshinodai, Chuou-ku, Sagamihara,
252-5210, Kanagawa, Japan}
\newcommand{\UBOLOGNA}{Dipartimento di Fisica e Astronomia, Universita’
di Bologna, via Gobetti 93/2, 40129 Bologna, Italy}
\newcommand{\MPIfR}{Max-Planck-Institut f\"ur Radioastronomie,
Auf dem H\"ugel 69, D-53121 Bonn, Germany}
\newcommand{\YAMAGUCHIU}{Graduate School of Science and Engineering,
Yamaguchi University, 753-8511, Yamaguchi, Japan}

\shorttitle{Parabolic Jets in M87}
\shortauthors{Nakamura \etal}

\begin{document}

\title{Parabolic Jets from the Spinning Black Hole in M87}

\author{Masanori Nakamura\altaffilmark{1},
Keiichi Asada\altaffilmark{1},
Kazuhiro Hada\altaffilmark{2},
Hung-Yi Pu\altaffilmark{3,1},
Scott Noble\altaffilmark{4,5},
Chihyin Tseng\altaffilmark{6,1},
Kenji Toma\altaffilmark{7},
Motoki Kino\altaffilmark{8,9},
Hiroshi Nagai\altaffilmark{9},
Kazuya Takahashi\altaffilmark{10}
Juan-Carlos Algaba\altaffilmark{11},
Monica Orienti\altaffilmark{12},
Kazunori Akiyama\altaffilmark{13,14,2},
Akihiro Doi\altaffilmark{15},
Gabriele Giovannini\altaffilmark{16,12},
Marcello Giroletti\altaffilmark{12},
Mareki Honma\altaffilmark{2},
Shoko Koyama\altaffilmark{1},
Rocco Lico\altaffilmark{17,16,12},
Kotaro Niinuma\altaffilmark{18},
Fumie Tazaki\altaffilmark{2}
}
\affil{
$^1$\ASIAA}
\affil{
$^2$\NAOJM}
\affil{
$^3$\PITP}
\affil{
$^4$\UTULSA}
\affil{
$^5$\GODDARD}
\affil{
$^6$\NTU}
\affil{
$^7$\UTOHOKU}
\affil{
$^8$\KOGAKUINU}
\affil{
$^9$\NAOJ}
\affil{
$^10$\YITP}
\affil{
$^{11}$\SNU}
\affil{
$^{12}$\INAF}
\affil{
$^{13}$\NRAO}
\affil{
$^{14}$\MIT}
\affil{
$^{15}$\ISAS}
\affil{
$^{16}$\UBOLOGNA}
\affil{
$^{17}$\MPIfR}
\affil{
$^{18}$\YAMAGUCHIU
\\
{\tt nakamura@asiaa.sinica.edu.tw}
}

\begin{abstract}
The M87 jet is extensively examined by utilizing general relativistic
magnetohydrodynamic (GRMHD) simulations as well as the steady
axisymmetric force-free electrodynamic (FFE) solution. Quasi-steady
funnel jets are obtained in GRMHD simulations up to the scale of $\sim
100$ gravitational radius ($r_{\rm g}$) for various black hole (BH)
spins. As is known, the funnel edge is approximately determined by the
following equipartitions; i) the magnetic and rest-mass energy densities
and ii) the gas and magnetic pressures. Our numerical results give an
additional factor that they follow the outermost parabolic streamline of
the FFE solution, which is anchored to the event horizon on the
equatorial plane. We also identify the matter dominated, non-relativistic
corona/wind play a dynamical role in shaping the funnel jet into the
parabolic geometry. We confirm a quantitative overlap between the
outermost parabolic streamline of the FFE jet and the edge of jet sheath
in VLBI observations at $\sim 10^{1}$--$10^{5} \, r_{\rm g}$, suggesting
that the M87 jet is likely powered by the spinning BH. Our GRMHD
simulations also indicate a lateral stratification of the bulk
acceleration (i.e., the spine-sheath structure) as well as an emergence
of knotty superluminal features. The spin characterizes the location of
the jet stagnation surface inside the funnel. We suggest that the
limb-brightened feature could be associated with the nature of the
BH-driven jet, if the Doppler beaming is a dominant factor. Our findings
can be examined with (sub-)mm VLBI observations, giving a clue for the
origin of the M87 jet.
\end{abstract}


\section{Introduction}
\label{sec:INT}

Active galactic nuclei (AGN) jets are widely believed to be initiated in
the vicinity of supermassive black holes (SMBHs)---with masses $M \simeq
10^{7}$--$10^{10}\,M_{\odot}$---at around the gravitational radius
$r_{\rm g}$ ($\lesssim$ milliparsec) and extend up to a scale of $\sim$
megaparsec (Mpc) as giant radio lobes. Force-free electrodynamic (FFE)
and/or magnetohydrodynamic (MHD) mechanisms are frequently invoked to
extract energy and momentum from a star, compact object, or an accretion
disk around either \citep[e.g.][]{BZ77, BP82, US85, L87, MKU01, B10,
DLM12}. Key issues to be answered are the mechanism for bulk
acceleration up to the relativistic regime as is inferred from the
superluminal motions $\lesssim 40 \, c$ (the speed of light) and high
brightness temperatures observed \citep[][]{Lis13}, as well as the huge
amount of energy $\lesssim 10^{60}$--$10^{61}$ erg deposited into the
intracluster medium in the magnetic \citep[e.g.][]{K01} and/or kinetic
\citep[e.g.][]{G14} forms during their duty cycles of $\sim
10^{7}$--$10^{8}$ years. The power of relativistic jets may be larger
than the accretion luminosity, implying that a rotating black hole may
play a role \citep[][]{G14}.

Special or general relativistic MHD (SRMHD or GRMHD) flows with a
generalized parabolic geometry [where $z \propto R^{\epsilon}$ in the
cylindrical coordinates $(R, z)$ with $\epsilon > 1$] could be
accelerated due to the so-called ``magnetic nozzle effect''\footnote{An
effective separation between neighboring poloidal field lines (faster
than rate at which their cross-sectional radius increases) causes an
efficient conversion from the Poynting to matter energy flux along a
streamline.}  \citep[]{C87, LCB92, BL94} in the trans-to-super fast
magnetosonic regime. The Lorentz factor $\Gamma \gtrsim 10$ is confirmed
at large distances of $z/r_{\rm g} \gtrsim 10^{3}$ by utilizing
semi-analytical steady solutions and numerical simulations
\citep[e.g.][]{LCB92, BL94, C95, VK03a, VK03b, TT03, FO04, BN06, BN09,
KBVK07, KVKB09, TMN09, L09, L10, TT13}. An even higher $\Gamma$ is
obtained in semi-analytical/numerical studies of FFE jets
\citep[]{NMF07, TMN08}\footnote{An infinitely magnetized (i.e.,
force-free) fluid could have the same speed as the drift speed $V_{\rm
d}$ of the electromagnetic ($\bolE$ and $\bolB$) fields; $V_{\rm
d}/c=|\bolE \times \bolB|/B^{2}=E/B$ ($\bolE \cdot \bolB$=0) and thus
the Lorentz factor from the drift speed is $\Gamma^{2} \approx
1/(1-V_{\rm d}^{2}/c^{2})= B^2/(B^2-E^2)$.}.

The value $\Gamma$ of a cold, relativistic MHD (Poynting flux-dominated:
PFD) outflow is related to $\mu$, the total [matter (kinetic plus
rest-mass) + electromagnetic] energy flux per unit rest-mass energy flux
\citep[e.g.][]{TT13}:
\beq
\label{eq:TOTAL-TO-MATTER-ENG.FLUX} \mu/\Gamma=1+\sigma,
\eeq
where $\sigma$ is the Poynting flux per unit matter energy flux (i.e.,
so-called ``magnetization''), and $\mu$ is constant along a streamline
(poloidal magnetic field line) in a steady axisymmetric ideal-MHD
flow. Therefore, $\Gamma$ approaches its maximum value $\Gamma_{\infty}
\simeq \mu$ with $\sigma_{\infty} \simeq 0$, when a full conversion of
electromagnetic energy to matter kinetic energy occurs. It is, however,
still unknown how/where the MHD bulk acceleration is terminated in the
realistic galactic environment, or what value is
$\sigma_{\infty}$. Also, the radial ($R$) expansion of MHD jets
naturally produces a lateral stratification of $\Gamma$ in the jet
interior with a different evolution of $\epsilon$ and $\sigma$
\citep[]{M06, KBVK07, KVKB09, TMN09}. Recently, values of $\mu \lesssim
30$ are suggested, which implies $\sigma_{\infty} \lesssim 1$, although
$\mu \sim 10$-10$^{3}$ could be expected in the MHD regime for
relativistic outflows \citep[]{NBKZ15}.

M87 is one of the nearest active radio galaxies \citep[16.7 Mpc;
][]{B09} that exhibits subliminal to superluminal motions (see, Figure
\ref{fig:U} and references therein).
%\citep[e.g.][]{R89, B95, B99, C07, K07, L07, G12, M13, A14, H16,
%M16}.
With its proximity, the black hole mass $M$ is estimated to be in a
range of $(3.3 - 6.2) \times 10^{9} M_{\sun}$ \citep[e.g.][]{M97, GT09,
G11, W13}. The largest mass of $6.2 \times 10^{9} M_{\sun}$ gives an
apparent angular size $\sim$ 3.7 $\mu$as$/r_{\rm g}$. This galaxy
therefore provides a unique opportunity to study a relativistic outflow
with the highest angular resolution in units of $r_{\rm g}$. Global mm
VLBI observations, known as the Event Horizon Telescope (EHT) project,
is expected to resolve the black hole shadow in M87
\citep[]{D12}. Therefore, we also expect to resolve the jet launching
region in the coming years.

Extended synchrotron emission of the one-sided jet, emerging from the
nucleus, has been the target of multi-wavelength studies, from radio
(see, Figure \ref{fig:R-z} and references therein) to X-ray bands for
decades,
%\citep[e.g.][]{R89, O89, S96, P99, J99, M02, WY02, PW05, H06, D06,
%L07, K07, M09, H09, AN12, H16, ANP16}
which cover the scale of $\sim0.2$ mas -- 14 arcsec with the viewing
angle $\theta_{\rm v}=14^{\circ}$ \citep[]{WZ09}, corresponding to $\sim
2.3 \times 10^{2} \, r_{\rm g}$ -- $1.6 \times 10^{7}\, r_{\rm g}$ in
de-projection. VLBI cores are considered to be the innermost jet
emission at observed frequencies \citep[]{BK79}. Observations at cm to
mm wavelengths \citep[]{H11} therefore may be used to explore the jet
further upstream $\lesssim 200\, r_{\rm g}$ \citep[]{H13, NA13}
including the VLBI cores at 230 GHz by EHT observations \citep[]{D12,
A15}. Examinations of VLBI cores in M87 suggest a strongly magnetized
jet in the vicinity of the SMBH \citep[]{K15}, challenging the classical
equipartition paradigm.

From low to high frequency VLBI observations,
%\citep[]{R89, J99, K07, L07, H11, AN12, NA13, H13, H16, ANP16},
``limb-brightened'' features (dominated by the ``jet sheath'' emission)
are widely confirmed on the scale $\sim 200$--$10^{5}\,r_{\rm g}$ (in
de-projection). Despite the fact various models are proposed for AGN
jets in general \citep[see, discussions by][]{K07}, readers can refer
specific models to the M87 jet on large scales $\gtrsim 10^{3}\,r_{\rm
g}$; either a concentration of the magnetic flux at the outer boundary
of the relativistic jet, which is confined by non-relativistic disk wind
\cite[]{G09}, or a pileup of the material along the edge of the jet
under the pressure equilibrium in the lateral direction
\citep[]{Z08}. These models nicely reproduce the synthetic synchrotron
map on pc scales ($\sim 10^3$--$10^4\, r_{\rm g}$), but both models
suggest a relatively high $\Gamma \sim 10$--15 on this spatial
scale. Furthermore, the recent discovery of the ``ridge-brightened''
features (dominated by the ``jet spine'' emission) \citep[]{ANP16, H17G}
sheds light on the complex structure in the M87 jet at $\gtrsim$ mas
($\sim 10^{3}\, r_{\rm g}$) scales. This may be a direct confirmation of
the jet ``spine-sheath'' structure in AGNs, but the emission mechanism
there is not understood sufficiently well to provide a robust prediction
of the ``ridge+limb-brightened'' feature.

One of the feasible ways to estimate the jet's global structure is to
measure the Full Width Half Maximum (FWHM) of the transverse intensity
as a diameter\footnote{To evaluate the jet's width with a
limb-brightened feature, two Gaussians are fitted to the slice profile
of the jet and one can measure the separation between outer sides of the
Half-maximum point of each Gaussian.} at different frequencies and plot
its radius (FWHM/2) as a function of the jet's axial distance
(de-projected) in units of $r_{\rm g}$. This gives a proper sense
how/where the jet streamline could be and where it originates in the
vicinity of the SMBH. A linear fit on the log-log plot is very useful to
investigate the jet structure in two-dimensional space \citep[]{AN12,
NA13, H13}. There are several preceding studies on the M87 jet
\citep[e.g.][]{BrLo09, De12, MFS16} that investigate the horizon scale
structure, but it is essential to conduct a direct comparison of the jet
global structure in observations with theory and numerical simulations.

An accreting black hole plays a dynamically important role in producing
relativistic jets, which has been demonstrated in GRMHD simulations
during the past decade; a radiatively inefficient accretion flow (RIAF)
with a poloidal magnetic flux and a spinning black hole are key
ingredients for producing PFD funnel jets \citep[e.g.][]{GMT03, DHK03,
DHKH05, H04, MG04, HK06, M06, BHK08}. The system can be directly
applicable to low-luminosity AGNs (LLAGNs) such as M87. It has been
examined that the M87 jet (sheath) is slowly collimated from a full
opening angle of $\sim 60^{\circ}$ near the black hole to $\sim
10^{\circ}$ at large distances \citep[$\gtrsim 10$ pc; ][]{J99}. We note
that the opening angle in \citet[]{J99} is an {\em apparent} value in
the sky projection. \citet[]{M06} suggests that this wider sheath
emission could be due to a RIAF wind (outside of a well-collimated
relativistic cold PFD jet).

Regarding a co-existence of the PFD funnel jet and coronal wind from the
RIAF, \citet[]{DHK03, DHKH05} observed there to be a region of unbound
mass flux at the boundary between the evacuated funnel and the coronal
wind, referred to as the ``funnel-wall'' jet. The driving force could be
a high-pressure (gas $+$ magnetic) corona squeezing material against an
inner centrifugal wall, implying that the magneto-centrifugal mechanism
\citep[][hereafter BP82]{BP82} does play a minor role. \citet[]{HK06}
concluded that the precise shape and collimation of the entire outflow
(PFD jet + funnel-wall jet + coronal wind) are uncertain for two
reasons: i) the outer boundary of the matter-dominated funnel-wall jet
is somewhat indistinct and ii) there is a smooth transition as a
function of polar angle between mildly relativistic unbound matter and
slightly slower but bound coronal matter. On the other hand, the
boundary between the low-density PFD funnel jet interior and the
high-density funnel-wall jet is sharp and clear. Properties of the
coronal wind are investigated in GRMHD simulation with various black
hole spins and different magnetic configurations \citep[e.g.][]{N12,
S13, Y15}, but there is no unique way to discriminate the boundary
\citep[][]{S13}.

Comparisons of GRMHD simulations with steady solutions of the
axisymmetric force-free disk wind \citep[]{MN07a} provide a fundamental
similarity of the PFD funnel jet. In the fiducial GRMHD simulation, the
vertically (height) integrated toroidal current, which is enclosed
inside a radius, follows a remarkably similar power-law profile with the
parabolic (or simply we use parabolic throughout this paper) solution
($\epsilon=1.6$) of the disk wind (BP82), whereas the split-monopole
($\epsilon=1$) or genuine paraboloidal ($\epsilon=2$) solutions are
well-known \citep[][hereafter BZ77]{BZ77}. This scaling is found to be
maintained in a time-averaged sense, but also at each instant of
time. It is also independent of the black hole spin. As a consequence,
the poloidal magnetic field of the PFD jet in the GRMHD simulation
agrees well with the force-free solution of a non-rotating thin disk
having the parabolic geometry. \citet[]{MN07b} performed general
relativistic FFE (GRFFE) simulations of the disk wind. The magnetosphere
of their GRFFE simulation with parabolic geometry also matches
remarkably well to the PFD funnel jet in the fiducial GRMHD simulation,
but no better than with the non-rotating force-free thin disk solution
with the BP82-type parabolic geometry. It suggests that a rotation of
the magnetic field leads to negligible ``self-collimation''.

Notable agreement of the BP82-type parabolic shape of the PFD funnel jet
between GRMHD simulations and force-free (steady/time-dependent and/or
non-rotating/rotating) models indicates that gas plus magnetic pressure
of the wind/corona in GRMHD simulations is similar to the magnetic
pressure in the FFE disk wind outside the funnel region. Note that
\citet[]{MN07b} considered only the portion of i) the steady solution of
the axisymmetric FFE disk wind and ii) the GRFFE simulation of the disk
wind (both winds are in the parabolic shape) that overlap the funnel jet
region in the GRMHD simulation. So far, the boundary condition and the
shape of the funnel edge are poorly constrained. It is also unclear
where the footpoint of the outermost streamline of the PFD funnel jet
will be anchored in the quasi-steady states of GRMHD simulations.

The collimation of the PFD funnel jet is still the issue. GRMHD
simulations in the literature exhibit jet collimation ceasing at $\sim
50\, r_{\rm g}$ \citep[]{HK06}. The largest simulations to date extend
up to $r = 10^{4}\, r_{\rm g}$ \citep[][]{M06} and show $\Gamma_{\infty}
\lesssim 10$ saturated beyond $\sim$ a few of 100 $r_{\rm g}$ (despite
$b^{2}/\rho \gg 1$), where the jet collimation terminates, following a
conical expansion downstream. Global SRMHD or (GR)FFE simulations with a
``fixed'' curvilinear boundary wall \citep[i.e., parabolic;][]{KBVK07,
KVKB09, TMN08, TNM10} show bulk acceleration up to
$\Gamma_{\infty}\sim10^{1}$--$10^{3}$, whereas it is still unclear how
such a highly relativistic flow can be stably confined in a realistic
environment.  A recent semi-analytical model shows that the collimation
of PFD jets may take place by the wind in RIAFs, if the total wind power
$P_{\rm wind}$ exceeds about 10\% of the jet power $P_{\rm jet}$
\citep[]{GL16}, while $P_{\rm wind}/\dot{M} c^{2} \approx 10^{-3}$
(where $\dot{M}$ denotes the mass accretion rate at the horizon) is
obtained by a GRMHD simulation around the Schwarzschild black hole
\citep[]{Y15}.

In this paper, we examine the structure of the PFD funnel jet with GRMHD
simulations. The funnel edge is compared with steady self-similar
solutions of the axisymmetric FFE jet and we derive the physical
conditions of the boundary between the funnel jet and outside
(wind/corona). Results are compared with the M87 jet sheath in VLBI
observations. Methods and results for examining a parabolic jet
streamline are presented in Sections 2 and 3. Comparison with VLBI
observations is given in Section 4. Based on our results, Section 5
assigns topical discussions and prospects for exploring the origin of
the M87 jet with mm/sub-mm VLBI observations in the near
future. Conclusions are provided in Section 6.

\section{Methods}
\label{sec:METHODS}
We conduct a direct comparison between the observed jet geometry in M87
and theoretical/numerical models. The present paper investigates
especially the part of parabolic streams inside the SMBH's sphere of
influence. Quasi-steady black hole ergosphere-driven jets are
self-consistently generated from GRMHD simulations, and their connection
to mm/cm VLBI images is examined by utilizing steady axisymmetric FFE
jet solutions.

\subsection{Funnel Jet Boundary in the FFE Approximation}
\label{sec:FFE}
According to a steady self-similar solution of the axisymmetric
FFE jet \citep[][hereafter NMF07, TMN08]{NMF07, TMN08}, we 
consider here an approximate formula of the magnetic stream function
$\Psi (r, \theta)$ in polar $(r, \theta)$ coordinates in the
Boyer-Lindquist frame
\citep[]{TNM10}:
\beqn
\label{eq:PSI}
\Psi (r,\theta)=\left(\frac{r}{r_{\rm H}}\right)^{\kappa}(1-\cos{\theta}),
\label{eq:Psi}
\eeqn
where $r_{\rm H}=r_{\rm g} (1+\sqrt{1-a^{2}})$ is the radius of the
black hole horizon, and the dimensionless Kerr parameter $a=J/J_{\rm
max}$ describes the black hole spin.  $J$ is the black hole angular
momentum and its maximum value is given by $J_{\rm max}=r_{\rm g} M c =
G M^2/c$, where $G$ is the gravitational constant. $0 \leq \kappa \leq
1.25$ and $0 \leq \theta \leq \pi/2$ are adopted in TMN08.  $\Psi$ is
conserved along each field (stream-) line in a steady solution of the
axisymmetric MHD outflow\footnote{An asymptotic flow ($r/r_{\rm H} \gg
1$) follows $z \propto R^{\epsilon}$, where $\epsilon=2/(2-\kappa)$,
which includes conical and parabolic shapes ($1 \leq \epsilon \leq
2.67$).}. If $\kappa=0$ is chosen, the asymptotic streamline has a
split-monopole (radial) shape $z \propto R$ (where, $R=r\, \sin \theta$
and $z=r\, \cos \theta$), whereas if $\kappa=1$ is chosen, the
streamline has the (genuine) parabolic shape $z \propto R^{2}$ at $R \gg
r_{\rm g}$ (BZ77). $\kappa=0.75$ is the case of the parabolic shape $z
\propto R^{1.6}$ (BP82), which is important in this paper.

In magnetized RIAF simulations about a non-spinning black hole
\citep[]{I08}, the poloidal magnetic field distribution takes the shape
of an ``hourglass'' shape and has an insignificant vertical component on
the equatorial plane outside the black hole. This feature becomes more
prominent in the case of spinning black holes as is shown in GRMHD
simulations of the ergospheric disk with a vertical magnetic flux
\citep[the Wald vacuum solution;][]{W74}; as the poloidal magnetic flux
and mass accretes onto the black hole, all magnetic lines threading the
ergospheric disk develop a turning point in the equatorial plane
resulting in an azimuthal current sheet~\citep[]{K05}.

Due to strong inertial frame dragging inside the ergosphere, all plasma
entering this region is forced to rotate in the same sense as the black
hole. Thus, the poloidal field lines around the equatorial plane are
strongly twisted along the azimuthal direction. The equatorial current
sheet develops further due to the vertical compression of the poloidal
field lines caused by the Lorentz force acting toward the equatorial
plane at both upper and lower ($z\gtrless0$) directions. Magnetic
reconnection (although numerical diffusion is responsible for activating
the event in an ideal-MHD simulation) will change the field topology;
all poloidal field lines entering the ergosphere penetrate the event
horizon. A similar result is obtained in GRFFE simulations
\citep[]{KM07}.

We speculate that the situation is qualitatively unchanged even if the
weakly magnetized RIAF exists in the system. Strong poloidal fields in
the ergosphere compress the innermost black hole accretion flow
vertically and reduce the disk thickness down to $H/R \simeq 0.05$,
while $H/R \gtrsim 0.3$ ($H$: the vertical scale height) is the
reference value of the disk body outside the plunging region
\citep[e.g.][]{T15}.

\begin{figure}
 \centering\includegraphics[scale=0.45, angle=0]{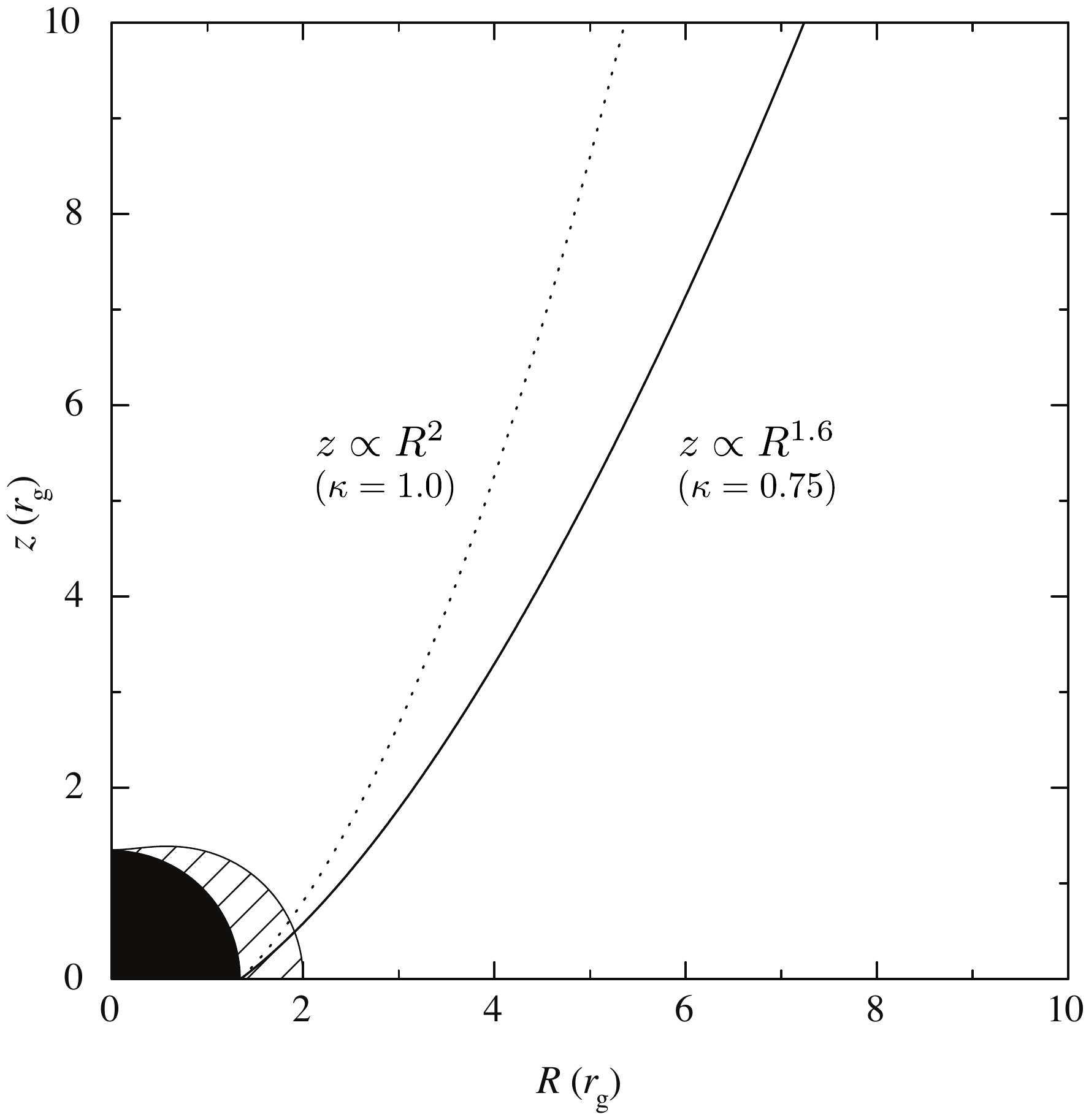}
 \caption{\label{fig:Ergo} Outermost streamlines of the steady
 axisymmetric solution of the FFE jet (NMF07, TMN08), which are anchored
 to the event horizon ($r=r_{\rm H}$) with the maximum colatitude angle
 at the footpoint $\theta_{\rm fp}=\pi/2$. A typical value of $a=0.9375$
 \citep[in GRMHD simulations;][]{GSM04, MG04, M06} is specified as a
 reference. The dotted line show the genuine paraboloidal streamline
 with $\kappa=1.0$ \citep[$z \propto R^{2}$ at $R \gg r_{\rm g}$;
 e.g.][]{BZ77}, while the solid line show the paraboloidal streamline
 with $\kappa=0.75$ \citep[$z \propto R^{1.6}$ at $R \gg r_{\rm g}$;
 e.g.][]{BP82}.  The black hole and the ergosphere are represented as
 the filled and the hatched areas.}
\end{figure}

Based on the physical picture above, we assume no poloidal magnetic flux
penetrates the equatorial plane at $R > r_{\rm H}$. Therefore, the
outermost field line, which is anchored to the event horizon with the
maximum colatitude angle at the footpoint $\theta_{\rm fp}=\pi/2$, can
be defined as
\beqn
\Psi (r_{\rm H}, \pi/2) = 1
\label{eq:Psi_0}
\eeqn
in Equation \citep[\ref{eq:Psi}; e.g., ][]{TMN08}. Figure \ref{fig:Ergo}
shows the outermost streamlines of $\Psi (r, \theta)=1$ with different
$\kappa$ ($\kappa=1$: BZ77 or $\kappa=0.75$: BP82) with a fiducial black
hole spin \citep[$a=0.9375$:][]{MG04, M06}. Let us compare the outermost
streamline of the funnel jet in GRMHD simulations with Equation
(\ref{eq:Psi_0}) at a quasi-steady state.

\subsection{Our Prospective in GRMHD Simulations}
\label{sec:GRMHD}

The public version of the two-dimensional (2D) axisymmetric GRMHD code
\texttt{HARM} \citep[]{GMT03, NGMDZ06} is used in our examinations.  The
code adopts dimensionless units $GM=c=1$. We, however, occasionally
reintroduce factors of $c$ for clarity. Lengths and times are given in
units of $r_{\rm g} \equiv GM/c^{2}$ and $t_{\rm g} \equiv GM/c^{3}$,
respectively.  We absorb a factor of $\sqrt{4 \pi}$ in our definition of
the magnetic field. \texttt{HARM} implements so-called modified
Kerr-Schild coordinates: $x_{0},\, x_{1},\, x_{2},\, x_{3}$, where
$x_{0}=t, \ x_{3}=\phi$ are the same as in Kerr-Schild coordinates, but
the radial $r (x_{1})$ and colatitude $\theta (x_{2})$ coordinates are
modified \citep[]{MG04}. The computational domain is axisymmetric,
expanding in the $r$-direction from $r_{\rm in}=0.98\, r_{\rm H}$ to
$r_{\rm out}=40\,r_{\rm g}$ and the $\theta$-direction from $\theta=0$
to $\theta=\pi$. An ‘‘outflow’’ boundary condition is imposed at
$r=r_{\rm out}$; all primitive variables are projected into the ghost
zones. The inner boundary condition is identical at $r=r_{\rm in} <
r_{\rm H}$ (no back flow into the computational domain). A reflection
boundary condition is used at the poles ($\theta=0,\, \pi$).

Typical 2D axisymmetric GRMHD simulations \citep[e.g.][]{GMT03, MG04,
M06} adopt a dense ``Polish Doughnut''-type torus \citep[]{FM76, A78},
which is in a hydrodynamic equilibrium supported by the centrifugal and
gas pressure ($p$) gradient forces. The torus is surrounded by an
insubstantial, but dynamic, accreting spherical atmosphere [the
rest-mass density $\rho$ and the internal energy density $u$ are
prescribed in power-law forms as $\rho_{\rm min} = 10^{-4} (r/r_{\rm
in})^{-3/2}$ and $u_{\rm min} = 10^{-6} (r/r_{\rm in})^{-5/2}$] that
interacts with the torus. This is the so-called ``floor model'' that
forces a minimum on these quantities in the computational domain to
avoid a vacuum. The initial rest-mass density $\rho_{0}$ in the system
is normalized by the maximum value of the initial torus $\rho_{0, \rm
max}$ on the equator.

A poloidal magnetic loop, which is described by the toroidal component
of the vector potential as a function of the density $A_{\phi} \propto
\max \,(\rho_{0}/\rho_{0, \rm max}-0.2,\, 0)$, is embedded in the
torus. The field strength is normalized with the ratio of gas to
magnetic pressure [the so-called plasma-$\beta$, hereafter $\beta_{\rm
p}\equiv 2(\gamma-1)u/b^{2}$, where $b^{2}/2=b^{\mu}b_{\mu}/2$ is the
magnetic pressure measured in the fluid frame]. The inner edge of the
torus is fixed at $(r,\, \theta)=(6 \, r_{\rm g},\, \pi/2)$ and the
pressure maximum is located at $(r,\, \theta)=(r_{\rm max},\, \pi/2)$,
where $r_{\rm max}=12 \, r_{\rm g}$ is adopted. $\beta_{\rm p0,
min}=100$, where $\beta_{\rm p0, min}$ denotes the minimum
plasma-$\beta$ at $t=0$, is chosen\footnote{The magnetic field strength
is normalized by $\beta_{\rm p0, min}$ by finding the global maxima of
$u$ and $b^{2}$ in the computational domain. They lie inside the torus,
but not at the same grid point.  The magnetic ``O-point'' is located at
the gas pressure maximum on the equatorial plane.} for our fiducial
run. An ideal gas equation of state $p=(\gamma-1)u$ is used and the
ratio of specific heats $\gamma$ is assumed to be 4/3.  If not otherwise
specified, a value of $a=0.9375$ is used \citep[]{M06}. For further
computational details, readers can refer to \citet[]{GMT03, MG04} which
adopt default parameters in \texttt{HARM}.

Given small perturbations in the velocity field, the initial state of a
weakly magnetized torus with a minimum value of $\beta_{\rm p0, min}
\gtrsim 50$ is unstable against the magneto-rotational instability
\citep[MRI;][]{BH91} so that a transport of angular momentum by the MRI
causes magnetized material to plunge from the inner edge of the torus
into the black hole. The turbulent region in the RIAF's body and
corona/wind gradually expands outward.  The inner edge of the torus
forms a relatively thin disk with the ``Keplerian'' profile on the
equator in both cases of non-spinning \citep[]{I08} and spinning
\citep[]{MG04} black holes. The turbulent inflow of the RIAF body
becomes laminar at the plunging region inside the innermost stable
circular orbit (ISCO) \citep[e.g.][]{KH02, DHK03, MG04, RF08}, whereas
there is no other specific signature in the flow dynamics across the
ISCO \citep[]{TM12}. During the time evolution of the system, PFD
(highly magnetized, low-density) funnel jets are formed around both
polar axes ($\theta = 0$ and $\pi$).

Our goal in using GRMHD simulations is to examine the quasi-steady
structure of the boundary between the low-density funnel interior (PFD
jet) and the high-density funnel exterior (corona/wind outside the RIAF
body). It can be directly compared with an outermost streamline of the
semi-analytical FFE jet solution, which is anchored to the horizon
(Figure \ref{fig:Ergo}), when the funnel enters a long, quasi-steady
phase.

\section{Results}
\label{sec:RESULTS}

\subsection{Fiducial Run: $a=0.9375$}
\label{sec:RESULTS-FID}

We first examine a high-resolution (fiducial) model in \citet[]{MG04};
default parameters in \texttt{HARM} (as introduced above) are adopted
with a fine grid assignment $N_{x_{1}} \times N_{x_{2}} =
456\times456$. The simulation is terminated at $t/t_{\rm g}=2000$, or
about 7.6 orbital periods at the initial pressure maximum on the
equator. The MRI turbulence in the RIAF body is not sustained at
$t/t_{\rm g} \gtrsim 3000$ due to our assumption of axisymmetry as
explained by the anti-dynamo theorem \citep[]{C34}. However, the decay
of the turbulence does not affect the evolution of the PFD funnel jet
\citep[]{M06}. Thus, we extended the time evolution up to $t/t_{\rm g} =
9000$ in order to examine whether the quasi-steady state of the PFD
funnel jet is obtained or not. This enables us to perform a direct
comparison between the steady axisymmetric FFE solution and axisymmetric
GRMHD simulations regarding the funnel shape. Constraining physical
quantities at the funnel edge is important for further understanding the
structure of relativistic jets in M87 and others in general.

\begin{figure}
 \centering\includegraphics[scale=.9, angle=0]{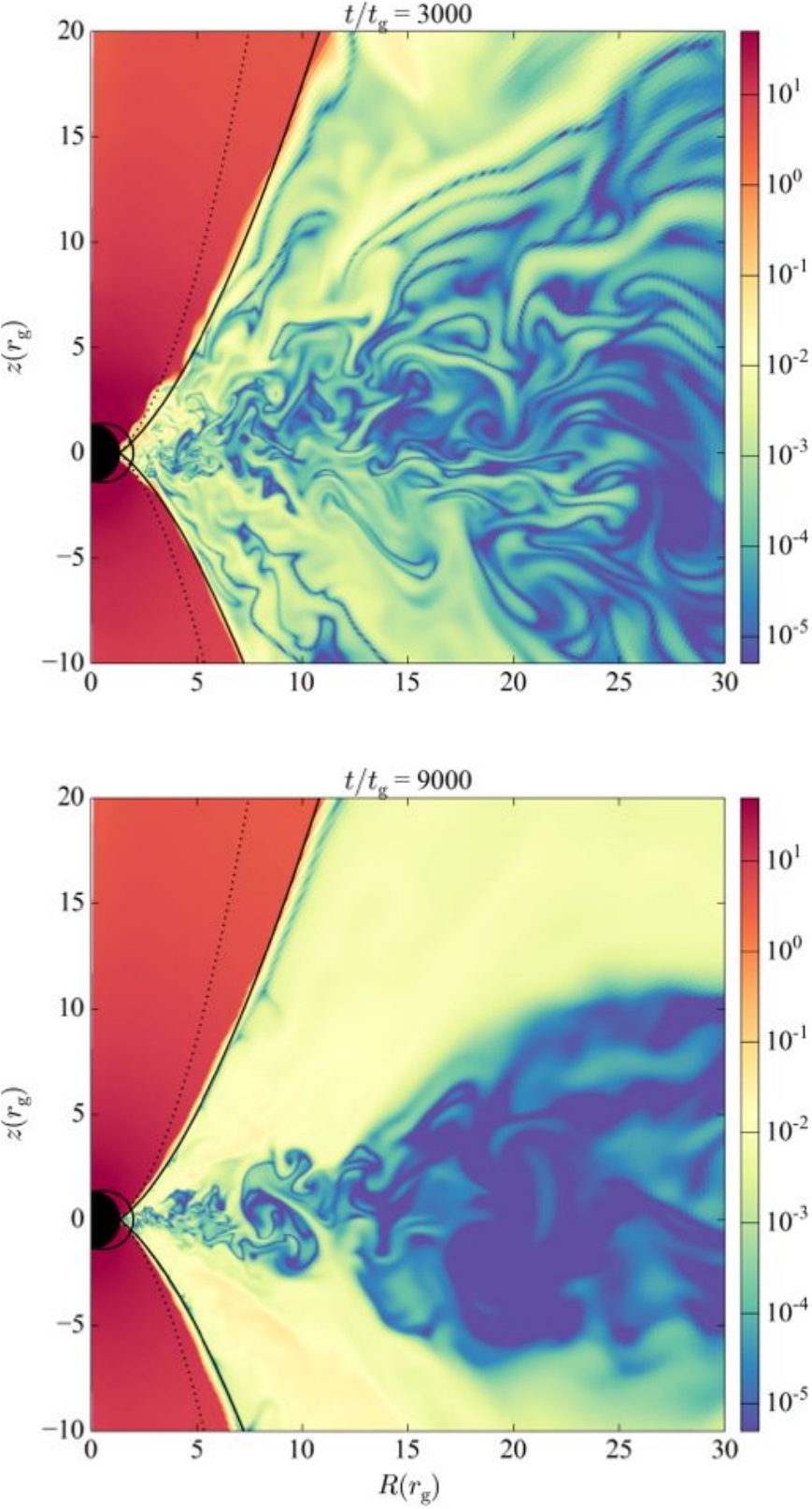}
 \caption{\label{fig:FID-MAG-EVO} Time evolution of the fiducial run
 ($a=0.9375$); $t/t_{\rm g}=3000$ ({\em top}) and 9000 ({\em bottom}),
 respectively. A color filled contour shows the magnetic energy per unit
 particle $b^{2}/\rho$, which is measured in the fluid frame. The black
 hole, the ergosphere (``not hatched''), and two outermost streamlines
 (genuine parabolic/parabolic), which are anchored to the event horizon,
 are displayed in the same manner as in Figure \ref{fig:Ergo}.}
\end{figure}

\begin{figure}
 \centering\includegraphics[scale=.9, angle=0]{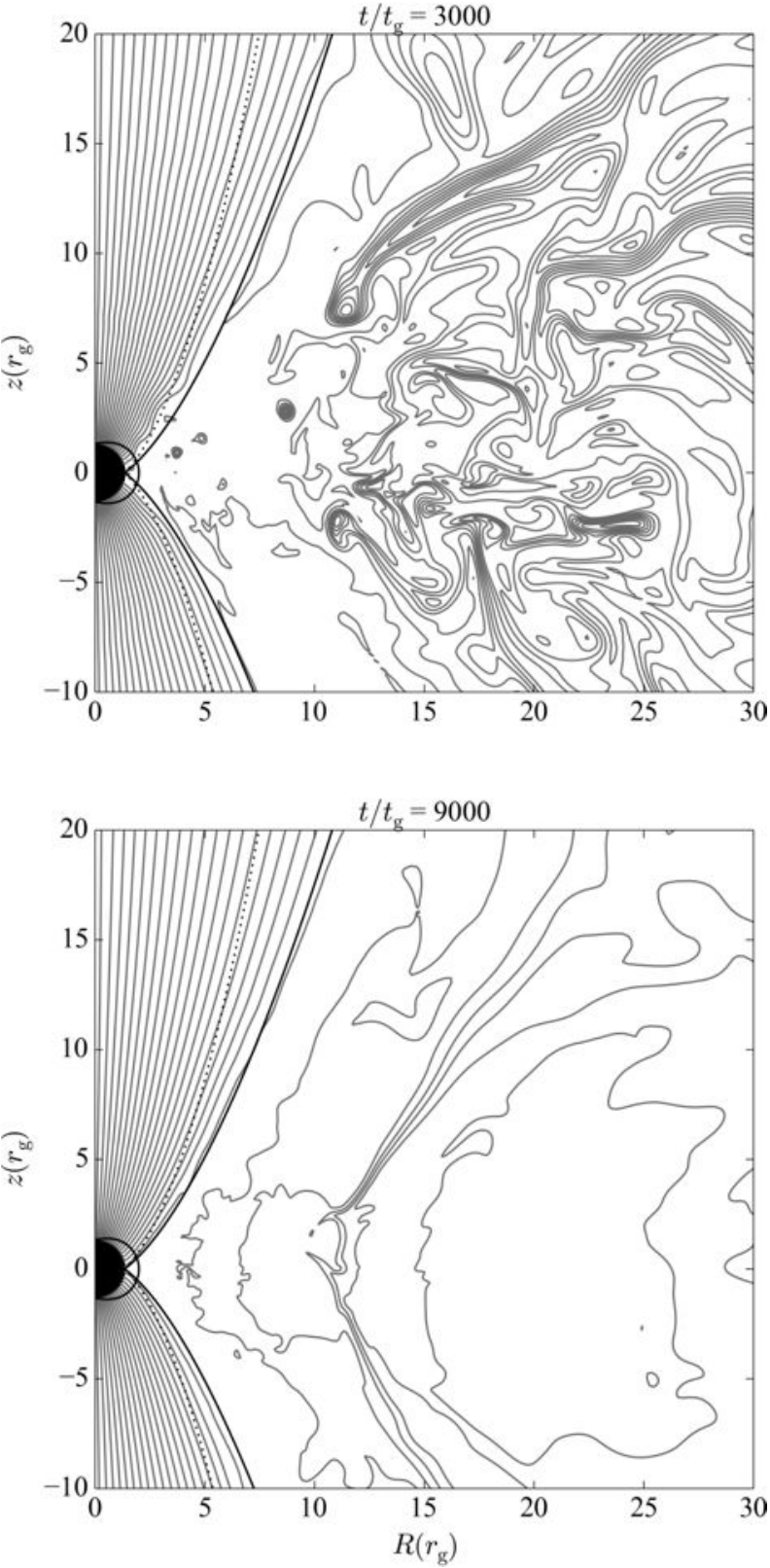}
 \caption{\label{fig:FID-B-LINE-EVO} Time evolution of the fiducial run
 ($a=0.9375$); $t/t_{\rm g}=3000$ ({\em top}) and 9000 ({\em bottom}),
 respectively. Contours (gray) represent poloidal magnetic field
 lines. Other components in the panels are identical to those in Figure
 \ref{fig:FID-MAG-EVO}.}
\end{figure}

\begin{figure}
 \centering\includegraphics[scale=.4, angle=0]{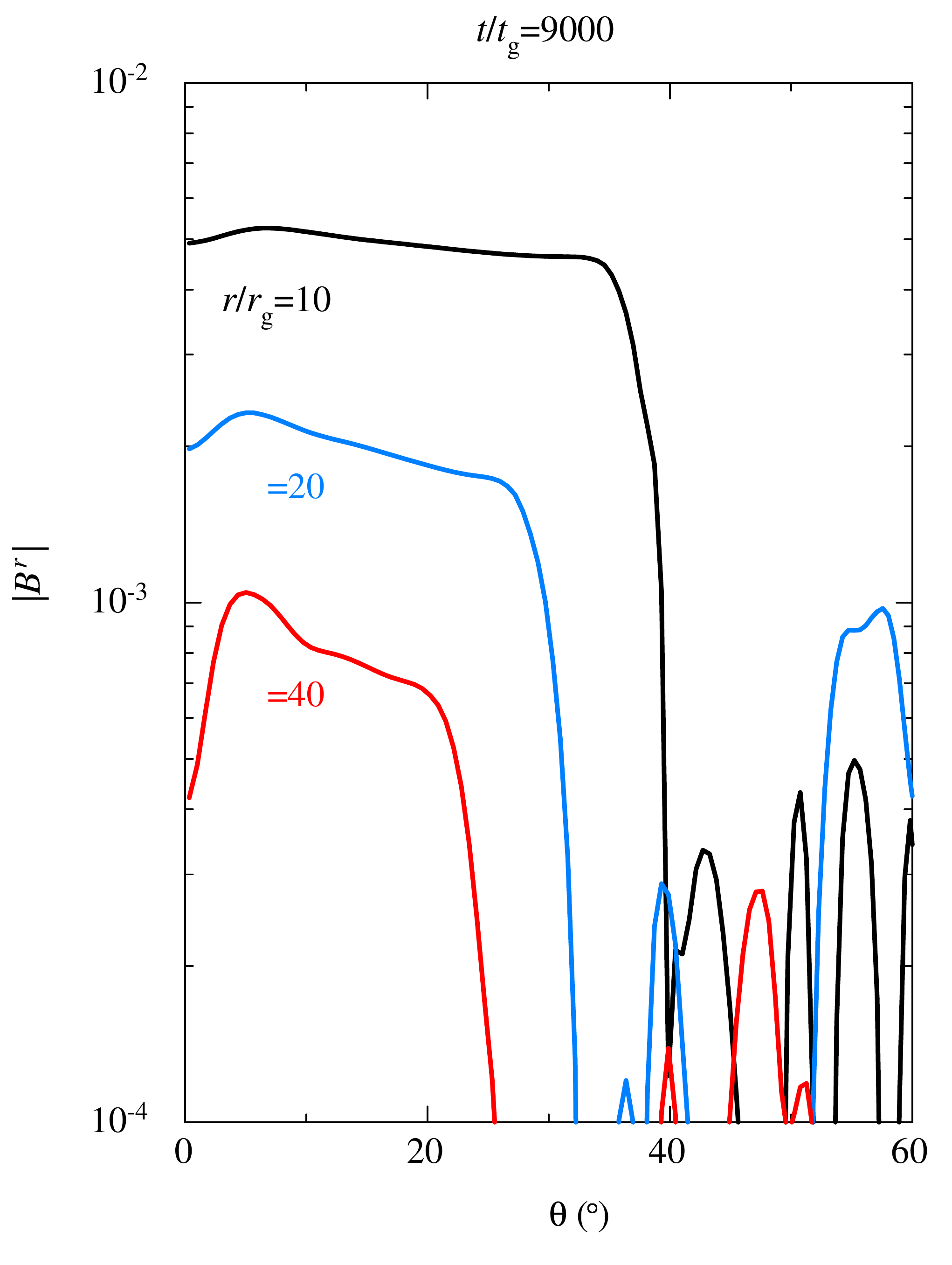}
 \caption{\label{fig:FID-B-FIELD-BUNCH} A $\theta$
 cross-section at $r/r_{\rm g}=$10 (black), 20 (blue), and 40 (red)
 showing the absolute value of the radial magnetic field $|B^{r}|$ in
 the fiducial run ($a=0.9375$) at $t/t_{\rm g}=9000$.}
\end{figure}

\begin{figure}
 \centering\includegraphics[scale=.37, angle=0]{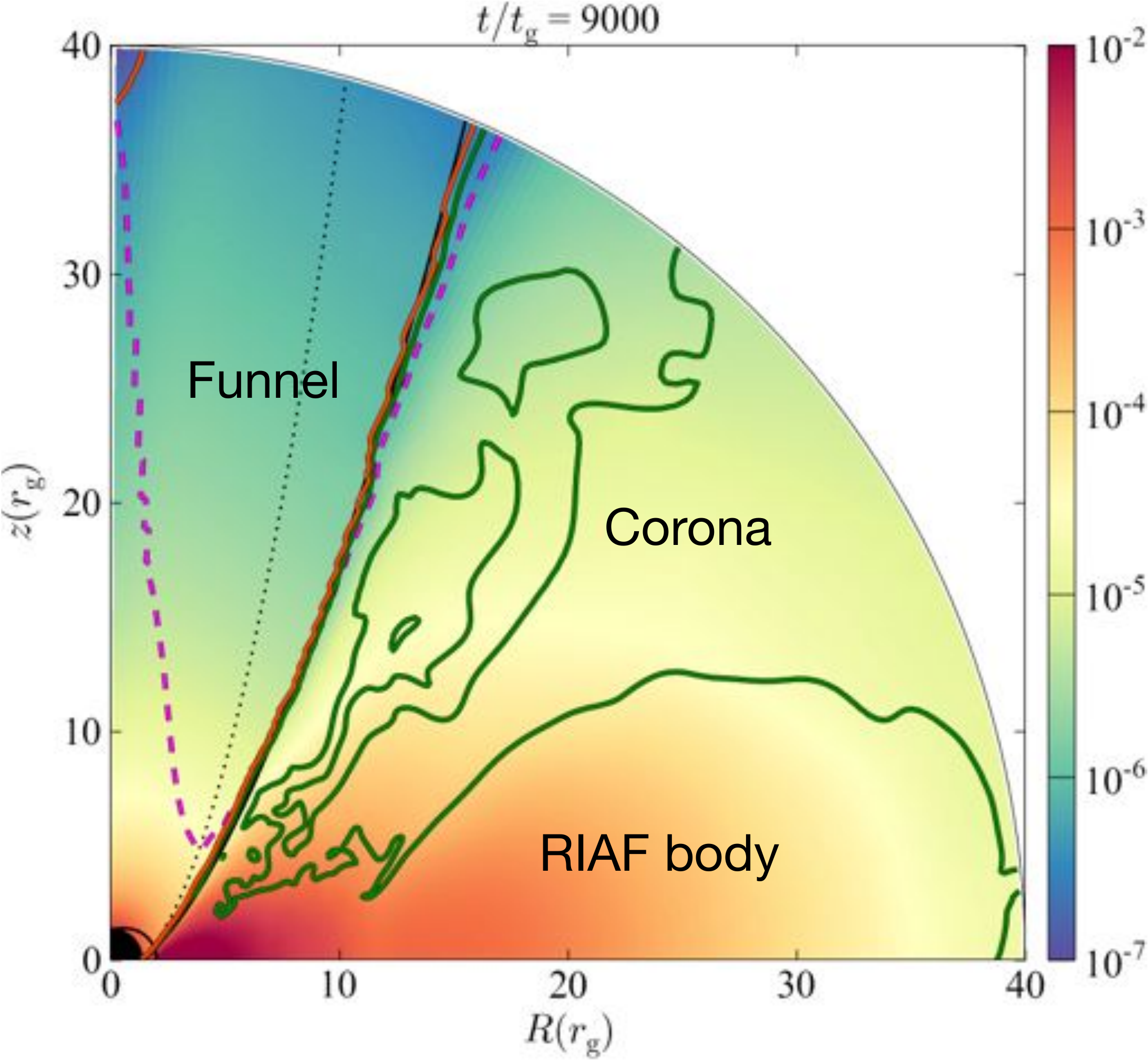}
 \caption{\label{fig:FID-PR_TOT-LINES} The final snapshot of the
 fiducial run ($a=0.9375$); $t/t_{\rm g}=9000$. A color filled contour
 shows the total pressure $p+b^{2}/2$ (in the fluid frame). Contours of
 equipartition quantities are exhibited (the upper computational domain;
 $0 \leq \theta \leq \pi/2$). An orange solid line shows $b^{2}/\rho=1$,
 while green solid lines show $\beta_{\rm p}=1$. $-u_{t}=1 \ (Be \approx
 0)$ is shown with a magenta dashed line. Other components are identical
 to those in Figure \ref{fig:FID-MAG-EVO}.}
\end{figure}

Figure \ref{fig:FID-MAG-EVO} shows the time sequence of the distribution
of the relative densities of magnetic and rest-mass energy
$b^{2}/\rho$. This figure provides a quantitative sense for the spatial
distribution of the PFD funnel jet, wind/corona, and RIAF body (see also
Figures \ref{fig:FID-PR_TOT-LINES} and \ref{fig:FID-MASSFLUX-U^r} for
details). Following some previous work \citep[e.g.][]{M06, De12}, we
confirmed that the funnel jet-wind/corona boundary can be identified to
be where $b^{2}/\rho \simeq 1$. At the stage $t/t_{\rm g}=3000$ ({\em
top} panel), the MRI is well developed and thus the magnetized material
in the RIAF body is swallowed by the black hole. A certain amount of the
poloidal flux, which falls into the ergosphere, is twisted along the
azimuthal direction and powerful PFD jets are formed toward the polar
directions. Consequently, the funnel expands laterally by the magnetic
pressure gradient and its outer boundary ($b^{2}/\rho \simeq 1$; orange
between red and yellow on the contour map) shapes a non-conical
geometry, which is quantitatively similar to the parabolic outermost
streamline (thick solid line) $z \propto R^{1.6}$ (BP82; see Figure
\ref{fig:Ergo}).

After this phase, the MRI-driven turbulence in the wind/corona region
above the RIAF body decays gradually. The {\em bottom} panel shows the
final stage ($t/t_{\rm g}=9000$) of the system; the turbulent structure
is saturated in the wind/corona region, but it survives near the equator
in the RIAF body. It is notable that the BP82-type parabolic structure
of the funnel jet-corona/wind boundary is still sustained until this
phase (unchanged at the quantitative level during $t/t_{\rm g} \simeq
3000$--9000), suggesting the PFD funnel jet is entering a quasi-steady
state, while the outside region (wind/corona and RIAF body) still
evolves dynamically. The distribution of $b^{2}/\rho$ can be divided
into the following three regions; i) the funnel ($\gtrsim 1$), ii) the
wind/corona ($\simeq 10^{-3}$--$10^{-1}$), and iii) the RIAF body
($\lesssim 10^{-3}$), respectively. Thus, we clearly identified that the
PFD funnel jet (orange to red) is outlined with the BP82-type parabolic
shape, rather than the genuine parabolic shape (BZ77: $z \propto
R^{2}$). It is also notable that the boundary of the funnel follows
$b^{2}/\rho \simeq 1$ during the whole time of the quasi-steady state at
$t/t_{\rm g} \gtrsim 3000$.

Figure \ref{fig:FID-B-LINE-EVO} displays the poloidal magnetic field
line distribution at the same times chosen for Figure
\ref{fig:FID-MAG-EVO}. We can see that the ordered, large-scale poloidal
magnetic flux only exists inside the PFD funnel jet region where
$b^{2}/\rho \gtrsim 1$ (Figure \ref{fig:FID-MAG-EVO}) during the
quasi-steady state $t/t_{\rm g} \gtrsim 3000$. There seems to be no such
coherent poloidal magnetic flux penetrating the equatorial plane at $R >
r_{\rm H}$. This is also examined in \citet[]{K05, KM07}. At the stage
$t/t_{\rm g}=3000$, a lateral alignment of the poloidal magnetic flux
ends at around the outermost parabolic streamline $z \propto
R^{1.6}$. This holds until the final stage of $t/t_{\rm g}=9000$. Thus,
the distribution of poloidal magnetic field lines also indicates the
funnel interior reaches a quasi-steady state with insignificant
deviation when $t/t_{\rm g} \gtrsim 3000$.

The density of contours in Figure \ref{fig:FID-B-LINE-EVO} directly
represents the poloidal field strength (it may be a quantitatively
reasonable interpretation at least in the funnel area). At the interior
of the funnel jet, the lateral spacing of each field line decreases
($R/r_{\rm g} \rightarrow 0$) at around the event horizon ($r/r_{\rm g}
\lesssim$ a few), suggesting an accumulation of the poloidal flux around
the polar axis ($z$). This is caused by the enhanced magnetic hoop
stress by the toroidal field component and it is prominent if the black
hole spin becomes large \citep[$a \rightarrow 1$;][]{TNM10}. On the
other hand, we may also see this effect in the downstream region
($r/r_{\rm g} \lesssim 20$) along the polar axis, but a concentration of
the poloidal flux is rather smooth and weak compared with the innermost
region around the event horizon. It indicates that the toroidal field
does not yet fully dominate the poloidal one on this scale and no
effective bunching of the poloidal flux takes place (cf. Figure
\ref{fig:SPIN-B-LINE} for a visible inhomogeneity further downstream).
Figure \ref{fig:FID-B-FIELD-BUNCH} confirms this quantitatively; a
concentration of the poloidal magnetic field becomes clear as $r$
increases (at $r/r_{\rm g} \gtrsim 20$ and $\theta \lesssim
20^{\circ}$). We can also identify the gradual decrease of $|B^{r}|$ as
a function of $\theta$, implying a differential bunching of the poloidal
flux. Further examinations are presented in Section
\ref{sec:RESULTS-PARA2} (behaviors at the downstream in our large domain
computations with different black hole spins).

No visible (but very weak) bunching of poloidal magnetic flux at a few
$\lesssim r/r_{\rm g} \lesssim 20$ (see {\em middle} and {\em bottom}
panels in Figure \ref{fig:FID-B-LINE-EVO}) indicates that the local
poloidal field can be approximately treated as a force-free system
\citep[][]{NLT09}. It is worth noting that the radial ($r$) distribution
of $b^{2}/\rho$ inside the funnel jet has a weak dependence on the
colatitude angle ($\theta$), as is shown in Figure
\ref{fig:FID-MAG-EVO}. This also implies no effective bunching of the
poloidal flux as well as no concentration of the mass density toward the
polar axis. We consider that a quasi-uniform stratification of
$b^{2}/\rho$ inside the funnel (near the event horizon) plays a critical
role in determining the dynamics of the GRMHD jet (the lateral
stratification of the bulk acceleration), as is treated in Sections
\ref{sec:RESULTS-PARA2}, \ref{sec:RESULTS-PARA3}, and
\ref{sec:SPINE-SHEATH}.

In order to confirm the boundary between the PFD funnel jet and the
wind/corona region at a quantitative level, we provide Figure
\ref{fig:FID-PR_TOT-LINES}; Contours of $b^{2}/\rho=1$ and $\beta_{\rm
p}=1$ are shown, at the final stage of $t/t_{\rm g}=9000$ during the
long-term, quasi-steady state ($t/t_{\rm g} \gtrsim 3000$) in our
fiducial run. It is similar to Figure 2 in \citet[]{MG04} and Figures 3
in \citet[]{M06}. Note that their snapshots correspond to $t/t_{\rm g}
\leq 2000$, which is before when we find the funnel reaches a
quasi-steady state. Notably, the equipartition of these quantities
($b^{2}/\rho=1$ and $\beta_{\rm p}=1$) is maintained at the funnel edge
along the outermost BP82-type parabolic streamline. There are other
contours of $\beta_{\rm p}=1$, which are distributed outside the funnel
and above the equatorial plane. Quantitatively, the latter corresponds
to the boundary between the wind/corona and the RIAF body
\citep[$\beta_{\rm p}=3$; e.g.][]{MG04, M06}.

We also identify the boundary between the PFD funnel jet and the
external area (wind/corona and RIAF body) regarding the unbound/bound
state of the fluid to the black hole by means of the Bernoulli parameter
\citep[e.g.][]{DHK03, MG04}: $Be=-h u_{t}-1$, where $h=1+(u+p)/\rho$ is
the relativistic specific enthalpy and $u_t$ is the covariant time
component of the plasma 4-velocity. In the fiducial run (and other runs
as well in the next section), we confirm $p/\rho \lesssim 10^{-1}$
throughout the computational domain. Thus, we can approximately use $h
\approx 1$ and $Be \approx -u_{t}-1$, which is adopted throughout this
paper. A fluid is gravitationally unbound and can escape to infinity for
$Be>0$, and vice versa. The contour of $-u_{t}=1 \ (Be \approx 0)$ is
shown in Figure \ref{fig:FID-PR_TOT-LINES}, forming ``{\sf V}''-shaped
geometry \citep[originally found in][]{MG04}. We can clearly find the
bound region $Be < 0$ inside the PFD funnel (close to the black hole)
and whole outside (throughout the computational domain $r/r_{\rm g} \leq
40$). It is a well-known issue, though we emphasize here, that the
outgoing mass flux in the PFD funnel jet does {\em NOT} come from the
event horizon, whereas the Poynting flux is extracted from there via the
Blandford-Znajek process \citep[]{BZ77} as is widely examined
\citep[e.g.][]{MG04, M06, GL13, PNHMWA15}.

As investigated above, our fiducial run provides the boundary condition
of the funnel edge (at a quantitative level);
\beqn
\label{eq:BND_FUNNEL}
b^{2}/\rho \simeq 1, \
\beta_{\rm p} \simeq 1, \
{\rm and} \
-u_{t} \simeq 1 \ (Be \approx 0)
\eeqn
along the outermost BP82-type parabolic ($z \propto r^{1.6}$) streamline
(the ordered, large-scale poloidal magnetic field line), which is
anchored to the event horizon with the maximum colatitude angle
$\theta_{\rm fp} \simeq \pi/2$ (at the footpoint). There is a discrepant
distribution of these quantities in previous results \citep[]{MG04,
M06}. We speculate this may be just because the funnel does not reach a
quasi-steady state. Note that an alignment of the funnel edge along the
specific streamline depends weakly on the initial condition
(such as $\beta_{\rm p0, min}$); formation of a BP82-type parabolic
funnel is confirmed for a reasonable range of $50 \lesssim \beta_{\rm
p0, min} < 500$ under the fixed Kerr parameter ($a=0.9$) up to
$r_{\rm out}/r_{\rm g}=100$ (see Appendix A). We also note that a
qualitatively similar structure of the low-density funnel edge of
$b^{2}/\rho \simeq 1$ , which is anchored to the event horizon with a
high inclination angle $\theta > 80^{\circ}$, is identified in
three-dimensional (3D) GRMHD simulations with $a=0.5$
\citep[e.g.][]{R17}. Thus, we suggest our finding may not depend on the
dimensionality (2D or 3D).

\begin{figure}
 \centering\includegraphics[scale=.8, angle=0]{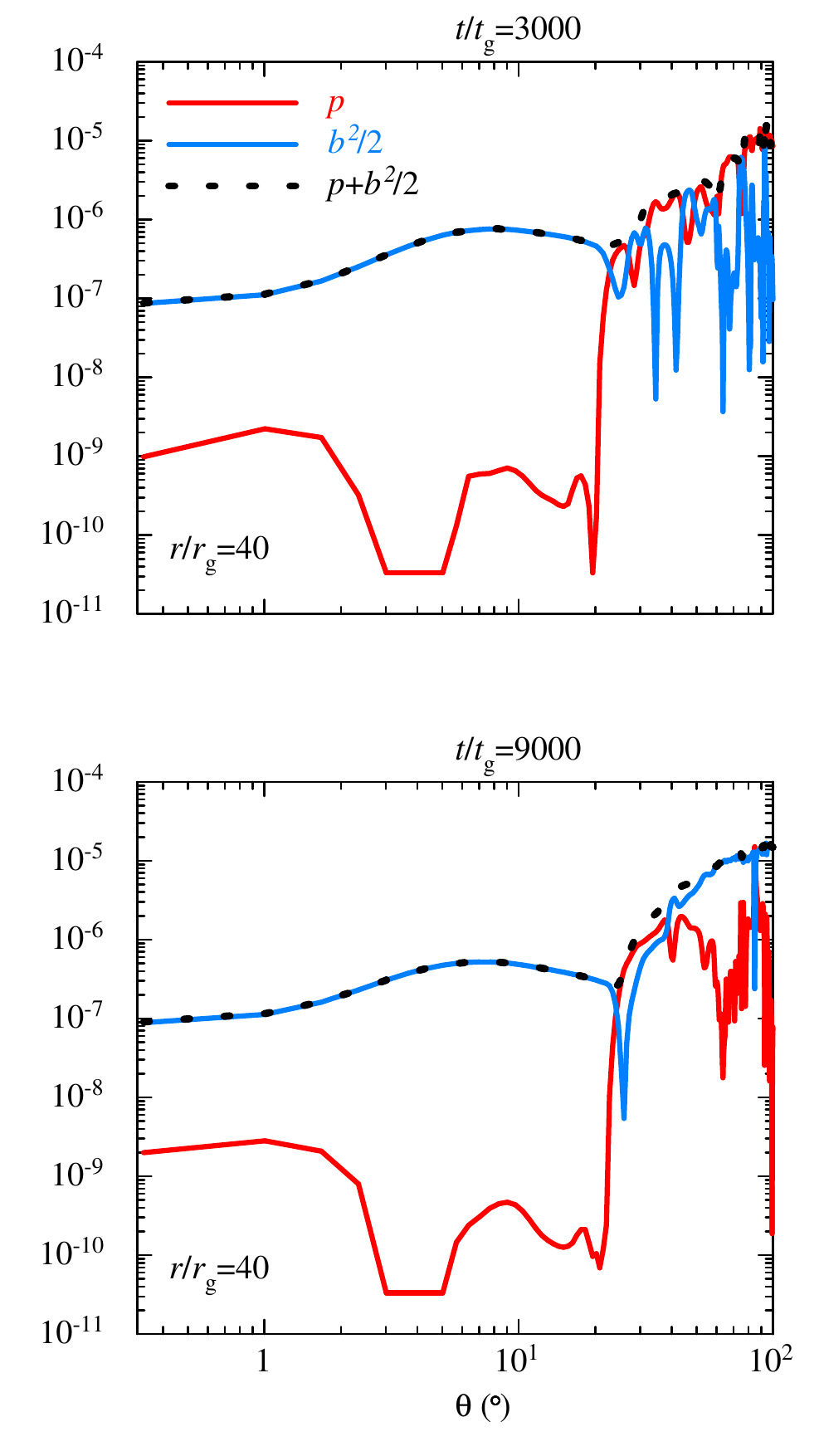}
 \caption{\label{fig:FID-PRS}
 A $\theta$ cross-section at $r/r_{\rm g}=40$ showing the
 gas pressure (red solid line), the magnetic pressure (blue solid line),
 and their sum (the total pressure: black dotted line) in the fiducial
 run ($a=0.9375$); $t/t_{\rm g}=3000$ ({\em top}) and 9000 ({\em
 bottom}).}
\end{figure}

We further examine the distribution of the total (gas + magnetic)
pressure, which is underlaid in Figure \ref{fig:FID-PR_TOT-LINES}. It
gives a good sense of the jet confinement by the ambient medium.
External coronal pressure outside the funnel, which
consists of both gas and magnetic contributions ($\beta_{\rm p} \simeq
1$), is surely over-pressured with respect to the magnetic
pressure-dominated region ($\beta_{\rm p} \ll 1$) inside the funnel. The
situation seems unchanged even after the MRI-driven MHD
turbulence saturates as is shown in Figure \ref{fig:FID-PRS}.  In the
vicinity of the black hole ($r/r_{\rm g} \lesssim 5$), the total
pressure of the RIAF body, which is dominated by the gas pressure
($t/t_{\rm g} \lesssim 3000$) or in an equipartition ($\beta_{\rm p}
\sim 1$; $t/t_{\rm g} \gtrsim 3000$). The important point of Figure
\ref{fig:FID-PRS} is that the funnel pressure (magnetically dominated)
and the coronal pressure ($\beta_{\rm p} \simeq 1$) are almost unchanged
during $t/t_{\rm g} \simeq 3000$--9000, suggesting the quasi-steady
parabolic shape of the funnel is sustained. Some convenient
terminology is provided in Figure \ref{fig:FID-PR_TOT-LINES} for
dividing the domain into the funnel ($b^{2}/\rho \gg 1, \ \beta_{\rm p}
\ll 1$), corona ($b^{2}/\rho \simeq 10^{-2}, \ \beta_{\rm p} \simeq 1$),
and RIAF body ($b^{2}/\rho \ll 10^{-2}, \ \beta_{\rm p} \gtrsim 1$),
following the literature \citep[][]{DHK03, MG04, HK06, S13}.

We realize, {\em however}, that the ram pressure of the accreting gas
becomes even higher than the total pressure in the RIAF body
on the equatorial plane (by almost an order of
magnitude). This is conceptually similar to the so-called ``magnetically
arrested disk'' \citep[MAD: e.g.][]{NIA03, I08, TNM11} although the
accretion flow in our \texttt{HARM} fiducial run is identified as the
``standard and normal evolution'' \citep[SANE: ][]{N12} without having
an arrested poloidal magnetic flux on the equatorial plane at $R >
r_{\rm H}$ (see Figure \ref{fig:FID-B-LINE-EVO}). Anyhow, as a
consequence of this combination of effects, we could expect that the PFD
jet to be deformed into a non-conical geometry.  Note that the funnel
structure becomes radial (i.e., conical) if the magnetic pressure in the
funnel is in equilibrium with the total pressure in the corona and the
RIAF body in 3D GRMHD simulations \citep[e.g.][]{H04}. We also remark
that the low-density funnel edge with $b^{2}/\rho \simeq 1$ is
established even in the MAD state with 3D runs \citep[]{R17}.

\begin{figure}
 \centering\includegraphics[scale=.37, angle=0]{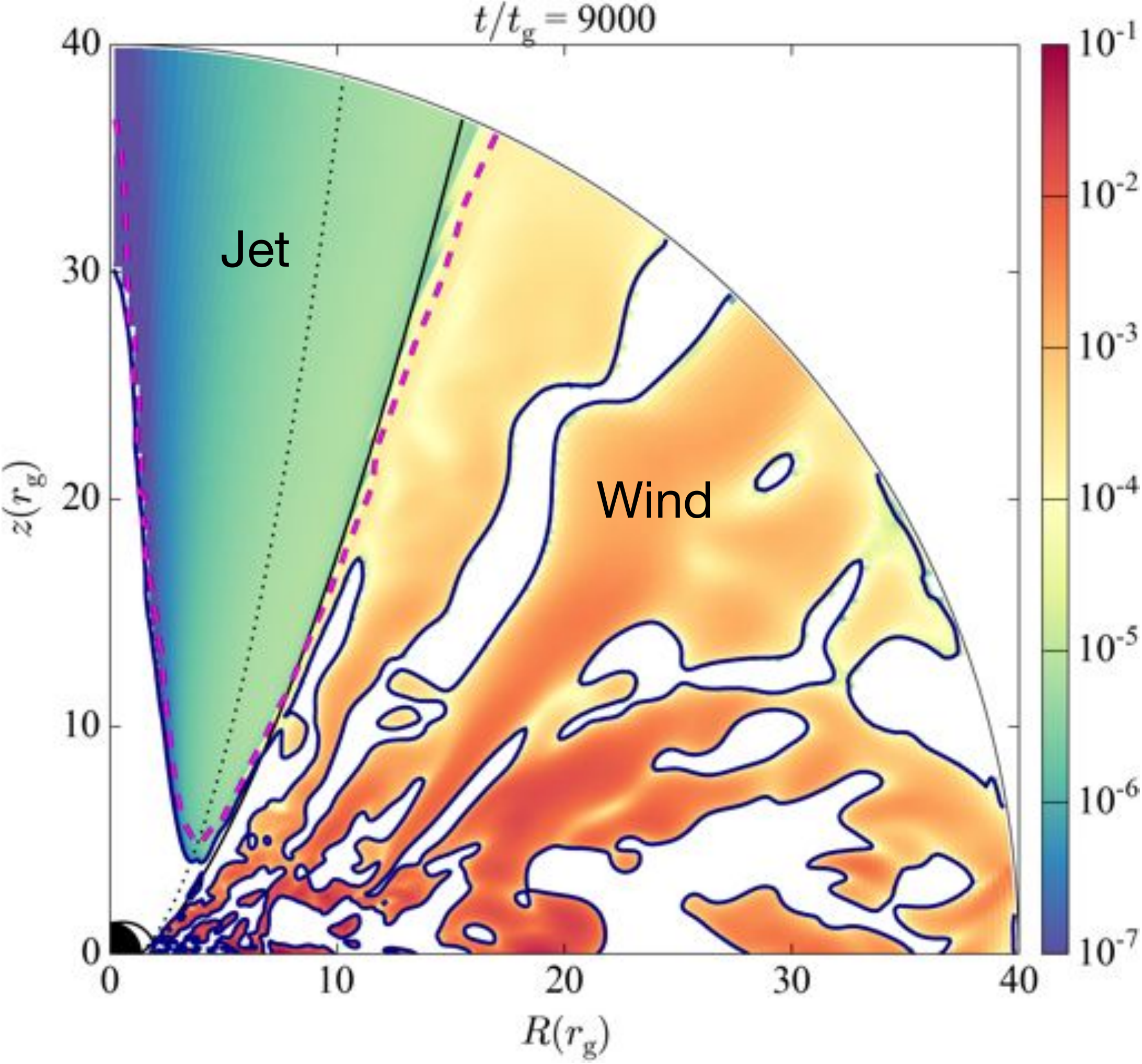}
 \caption{\label{fig:FID-MASSFLUX-U^r} The final snapshot of the
 fiducial run ($a=0.9375$); $t/t_{\rm g}=9000$. The magnitude of the
 outgoing radial mass flux density $\sqrt{-g}\rho u^{r} (>0)$, where
 $u^{r}$ is the radial component of the four-velocity, $\sqrt{-g}=\Sigma
 \sin \theta$ is the metric determinant, and $\Sigma\equiv r^{2}+a^{2}
 \cos^{2} \theta$, is shown with a color filled contour (the upper
 computational domain; $0 \leq \theta \leq \pi/2$). Contours with navy
 solid lines show $u^{r}=0$, while ``whiteout'' regions indicate the
 magnitude of the ingoing radial mass flux density $\sqrt{-g}\rho u^{r}
 (<0)$. The jet stagnation surface is clearly displayed inside the PFD
 funnel ($u^{r}=0$). $-u_{t}=1 \ (Be \approx 0)$ is shown with a magenta
 dashed line. Other components are identical to those in Figure
 \ref{fig:FID-MAG-EVO}.}
\end{figure}

Finally, Figure \ref{fig:FID-MASSFLUX-U^r} provides further examination
of outflows (and inflows as well) in the system. The outgoing radial
mass flux density exists both inside and outside the funnel, but that in
the corona is significantly higher than the funnel with $\sim 1$--3
orders of magnitude, which is quantitatively consistent with other 3D
simulations in the SANE state \citep[e.g.][]{S13,Y15}. Aside from the
terminology in \citet[]{MG04}, we consider outflows in the funnel as
jets, while those in the corona as winds \citep[e.g.][]{S13}, as is
labeled. The former is highly magnetized with considerable poloidal
magnetic flux and powered by the spinning black hole, but the latter is
not (see, Figures \ref{fig:FID-MAG-EVO} and \ref{fig:FID-B-LINE-EVO}). A
division boundary of these outflows lies on $Be \approx 0$ contour along
the outermost BP82-type parabolic streamline. That is, the black
hole-driven PFD funnel jet is unbound, but the RIAF-driven coronal wind
is bound (at least in our simulation up to $r_{\rm out}/r_{\rm g}=100$).

In the coronal wind (bound), a considerable mass is supplied from the
RIAF and thus the outflow does not possess sufficient energy to overcome
the gravitational potential. On the other hand, the funnel jet
(unbound), which carries very little mass, becomes relativistic quite
easily (due to the magnetic acceleration along the poloidal magnetic
field) and could escape to infinity. However, the bound wind may not
exist permanently at large distance; $Be$ is presumably not a constant
(in the non-steady flow) and the sign can be positive with a 3D
turbulent environment \citep[]{Y15}. The jet stagnation surface in the
PFD funnel, at which the contravariant radial component of the plasma
4-velocity becomes zero \citep[$u^{r}=0$;][]{M06, PNHMWA15, BT15}, is
clearly identified.

In Figure \ref{fig:FID-MASSFLUX-U^r}, we can see that the jet stagnation
surface does reasonably coincide with the bound/unbound boundary: $Be
\approx 0 \ (-u_{t}=1)$ except for the polar region ($\theta \sim 0$) of
the {\sf V}-shaped geometry. Outside the PFD funnel jet, the coronal
wind can be identified with the outgoing radial mass flux density , but
there is an accreting gas (identified by the ingoing radial mass flux
density) around $\theta \simeq \pi/4$. On the other hand, there is the
outgoing gas in the RIAF body at $\pi/3 \lesssim \theta \lesssim 2/3\pi$
(see Figure \ref{fig:FID-PR_TOT-LINES}). This is not a wind, but
associated with turbulent motions. Thus, both inflows and outflows are
mixed up in the corona and RIAF body, suggesting the adiabatic
inflow-outflow solution \citep[ADIOS:][]{BB99, BB04}. Note that $u^{r}$
does not vanish along the jet/wind boundary and thus the wind plays a
dynamical role in confining the PFD jet (see also Figure
\ref{fig:SPIN-MASSFLUX-U^r}).

\begin{figure*}[!h]
 \centering\includegraphics[scale=.66, angle=0]{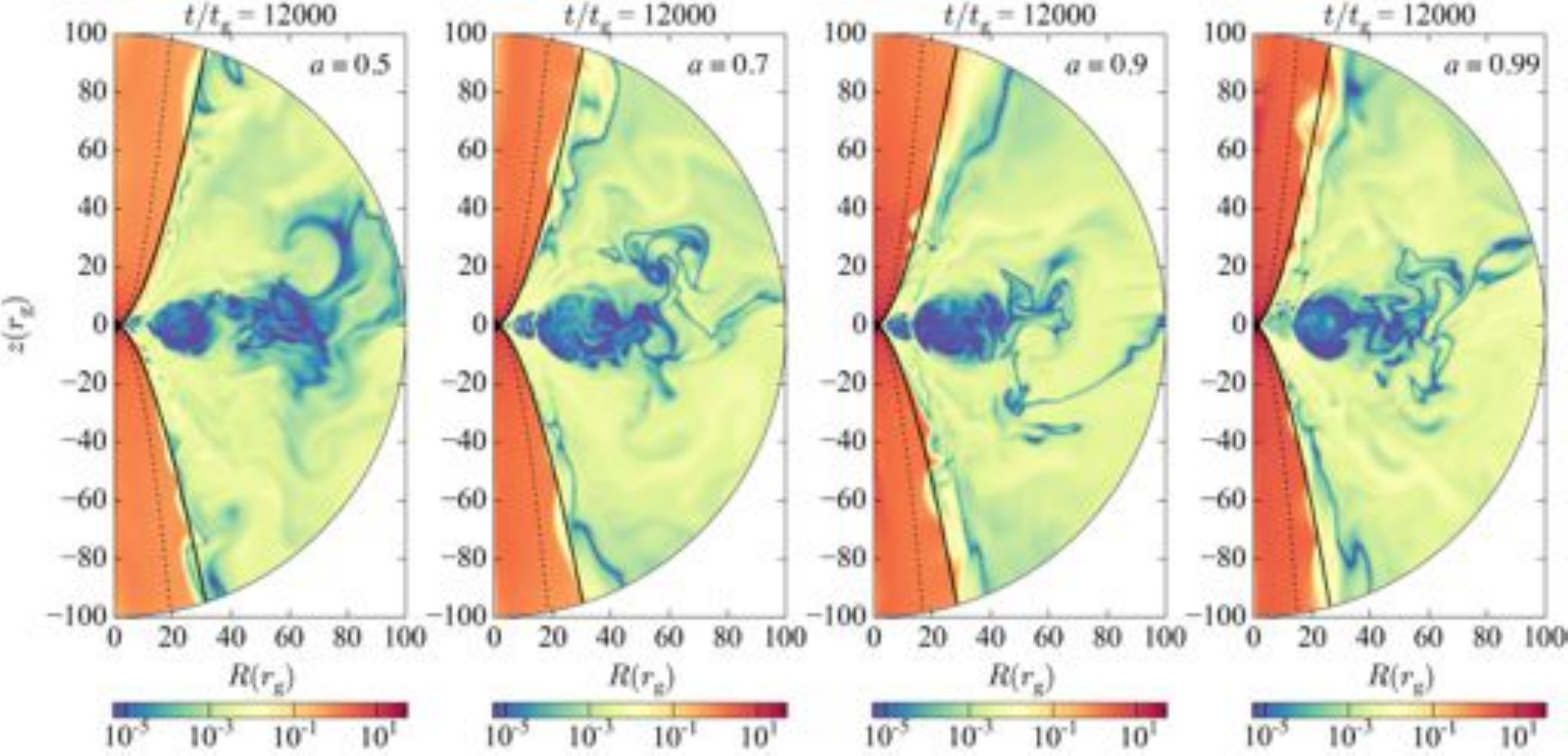}
 \caption{\label{fig:SPIN-MAG} A color filled contour of $b^{2}/\rho$
 for four different runs with different black hole spins (from {\em
 left} to {\em right}: $a=0.5,\, 0.7,\, 0.9$, and 0.99). The final
 snapshot ($t/t_{\rm g}=12000$) is displayed for each run with the whole
 computational domain $r/r_{\rm g} \leq 100$ and $0 \leq \theta \leq
 \pi$. Other components in panels are identical to those in Figure
 \ref{fig:Ergo}, but the black hole spin is adjusted.}
\end{figure*}

\begin{figure*}[!h]
 \centering\includegraphics[scale=.65, angle=0]{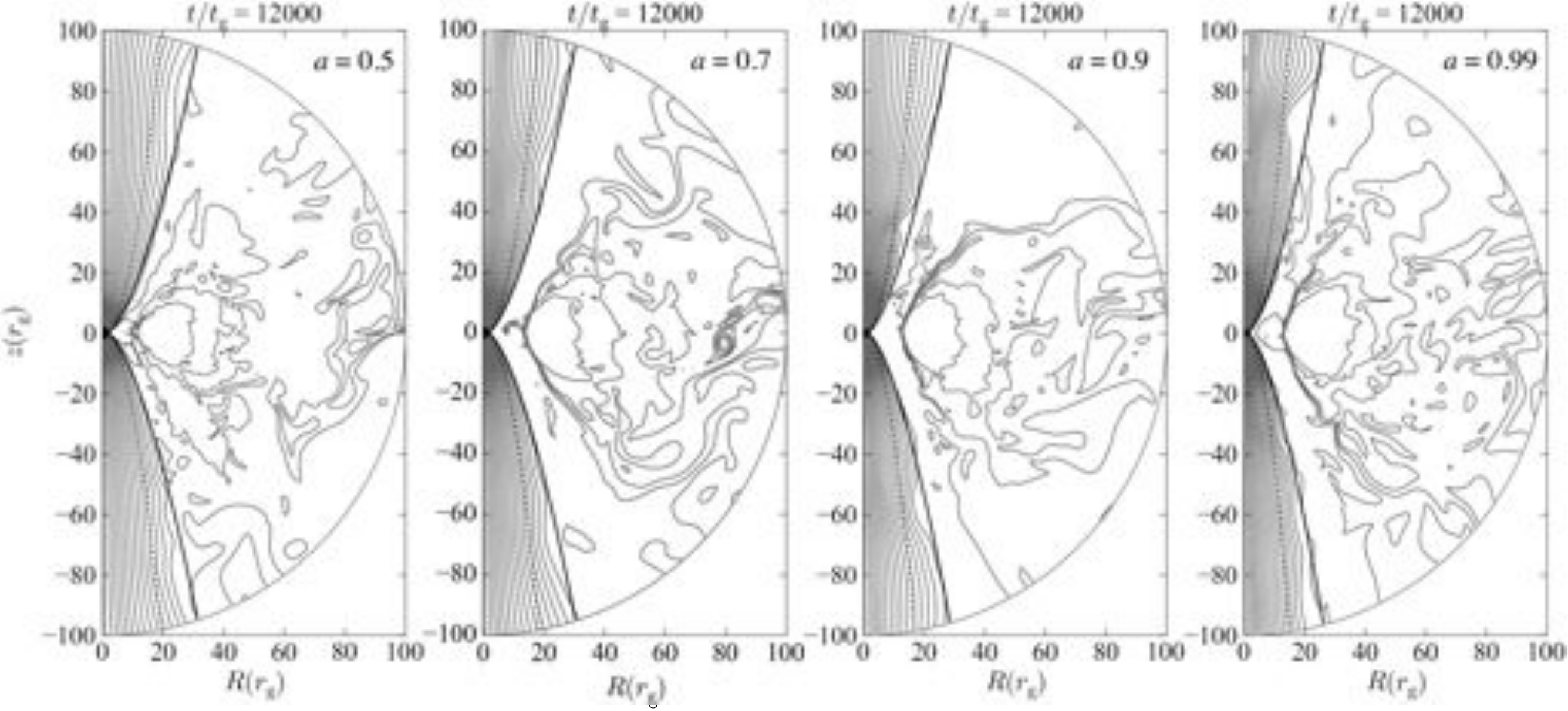}
 \caption{\label{fig:SPIN-B-LINE} Contours (gray) show poloidal magnetic
 field lines for four different runs with different black hole spins
 (from {\em left} to {\em right}: $a=0.5,\, 0.7,\, 0.9$, and 0.99). The
 final snapshot ($t/t_{\rm g}=12000$) is displayed for each run with the
 whole computational domain $r/r_{\rm g} \leq 100$ and $0 \leq \theta
 \leq \pi$. Other components in panels are identical to those in Figure
 \ref{fig:SPIN-MAG}.}
\end{figure*}

The origin of the wind is beyond the scope of our consideration here,
but the magneto-centrifugal mechanism (BP82) would be unlikely to
operate; it is because of the absence of a coherent poloidal magnet
field outside the PFD funnel (see Figure \ref{fig:FID-B-LINE-EVO}),
where the toroidal magnetic field is dominant and the plasma is not
highly magnetized ($b^{2}/\rho \ll 1$ and $\beta_{\rm p} \gtrsim 1$; see
Figures \ref{fig:FID-MAG-EVO} and \ref{fig:FID-PR_TOT-LINES}). Note that
a dominant toroidal magnetic field may be also true in the SANE state
with 3D runs \citep[e.g.][]{H04}. Regarding the funnel-wall jet
\citep[]{HK06}, which could be driven by a high-total pressure corona
squeezing material against an inner centrifugal wall, does not seem to
appear in our fiducial simulation. We do not find any significant
evidence of it; no pileup of the mass flux and/or the total pressure at
the funnel edge along the BP82-type parabolic outermost
streamline. There is a key difference between the coronal wind and the
funnel-wall jet; the former is bound, while the later is unbound at
least in the vicinity of the black hole \citep[]{DHK03}. The outer
boundary of the matter-dominated coronal wind ($b^2/\rho < 0.1$; see
Figures \ref{fig:FID-MAG-EVO} and \ref{fig:FID-MASSFLUX-U^r}) is
somewhat indistinct as is indicated by \citet[]{HK06}.

\begin{figure}[!h]
\centering\includegraphics[scale=.8, angle=0]{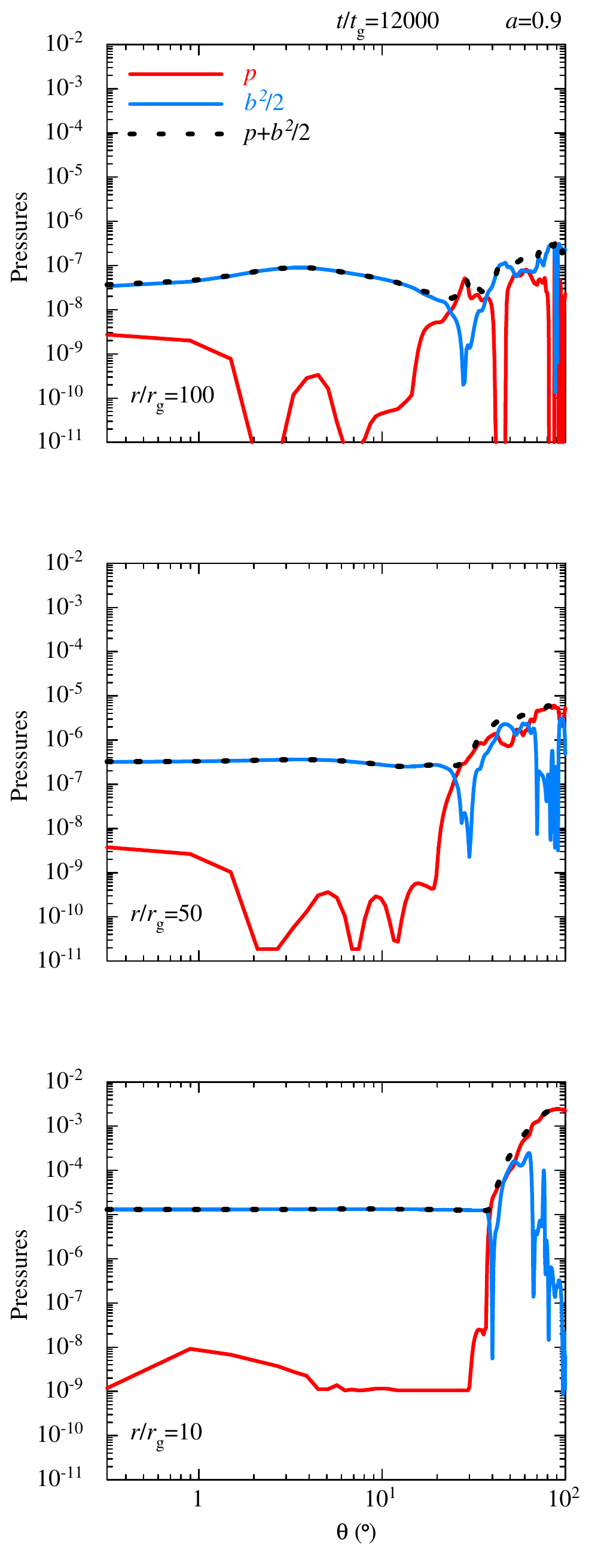}
\caption{\label{fig:SPIN-PRS} A $\theta$ cross-section
showing the gas pressure (red solid line), the magnetic pressure (blue
solid line), and their sum (the total pressure: black dotted line) at
the final stage ($t/t_{\rm g}=12000$) for $a=0.9$; $r/r_{\rm g}=100$
({\em top}), 50 ({\em middle}), and 10 ({\em bottom}).}
\end{figure}

\subsection{Parameter Survey: (In)dependence on Black Hole Spins}
\label{sec:RESULTS-PARA}
Based on our fiducial run, we further examine the BP82-type parabolic
structure ($z \propto R^{1.6}$) of the PFD funnel jet against the
varying black hole spin.  Different Kerr parameters are examined with
same value of $\beta_{\rm p0,min}=100$ in the extended computational
domain $r_{\rm out}/r_{\rm g} = 100$ (with a grid assignment $N_{x_{1}}
\times N_{x_{2}} = 512\times512$). We prescribe $r_{\rm max}/r_{\rm
g}$=12.95, 12.45, 12.05, and 11.95, for $a=0.5, 0.7, 0.9$, and 0.99,
respectively.

\subsubsection{Funnel Structure}
\label{sec:RESULTS-PARA1}
Figure \ref{fig:SPIN-MAG} exhibits $b^{2}/\rho$ at the final stage
$t/t_{\rm g}=12000$ for various black hole spins. First of all, we
confirmed that the overall structure outside the funnel seems to be
unchanged with different spins; $b^2/\rho \simeq 10^{-3}$--$10^{-1}$ is
obtained in the wind/corona region, while the RIAF body sustains
$b^{2}/\rho \lesssim 10^{-3}$. The MRI-driven turbulence has decayed in
both the wind/corona and RIAF body by this phase. The funnel boundary
fairly follows the outermost BP82-type parabolic streamline, which is
anchored to $r=r_{\rm H}$, but shifts inward with increasing $a$ (the
funnel becomes ``slimmer''). We find that $b^{2}/\rho \simeq 1$ is
maintained along the funnel edge for sufficiently high spins ($a \ge
0.9$), while the value is somewhat smaller---around $b^{2}/\rho \simeq
0.5$---for lower spins ($a \le 0.7$) in the quasi-steady phase;
$b^{2}/\rho \leq 5$ is obtained at the jet stagnation surface (see
Figure \ref{fig:SPIN-B-LINE-IOSS-MAG} in the next section) so that the
downstream of the funnel jet does not hold a highly PFD state.

\begin{figure*}[!b]
 \centering\includegraphics[scale=.66, angle=0]{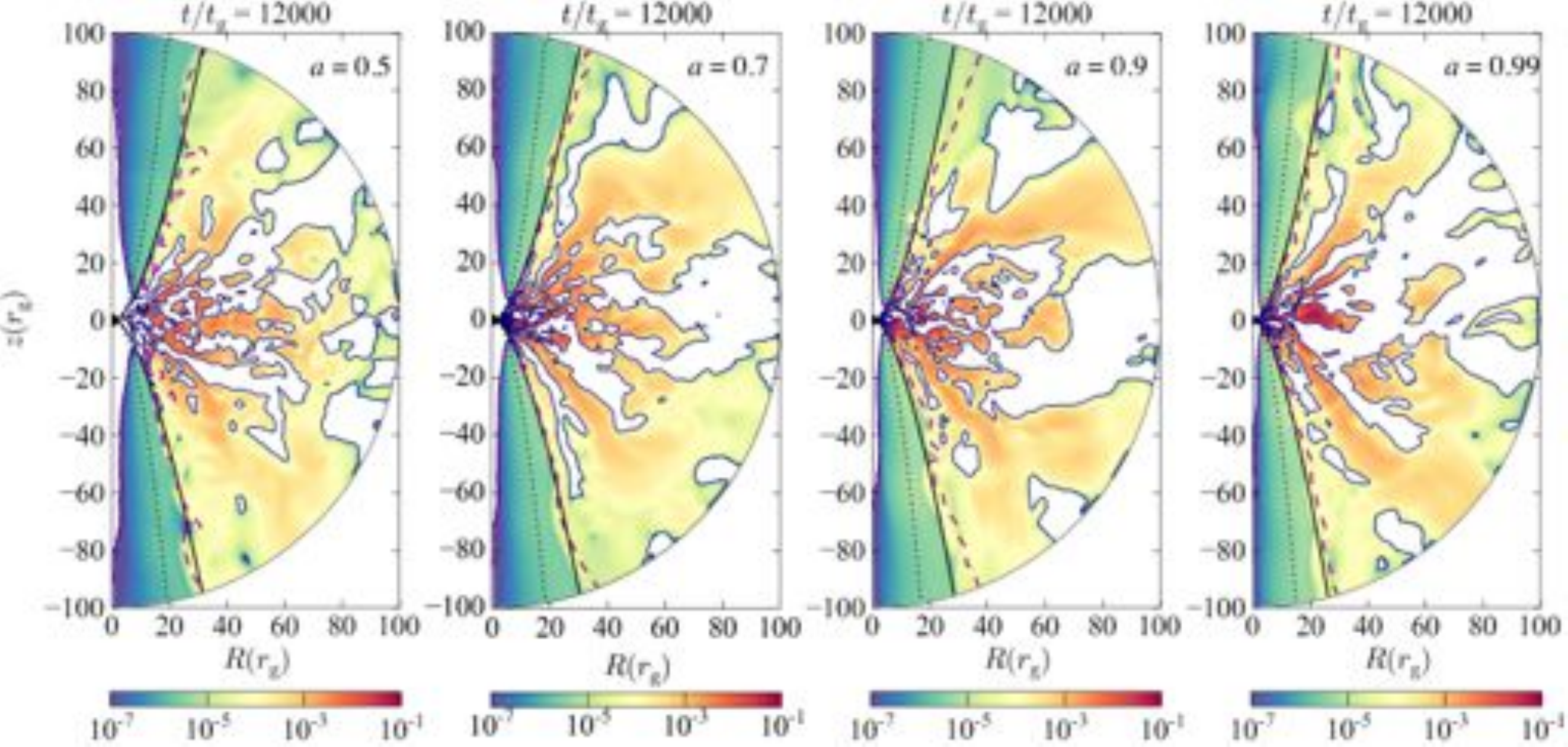}
 \caption{\label{fig:SPIN-MASSFLUX-U^r} A color filled contour of the
 magnitude of the outgoing radial mass flux density (similar to Figure
 \ref{fig:FID-MASSFLUX-U^r}) for four different runs with different
 black hole spins (from {\em left} to {\em right}: $a=0.5,\, 0.7,\,
 0.9$, and 0.99). The final snapshot ($t/t_{\rm g}=12000$) is displayed
 for each run with the whole computational domain $r/r_{\rm g} \leq 100$
 and $0 \leq \theta \leq \pi$. Navy solid line shows $u^{r}=0$, while
 ``whiteout'' regions indicate the magnitude of the ingoing radial mass
 flux density. The jet stagnation is clearly displayed inside the PFD
 funnel ($u^{r}=0$), and it shifts towards the black hole when $a$
 increases. $-u_{t}=1 \ (Be \approx 0)$ is shown with a magenta dashed
 line. Other components in panels are identical to those in Figure
 \ref{fig:SPIN-MAG}.}
\end{figure*}

\begin{figure*}[!t]
 \centering\includegraphics[scale=.66, angle=0]{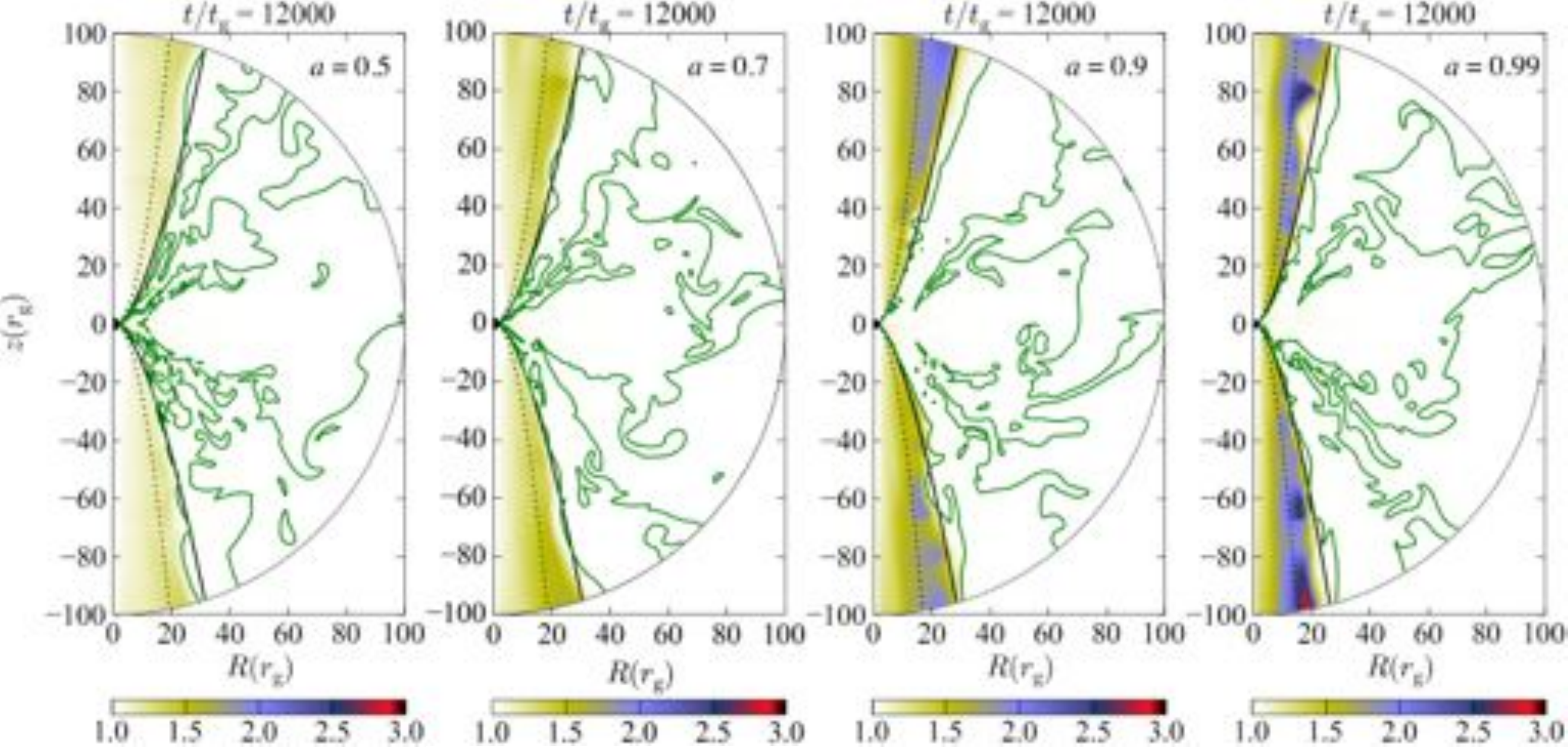}
 \caption{\label{fig:SPIN-LORENTZ} A color filled contour of the Lorentz
 factor $\Gamma$ (only where $u^{r}>0$) for four different runs with
 different black hole spins (from {\em left} to {\em right}: $a=0.5,\,
 0.7,\, 0.9$, and 0.99). The final snapshot ($t/t_{\rm g}=12000$) is
 displayed for each run with the whole computational domain $r/r_{\rm g}
 \leq 100$ and $0 \leq \theta \leq \pi$. Green solid lines show
 $\beta_{\rm p}=1$ (in the fluid frame). Other components in panels are
 identical to those in Figure \ref{fig:SPIN-MAG}.}
\end{figure*}

We report one obvious, but notable dependence on the black hole spin $a$
during the time evolution of the system in Figure \ref{fig:SPIN-MAG};
$b^2/\rho$ in the funnel grows larger when $a$ increases. This implies
that a large spin would be suitable for obtaining a large value of
$b^2/\rho$ at the jet launching region, which determines the maximum
Lorentz factor \citep[e.g.][see also Figure
\ref{fig:SPIN-B-LINE-IOSS-MAG} for reference values]{M06}. Thus, we may
consider the black hole spin dependence on the asymptotic Lorentz factor
at large distance although another factor (i.e., the magnetic nozzle
effect) may also affect this. Beneath the jet stagnation surface, there
is an inflow $u^{r} < 0 \ (Be < 0)$ along each streamline because the
fluid is strongly bound by the hole's gravity. Thus, a shift of the jet
stagnation can be expected, depending on the black hope spin; the fluid
can escape from a deeper gravitational potential well of the hole due to
an enhanced magneto-centrifugal effect in a cold GRMHD flow
\citep[]{TNTT90, PNHMWA15} if $a$ becomes large. See Section
\ref{sec:RESULTS-PARA3} with Figures \ref{fig:SPIN-B-LINE-IOSS-MAG} for
more details.

On the other hand, we confirmed that the shape of the funnel exhibits a
weak dependence on the black hole spin ($a \geq 0.5$). As $a$ increases,
the angular frequency $\Omega$ of a poloidal magnetic field line
\citep[]{F37} increases where frame dragging is so large inside the
ergosphere that it generates a considerably larger toroidal magnetic
field. Consequently, a magnetic tension force due to hoop stresses
(``magnetic pinch'') would be expected to act more effectively on
collimating the funnel jet. It, however, is not the case at the funnel
edge; hoop stresses nearly cancel centrifugal forces, suggesting that a
self-collimation is negligible \citep[]{MN07a, MN07b}. Note that a
self-collimation will be effective at the funnel interior as $a$
increases (see below).

As is introduced in Section \ref{sec:FFE}, the outermost BP82-type
parabolic streamline ($z \propto R^{1.6}$) of the steady axisymmetric
FFE jet solution (NMF07, TMN08), which is anchored to the event horizon
with the maximum colatitude angle $\theta_{\rm fp}=\pi/2$, can suitably
represent the boundary (Equation \ref{eq:Psi_0}) between the PFD funnel
jet and the wind/corona region. The funnel edge can be approximately
defined as the location where $b^2/\rho \simeq 0.5$--1 along the
specific curvilinear shape. This is, at least in our GRMHD simulations,
valid on the scale of $r/r_{\rm g} \lesssim 100$ with a range of Kerr
parameters $a=0.5$--$0.99$ (due to the limited space, we omit to show
the case of $a=0.998$, but we confirmed similar structure of the PFD
funnel jet as is shown in Figure \ref{fig:SPIN-MAG}). Again, we report a
formation of the BP82-type parabolic funnel with $50
\lesssim \beta_{\rm p0, min} < 500$ ($a=0.9$) up to $r_{\rm
out}/r_{\rm g}=100$ in our GRMHD runs (see Appendix A).

Figure \ref{fig:SPIN-B-LINE} displays the poloidal magnetic field lines,
corresponding to each panel of Figure \ref{fig:SPIN-MAG}. We obtained
qualitatively similar results with Figure \ref{fig:FID-B-LINE-EVO} at
the final stage ($t/t_{\rm g}=9000$) and the overall feature is
unchanged with varying the black hole spin as is similar to Figure
\ref{fig:SPIN-MAG}. Again, there is no coherent structure of the
poloidal magnetic field lines in both the corona and RIAF body for all
cases: $a=0.5$-0.99. It is likely that the MHD turbulence (due to the
MRI) saturates in our axisymmetric simulations, whereas there are some
features of poloidal field tangling. As is widely examined in a 3D
simulation with similar initial weak poloidal field loops lying inside
the torus, the evolved field is primarily toroidal in the RIAF body and
corona \citep[e.g.][]{H04}; the magnetic energy of the coronal field is,
on average, in equipartition with the thermal energy ($\beta_{\rm p}
\simeq 1$) and the corona does not become a magnetically dominated
force-free state, which is quantitatively consistent with our results.

By contrast, in the funnel region ($b^2/\rho \gtrsim 1$) the field is
essentially ``helical'' with a dominant radial component outside the
ergosphere and a toroidal component that becomes larger with increasing
black hole spin \citep[see also][]{H04}. In their result, the magnetic
pressure in the funnel is in equilibrium with the total pressure in the
corona and the inner part of the RIAF body.  This is, however, different
form our results; as is shown in our fiducial run (Figure
\ref{fig:FID-PR_TOT-LINES}), the total pressure in the PFD funnel (i.e.,
the dominant magnetic pressure) is smaller than the outer total pressure
(especially at $r/r_{\rm g} \lesssim 20$-30, where the curvature of the
funnel edge is large). We understand that this is one of the reasons why
our PFD funnel jet is deformed into a paraboloidal configuration, rather
than maintaining the radial shape.

Figure \ref{fig:SPIN-PRS} shows that an excess of the external coronal
pressure (compared with the funnel pressure) gradually decreases as $r$
increases (as the curvature of the parabolic funnel becomes small). They
are almost comparable at $r/r_{\rm g}=100$, indicating that the pressure
balance is established, where the magnetic pressure inside the funnel
decays about two orders of magnitude compared with that at $r/r_{\rm
g}=10$. By combining with Figure \ref{fig:SPIN-MASSFLUX-U^r}, it is
implied that a coronal wind plays a dynamical role in shaping the jet
into a parabolic geometry.

An inhomogeneous spacing of poloidal magnetic field lines in the lateral
direction ($R$) at $z/r_{\rm g} \gtrsim 50$ (the tendency becomes strong
with $a \geq 0.7$) is confirmed in Figure \ref{fig:SPIN-B-LINE} (the
density of contours is high around the polar axis, while it becomes low
near the funnel edge). This is widely seen in the literature and
interpreted as the self-collimation by the magnetic hoop stress, which
collimates the inner part of streamlines relative to the outer part
\citep[e.g.][]{NLL06, KBVK07, KVKB09, TMN08, TMN09, TNM10}. This is
named as ``differential bunching/collimation'' of the poloidal magnetic flux
\citep[]{TMN09, KVKB09}, leading to a sufficient opening of neighboring
streamlines for an effective bulk acceleration via the magnetic nozzle
effect \citep[e.g.][for recent progress and references
therein]{TT13}. In the next section, further quantitative examination
of the differential bunching of the poloidal magnetic flux is presented
(see also Figure \ref{fig:SPIN-B-FIELD-BUNCH}), which is associated to
the jet bulk acceleration.

\subsubsection{Outflows}
\label{sec:RESULTS-PARA2} Next, we examine the nature of the coronal
wind in Figure \ref{fig:SPIN-MASSFLUX-U^r}, corresponding to each panel
of Figure \ref{fig:SPIN-MAG}. Again, we obtain qualitatively similar
results with Figure \ref{fig:FID-MASSFLUX-U^r} (the magnitude of the
outgoing radial mass flux density of the coronal wind is higher than
that of the funnel jet by $\sim 1$--3 orders of magnitude) even as the
black hole spin varies. This is consistent with the SANE state of 3D
GRMHD simulations, in which the magnitude of the outgoing radial mass
flux density of the coronal wind does not exhibit a noticeable
dependence on the black hole spin \cite[]{S13}. We also confirm the
distribution of $Be \approx 0 \ (-u_{t}=1)$ forms a {\sf V}-shaped
geometry as mentioned in Section \ref{sec:RESULTS-FID}. The right side
of the {\sf V}-shaped distribution of $Be \approx 0$ reasonably follows
the outermost BP82-type parabolic streamline, which is anchored to the
event horizon, up to $r_{\rm out}/r_{\rm g}=100$ for $a=0.5$--0.99
(although there is some deviation near the outer radial boundary in
$a=0.9$), suggesting that the coronal wind is bound up to $r_{\rm
out}/r_{\rm g} \sim 100$. It is notable that the jet stagnation surface
($u^{r}=0$) inside the funnel shifts towards the black hole as $a$
increases \citep[e.g.][]{TNTT90, PNHMWA15}, but it coincides with the
left side of the {\sf V}-shaped distribution of $Be \approx 0$ except
for the polar region ($\theta \sim 0$; see also Figure
\ref{fig:FID-MASSFLUX-U^r}).

Finally, we examine the velocity of the funnel jet. We evaluate the
Lorentz factor $\Gamma$, which is measured in Boyer-Lindquist
coordinates \citep[e.g.][]{M06, PNHMWA15};
\beqn
\Gamma=\sqrt{-g_{tt}} u^{t},
\eeqn
where
\beqn
&&g_{tt} \equiv - (\Delta-a^{2}\sin^{2}{\theta})/\Sigma, \nonumber \\
&&\Delta \equiv r^{2}-2r+a^{2}, \ \Sigma\equiv r^{2}+a^{2} \cos^{2}
\theta. \nonumber
\eeqn
We report another visible dependence on the black hole spin $a$ during
the time evolution of the system in Figure \ref{fig:SPIN-LORENTZ}, which
exhibits $\Gamma$ at the final stage $t/t_{\rm g}=12000$, corresponding
to each panel of Figure \ref{fig:SPIN-MAG}. It is notable that a high
value of $\Gamma$ ($\lesssim 1.5$ in $a=0.7$, $\lesssim 2.3$ in $a=0.9$,
and $\lesssim 2.8$ in $a=0.99$, respectively) is distributed between two
outermost streamlines ($z \propto R^{2}$ and $z \propto R^{1.6}$) of the
semi-analytical FFE jet, which are anchored to the horizon (Figure
\ref{fig:Ergo}).

Qualitatively similar results ($\Gamma$ becomes large at the outer layer
in the funnel, while $\Gamma \simeq 1$ is sustained at the inner layer)
are obtained in SRMHD simulations with a fixed jet boundary and a
solid-body rotation at the jet inlet, which provides a good
approximation of the behavior of field lines that thread the horizon
\citep[]{KBVK07, KVKB09}. Furthermore, the trend can be seen in FFE
\citep[$a=1$;][]{TMN08}, GRFFE \citep[$a=0.9$--0.99;][]{TNM10}, and
GRMHD \citep[$a=0.7$--0.98;][]{P13} simulations with modified initial
field geometries in the torus \citep[]{N12}. Note that the inner
``cylinder-like'' layer with a lower $\Gamma < 1.5$ for all spins arises
from a suppression of the magnetic nozzle effect; an effective bunching
poloidal flux takes place near the polar axis due to the hoop stresses
(see Figure \ref{fig:SPIN-B-LINE}) so that enough separation between
neighboring poloidal field lines is suppressed \citep[see
also][]{KBVK07, KVKB09}.

Another notable feature is the inhomogeneous distribution of $\Gamma$
(blobs) at the outer layer in the funnel; throughout the time evolution,
knotty structures are formed with $\Gamma \gtrsim 2$ when $a \geq 0.9$
(two {\em right} panels in Figure \ref{fig:SPIN-LORENTZ}). Blobs appear
at around the funnel edge $R/r_{\rm g} \gtrsim 10$ ($z/r_{\rm g} \gtrsim
20$) with $\Gamma \gtrsim 1.5$, which could be observed as a
superluminal motion $\gtrsim c$ with a viewing angle $\lesssim
25^{\circ}$. Note that no proper motion has been detected within a scale
of $100\,r_{\rm g}$ in de-projection in M87, where we interpret VLBI
cores as an innermost jet emission. We discuss the formation of blobs
and their observational counterparts in Section \ref{sec:VLBI-CORES2}.

An evolution of the Lorentz factor in a highly magnetized MHD/FFE
outflow can be expressed with the approximated formula (in the so-called
``linear acceleration'' regime\footnote{In a less-collimated parabolic
stream with $1 < \epsilon < 2$, an outflow also follows the so-called
``power-law acceleration'' regime, which exhibits a slower growth of
$\Gamma \approx \sqrt{3/(\epsilon-1)}(\epsilon/\theta) \propto
z^{(\epsilon-1)/\epsilon}$ than the linear acceleration \citep[see
TMN08;][for details]{KVKB09}. A transition (from linear to power-law
acceleration) depends on $\epsilon=2/(2-\kappa)$ and $\theta_{\rm fp}$
(the colatitude angle at the footpoint; i.e., the event horizon in this
paper).}):
\beqn
\label{eq:Gamma_FFE}
\Gamma \approx \sqrt{1+(R \Omega)^{2}} \approx R \Omega
\propto z^{1/\epsilon},
\eeqn
where
\beqn
\label{eq:Omega_F}
\Omega \approx 0.5 \Omega_{\rm H} (a).
\eeqn
$\Omega_{\rm H}$ denotes the angular
frequency of the black hole event horizon as
\beqn
\label{eq:Omega_H}
\Omega_{\rm H} (a)=\frac{ac}{2 r_{\rm H}}.
\eeqn
This formula is valid at a moderately relativistic regime $\Gamma
\gtrsim 2$ \citep[NMF07, TMN08]{MN07b, KBVK07, KVKB09} and/or all the
way out to large distances \citep[the jet radius is large enough: $R
\Omega \gg 1$ where the curvature of the magnetic surface is
unimportant;][]{BN06, BN09}. This can be expected at $z/r_{\rm g}
\gtrsim 100$ as is exhibited in the steady axisymmetric GRMHD solution
\citep[]{PNHMWA15}. Numerical simulations of the FFE jets in both
genuine parabolic and the BP82-type parabolic shapes provide a
quantitative agreement with the analytical solution (TMN08).

\begin{figure}[!h]
\centering\includegraphics[scale=.45, angle=0]{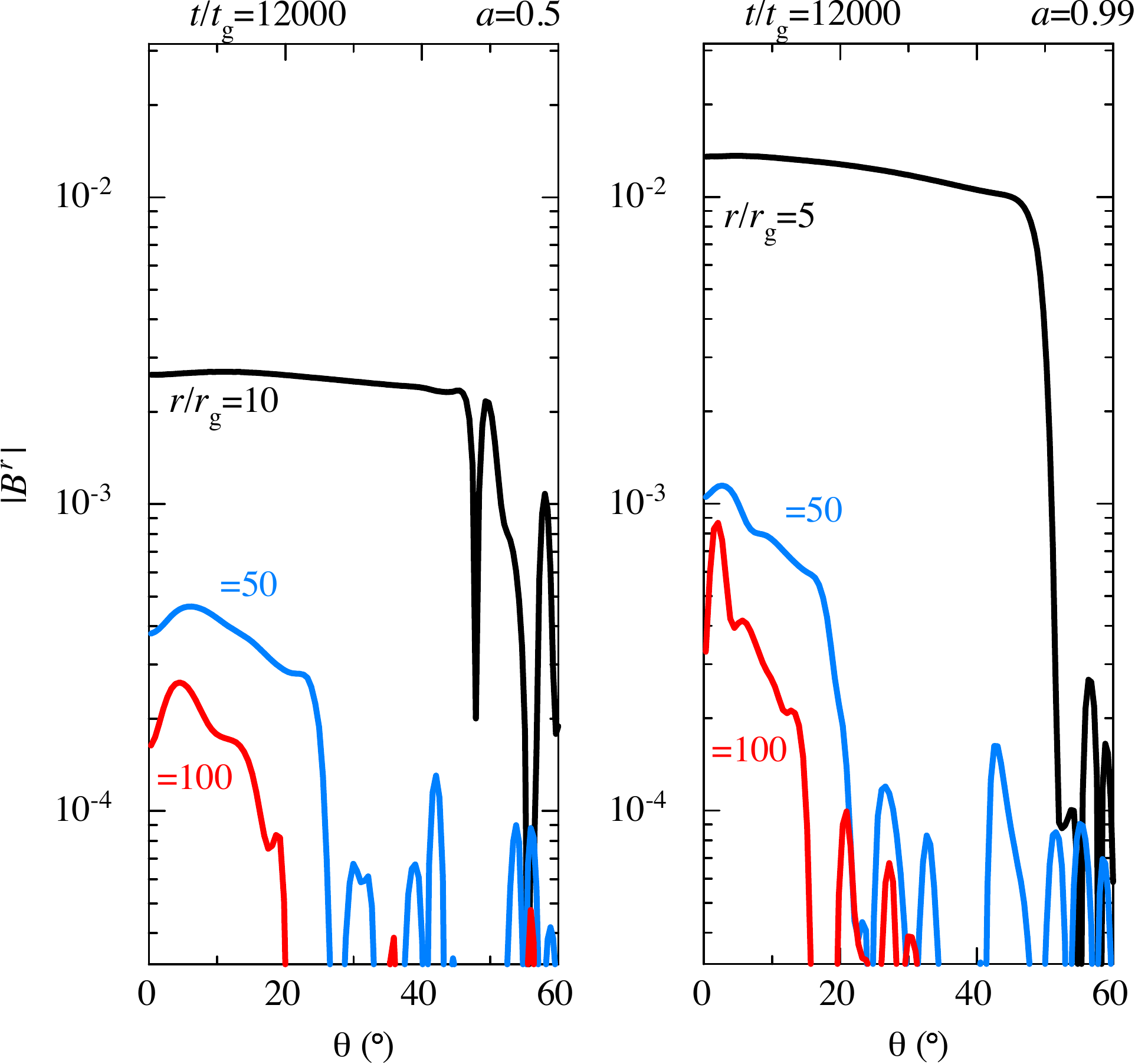}
\caption{\label{fig:SPIN-B-FIELD-BUNCH} A $\theta$ cross-section at
$r/r_{\rm g}=$10 (black, {\em left}) or 5 (black, {\em right}), 50
(blue), and 100 (red) showing the absolute value of the radial magnetic
field $|B^{r}|$ at $t/t_{\rm g}=12000$; $a=0.5$ ({\em left}) and
$a=0.99$ ({\em right}).}
\end{figure}

Based on our result shown in Figure \ref{fig:SPIN-LORENTZ}, at least up
to $r/r_{\rm g} = 100$, we could not find clear evidence of linear
acceleration in the funnel jet. This is presumably due to a lack of a
differential bunching/collimation of the poloidal magnetic flux as is
expected in a highly magnetized MHD/FFE regime \citep[e.g.][]{TMN09}
during the lateral extension of the funnel (in the regime of $R \Omega
\gg 1$). However, there are visible increases of distributed $\Gamma$ in
the funnel jet as $a$ increases. We can identify a physical reason in
Figure \ref{fig:SPIN-B-FIELD-BUNCH}; a concentration of the poloidal
magnetic field becomes strong as $r$ increases, but it is also enhanced
as a function of $a$ ($a=0.5 \rightarrow$ 0.99). As consequence of the
differential operation of the magnetic nozzle effect, a larger $\Gamma$
value in the funnel jet is obtained in the higher spin case. Note that
blobs (knotty structures) also appear near the funnel boundary at $a
\geq 0.9$ and we discuss this feature in Section \ref{sec:VLBI-CORES2}.

Contours of the equipartition $\beta_{\rm p}=1$ are also displayed in
Figure \ref{fig:SPIN-LORENTZ}. We can see that one of contours is
elongated near the outermost BP82-type parabolic streamline. Thus, our
boundary condition of the funnel edge (Equation \ref{eq:BND_FUNNEL}) is
moderately sustained up to $r_{\rm out}/r_{\rm g}=100$ (there is a
departure of $\beta_{\rm p}=1$ from the funnel edge in $a=0.9$, but this
seems to be a temporal and/or boundary issue). Figure
\ref{fig:SPIN-LORENTZ} also provides a clue of the velocity in the
corona/wind region. At the funnel exterior, no coherent poloidal
magnetic flux exists in the quasi-steady SANE state (Figure
\ref{fig:SPIN-B-LINE}). The weakly magnetized ($b^2/\rho \ll 1, \
\beta_{\rm p} \simeq 1$) coronal wind carries a substantial mass flux
compared with the funnel jet \citep[Figure \ref{fig:SPIN-MASSFLUX-U^r},
see also][]{S13, Y15}. Therefore it is unlikely that the coronal wind
could obtain a relativistic velocity; $\Gamma=1$ is sustained in all
cases with $a=0.5$--0.99 \citep[see also][with $a=0$]{Y15}. Note that
similar results are confirmed even in 3D simulations of the MAD state
\citep[e.g.][]{P13}, in which the coherent poloidal magnetic flux is
presumably arrested on the equatorial plane at $R > r_{\rm H}$
\citep[]{T15}.

With standard parameters in \texttt{HARM}, our GRMHD simulations provide
several interesting results, which does not depend on the black hole
spin when $a=0.5$-0.99. We, however, need further investigations to find
features such as a linear acceleration of the underlying flow with an
extended computational domain ($r_{\rm out}/r_{\rm g} > 100$).  Also,
one of the important issues is to investigate whether an unbound wind
could surely exist on large scales. If so, it indicates that $Be \simeq
0 \ (-u_{t} \simeq 1)$ will not hold at the jet/wind boundary. How the
BP82-type parabolic funnel jet could be maintained by an unbound wind
and/or other external medium? How the equipartition conditions $b^2/\rho
\simeq 0.5$--1 and $\beta_{\rm p} \simeq 1$ are maintained (or
modified)? These questions will be addressed in a forthcoming paper.

\subsubsection{Jet Stagnation Surface}
\label{sec:RESULTS-PARA3}

\begin{figure}[b]
\centering\includegraphics[scale=.59, angle=0]{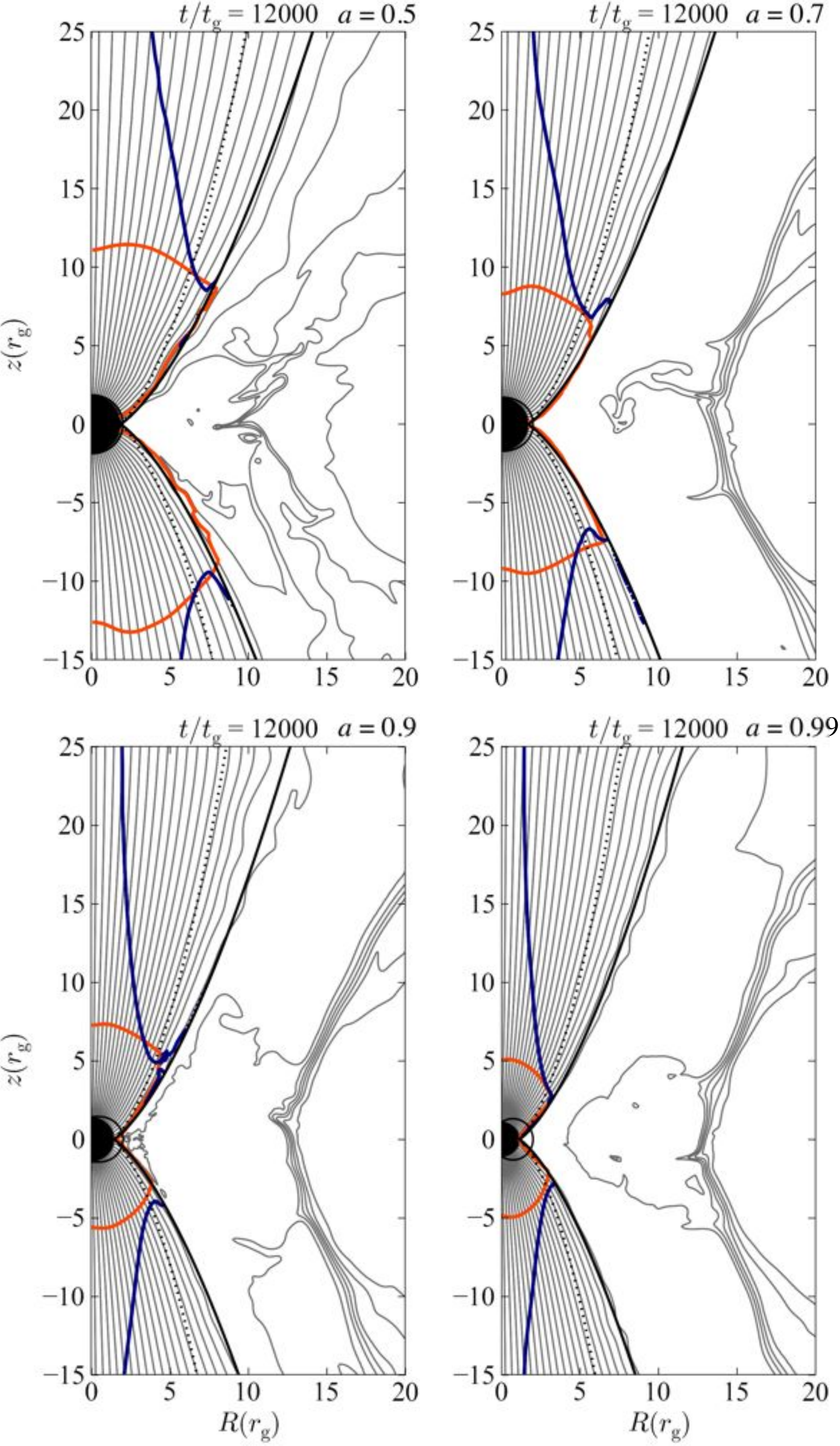}
\caption{\label{fig:SPIN-B-LINE-IOSS-MAG} Similar to Figure
\ref{fig:SPIN-B-LINE}, but the magnified view is shown for displaying
the poloidal magnetic field line around the black hole with the
computational domain $0 \leq R/r_{\rm g} \leq 20$ and $-15 \leq z/r_{\rm
g} \leq 25$. The jet stagnation surface: $u^{r}=0$ is drawn with navy
solid lines on each panel (from {\em upper-left} to {\em lower-right};
different black hole spins of $a=0.5, 0.7, 0.9$ and 0.99,
respectively). Orange solid lines on each panel show i) $b^{2}/\rho=2,
5, 10$, and 20, respectively (both at $z \lessgtr 0$).}
\end{figure}

We present Figure \ref{fig:SPIN-B-LINE-IOSS-MAG} to examine the jet
stagnation surface and the local value of $b^{2}/\rho$, with respect to
the black hole spin ($a$). As is also shown in Figure
\ref{fig:SPIN-MASSFLUX-U^r}, the jet stagnation surface ($u^{r}=0$
inside the funnel) shifts toward the black hole if $a$ increases
\citep[see, e.g.][for an analytical examination in the Kerr
space-time]{TNTT90} due to an increase of $\Omega$ (the outflow can be
initiated at the inner side), but qualitatively similar structures of
the surface are obtained ($a=0.5$--0.99) as is clearly seen in Figure
\ref{fig:SPIN-B-LINE-IOSS-MAG}.

Coherent poloidal magnetic field lines are regularly distributed inside
the funnel edge along the outermost parabolic streamline (BP82: $z
\propto r^{1.6}$), which is anchored to the event horizon. As the black
hole spin increases, the density of contours becomes high, indicating
that the poloidal field strength goes up. Especially, as is examined in
Figure \ref{fig:FID-B-LINE-EVO}, it is prominent at around the polar
axis ($z$) in the very vicinity of the event horizon ($r/r_{\rm g}
\lesssim$ a few) as $a \rightarrow 1$ \citep[e.g.][]{TNM10}. The closest
part (to the black hole) of the jet stagnation is located at around the
funnel edge with $a = 0.5$--0.99. During the quasi-steady state, the jet
stagnation surface is almost stationary in our GRMHD simulations.  Outer
edges of the stagnation surface in funnel jets ($r/r_{\rm g}\simeq
5$--10 in the case of $a=0.5$--0.99) give an (initial) jet half opening
angle of $\sim 40^{\circ}$--$50^{\circ}$ in de-projection (see Figure
\ref{fig:SPIN-B-FIELD-BUNCH} and \ref{fig:SPIN-B-LINE-IOSS-MAG}).

Contours of $b^2/\rho$ with selected values are also displayed on each
panel of Figure \ref{fig:SPIN-B-LINE-IOSS-MAG} for our reference.
In the vicinity of the black hole at $r/r_{\rm g} \lesssim 20$, the
value of $b^{2}/\rho$ decreases monotonically (approximately independent
on the colatitude angle $\theta$ inside the funnel) as $r$
increases. Again, this can be interpreted as a consequence of no visible
(but very weak) bunching of poloidal magnetic flux at $r/r_{\rm g}
\simeq 5$--10 (around the stagnation surface edges in the case of
$a=0.5$--0.99), as is shown in Figures \ref{fig:FID-B-FIELD-BUNCH},
\ref{fig:SPIN-B-FIELD-BUNCH}, and \ref{fig:SPIN-B-LINE-IOSS-MAG}. Thus,
we may not expect a significant concentration of the mass density toward
the polar axis at a few $\lesssim r/r_{\rm g} \lesssim 20$ as examined
with our fiducial run in Section \ref{sec:RESULTS-FID} (see also Figures
\ref{fig:FID-MAG-EVO} and \ref{fig:FID-B-LINE-EVO}). Depending on the
black hole spin ($a=0.5, 0.7, 0.9$ and 0.99), $b^{2}/\rho \simeq 2, 5,
10$ and 20 are identified at around the closest part (near the funnel
edge) on the jet stagnation surface. This is located between two
outermost streamlines ($z \propto r^{2}$ and $z \propto r^{1.6}$) of the
semi-analytical solution of the FFE jet.

As is seen in Figure \ref{fig:SPIN-LORENTZ}, a high value of $\Gamma$ is
distributed throughout an outer layer of the funnel between two
outermost streamlines ($z \propto R^{2}$ and $z \propto R^{1.6}$), which
are anchored to the event horizon. Having a high value of $b^{2}/\rho$
at the jet launching point is suggestive that the flow will undergo bulk
acceleration to relativistic velocities, as seen in Equation
(\ref{eq:TOTAL-TO-MATTER-ENG.FLUX}). This is a necessary, but not
sufficient condition, as the magnetic nozzle effect is also needed,
which can be triggered by a differential bunching of poloidal flux
toward the polar axis. As is suggested in \citet[]{TNTT90, PNHMWA15},
the location of the jet stagnation surface, where the
magneto-centrifugal force is balanced by the gravity of the black hole,
is independent of the flow property, such as the rate of mass loading,
because it is solely determined by $a$ and $\Omega$ (Equation
\ref{eq:Omega_F}). We point out that a departure of the jet stagnation
surface from the black hole at a higher colatitude ($\theta \rightarrow
0$) gives a prospective reason for the lateral stratification of
$\Gamma$ at large distances (where the sufficient condition for the bulk
acceleration may be applied).

The above issue could be associated with the so-called limb-brightened
feature in the M87 jet. Note we identified the value of $b^{2}/\rho
\simeq 0.5$--1 as the physical boundary at the funnel edge along the
outermost BP82-type parabolic streamline, which is anchored to the event
horizon. Thus, if this boundary condition holds even further downstream
($r/r_{\rm g} \gg 100$), the funnel edge is unlikely to be a site
exhibiting a relativistic flow, as is examined in Figure
\ref{fig:SPIN-LORENTZ}. A highly Doppler boosted emission may not be
expected there, but an alternative process, such as the in situ particle
acceleration may be considered at the edge of the jet sheath (as a
boundary shear layer) under the energy equipartition between the
relativistic particles and the magnetic field \citep[e.g.][]{SO02}. On
the other hand, limbs in the M87 jet have a finite width $\delta R$
inside their edges and $\delta R$ seems to increase in the downstream
direction \citep[e.g.][]{ANP16}, suggesting a differential bunching of
streamlines \citep[e.g.][]{K11}.

In this paper, we identify the outer jet structure (limbs) as the jet
sheath, while the inner jet structure (inside limbs) is identified as
the jet spine. In the next section, our results are compared with VLBI
observations, followed by related discussions in Section \ref{sec:DIS}.
Especially, we assign Section \ref{sec:SPINE-SHEATH} for discussion of a
limb-brightened feature in the context of MHD jets and Section
\ref{sec:VLBI-CORES2} to describe the origin of knotty structures.

\section{Comparison with VLBI observations}
\label{sec:COMP}
\subsection{Jet Morphology}

\label{sec:OBS-MOR}
\begin{figure*}
\centering\includegraphics[scale=0.72, angle=0]{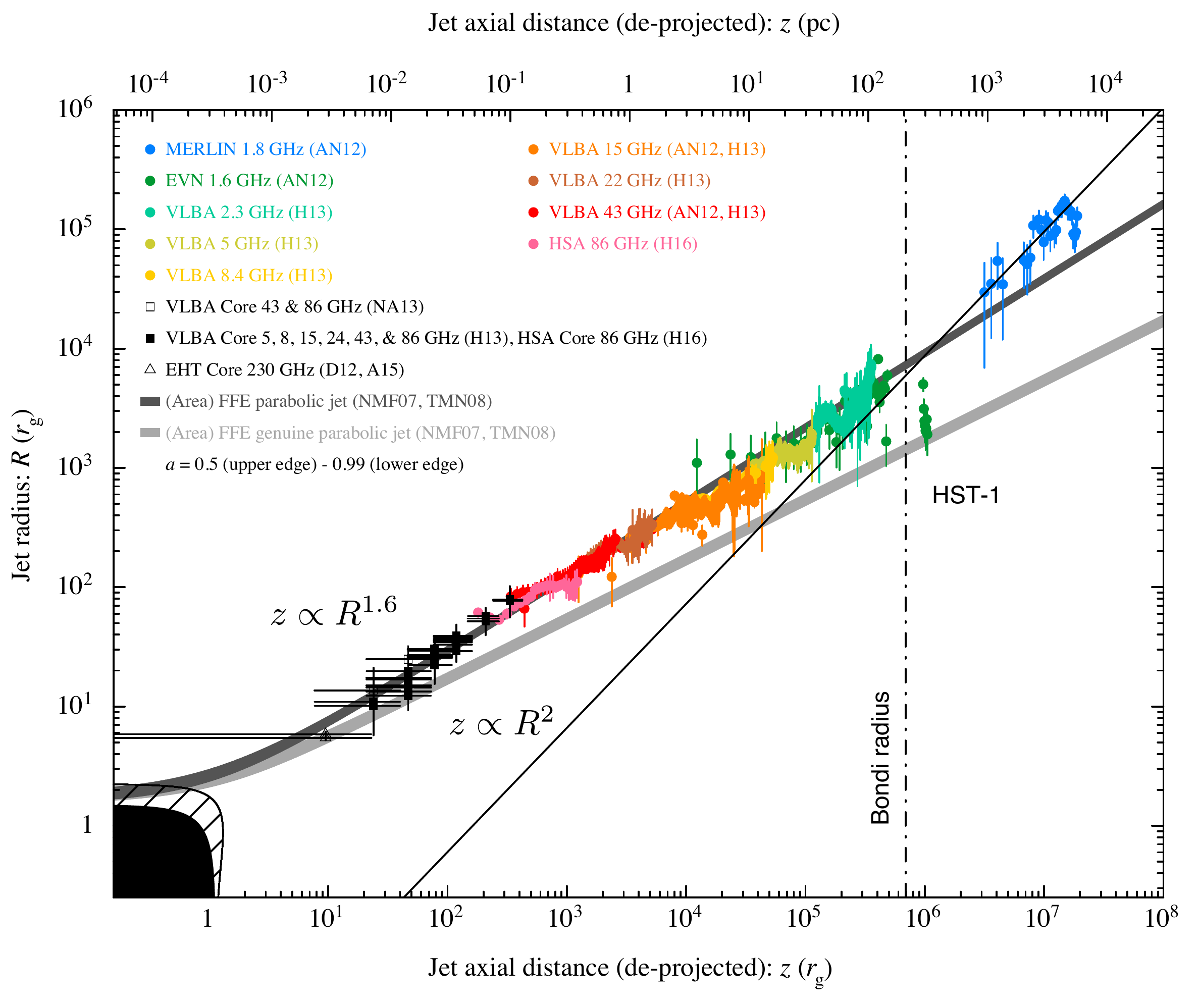}
\caption{\label{fig:R-z} Distribution of the jet radius $R$ as a
function of the jet axial distance $z$ (de-projected with $M=6.2 \times
10^{9} M_{\odot}$ and $\theta_{\rm v}=14^{\circ}$) from the SMBH in
units of $r_{\rm g}$ \citep[cf.][labeled as AN12, NA13, and H13,
respectively]{AN12, NA13, H13}. Additional data points are taken from
\citet[]{D12, A15, H16} (labeled as D12, A15, and H16,
respectively). The (vertical) dashed-dotted line denotes the Bondi
radius $r_{\rm B}$, located at $\simeq 6.9 \times 10^{5}\, r_{\rm g}$
and the HST-1 complex is around $10^{6}\,r_{\rm g}$. Filled black region
denotes the black hole (inside the event horizon), while the hatched
area represents the ergosphere for the spin parameter $a = 0.99$. The
light gray area denotes the approximate solution (e.g.  NMF07, TMN08) of
the FFE genuine parabolic jet (outermost BZ77-type streamline: $z
\propto R^{2}$ at $R/r_{\rm g} \gg 1$), while the dark gray area is the
case of the parabolic jet (outermost BP82-type streamline: $z \propto
R^{1.6}$ at $R/r_{\rm g} \gg 1$), respectively. In both of the outermost
streamlines, which are anchored to the event horizon with $\theta_{\rm
fp}=\pi/2$, a variation from $a=0.5$ (upper edge) to $a=0.99$ (lower
edge) is represented as a shaded area. The solid line is the linear
 least-square for data points of MERLIN 1.8 GHz, indicating the conical
 stream $z \propto R$ \citep[]{AN12}.}
\end{figure*}

Figure \ref{fig:R-z} overviews the geometry of the M87 jet by compiling
the data in the literature (see the caption for
references). Multi-wavelength observations by \citet[][hereafter
AN12]{AN12} revealed that the global structure of the jet sheath is
characterized by the parabolic stream $z \propto R^{1.73} $ at $z/r_{\rm
g} \sim 400$--$4 \times 10^{5}$ \citep[see also][]{H13}, while it
changes into the conical stream $z \propto R^{0.96}$ beyond the Bondi
radius of $r_{\rm B}/r_{\rm g} \simeq 6.9 \times 10^{5}$ \citep[$\sim
205$ pc:][]{R06}. \citet[]{H13} and \citet[]{NA13} examined the
innermost jet region ($z/r_{\rm g} \gtrsim 10$) by utilizing the VLBI
core shift \citep[]{H11}. \citet[]{H13} which suggests a possible
structural change toward upstream at around $z/r_{\rm g} \sim 300$,
where the VLBA core at 5 GHz is located. The innermost jet sheath
($z/r_{\rm g} \gtrsim 200$) is recently revealed with HSA 86 GHz
\citep[]{H16}. Based on our theoretical examinations presented in
previous sections, we overlay the outermost streamlines of the
semi-analytical FFE jet model (NMF07, TMN08) with varying Kerr
parameters ($a=0.5$--0.99) on data points in Figure \ref{fig:R-z} for
comparison.

There are notable findings: i) the inner jet radius (at $z/r_{\rm
g}\lesssim 100$), which is represented by VLBI cores at 15--230 GHz, are
traced by either outer parabolic or inner genuine parabolic streamlines
of FFE jets, which are anchored to $r=r_{\rm H}$ with $\theta_{\rm
fp}=\pi/2$. Within the uncertainties, we cannot distinguish the shape of
the streamline, but there is a tendency that the mean values shift
towards the genuine parabolic streamlines inside the funnel. Therefore,
we consider that the mm VLBI core at 230 GHz with EHT observations
\citep[][hereafter the EHT core]{D12, A15} presumably show the innermost
jet to be in a highly magnetized (PFD) regime $b^2/\rho\gg 1$ and
$\Gamma \lesssim 1.5$ (see, Figures \ref{fig:SPIN-MAG} and
\ref{fig:SPIN-LORENTZ}) inside the funnel. Note that the dominant
magnetic energy of the VLBI core at 230 GHz is originally proposed by
\citet[]{K15}. This may reflect the jet spine, however, rather than the
jet sheath at the funnel edge (see Section \ref{sec:VLBI-CORES1} for
discussions).

ii) At the scale of $100 \lesssim z/r_{\rm g} \lesssim 10^{4}$, we
identify a clear coincidence between the radius of the jet sheath and
the outermost BP82-type streamline of the FFE jet solution. We also
confirm a reasonable overlap between the VLBI cores at 5 and 8 GHz and
an extended emission of the jet sheath at VLBA 43 and HSA 86 GHz
\citep[see also][]{H13}. Therefore, the hypothesis of the VLBI core as
the innermost jet emission \citep[]{BK79} is presumably correct at this
scale although a highly magnetized state of VLBI cores, suggested by
\citet[]{K14, K15} and the frequency ($\nu$) dependent VLBI core shift
$\Delta z (\nu) \propto 1/\nu$, taking place in the non-conical jet
geometry in M87 \citep[]{H11} may conflict with original ideas in
\citet[][where an equipartition between the magnetic and synchrotron
particle energy densities and a constant opening angle and constant
velocity jet is considered]{BK79}. We also remind readers about our
recent result on the jet geometry of blazars which examined VLBI cores
\citep[]{ANAL17}, suggesting that non-conical structures may exist
inside the sphere of influence (SOI) $r_{\rm SOI} \sim
10^{5}$--$10^{6}\, r_{\rm g}$.

iii) At around $z/r_{\rm g} \simeq 10^{4}$--$10^{5}$, it is visible that
data points (the radius of the jet sheath) start to deviate slightly
from $z \propto R^{1.6}$, but a parabolic shape is sustained. This may
indicate a new establishment of the lateral force-equilibrium between
the funnel edge and the outer medium (wind/corona above the RIAF), {\em
or} the jet sheath starts to be Doppler de-boosted (see Figures
\ref{fig:U} and \ref{fig:U-BEAMING} as well as our discussion in Section
\ref{sec:SPINE-SHEATH}). Previous GRMHD simulations exhibit a conical
shape of the funnel edge at $z/r_{\rm g} \gtrsim$ a few of 100
\citep[e.g.][]{M06}, implying the jet is over-pressured against the
outer medium. This however could not be the case in M87. The intrinsic
half opening angle ($\theta_{\rm j}$) of the jet sheath attains the
level of $\theta_{\rm j} \simeq 0.5^{\circ}$ at around $z/r_{\rm g}
\simeq 4 \times 10^{5}$.

iv) Data points are clearly deviated from the outermost BP82-type
parabolic streamlines of the FFE jet solution beyond $r_{\rm B}$. As is
originally suggested by \citet[]{C07}, a structured complex known as
``HST-1'' \citep[]{B99} is located just downstream of $r_{\rm B}$ at
around $10^{6}\, r_{\rm g}$. AN12 suggest a geometrical transition of
the M87 jet as a consequence of the over-collimation of the highly
magnetized jet at the HST-1 complex, which can be initiated by forming
the HST-1 complex, at $\simeq r_{\rm B}$. The jet exhibits the conical
geometry with $\theta_{\rm j} \simeq 0.5^{\circ}$ (const) at the kpc
scale ($z/r_{\rm g} \gtrsim 3 \times 10^{6}$), while $\theta_{\rm j}
\simeq 0.1^{\circ}$ is obtained at around HST-1. We can refer to another
sample of the AGN jet structural transition at $\simeq r_{\rm SOI}$ in
NGC 6251 \citep[][see also Section \ref{sec:N6251}]{T16}.

Note the magnetic pressures at HST-1 and several knots in the downstream
\citep[$\simeq 10^{-9}$--$10^{-8}$ dyn cm$^{-2}$;][]{O89, P03, H09} are
highly over-pressured against the ISM pressure of $\simeq
10^{-11}$--$10^{-10}$ dyn cm$^{-2}$ \citep[]{MBFB02}. Thus, the lateral
pressure equilibrium between the conical jet sheath and the ambient
medium does not seem to be sustained beyond $r_{\rm B}$.  The inner part
of highly magnetized jets can be heavily over-pressured with respect to
the outer part due to the hoop stress as is examined in numerical
simulations and self-similar steady solutions \citep[]{NLL06, Z08}. A
conical expansion of the highly magnetized (with a dominant toroidal
field component), over-pressured jet sheath against the uniform ISM
environment is reproduced in numerical simulations
\citep[e.g.][]{CNB86}.

As a summary of this section, we conclude that the edge of the jet
sheath in M87 upstream of $r_{\rm B}$ can be approximately described as
the outermost BP82-type streamline of the FFE jet solution with the Kerr
parameter $a > 0$, which is anchored to the event horizon. Thus, we
suggest the parabolic jet sheath in M87 is likely powered by the
spinning black hole. Recent theoretical arguments clarified that the
outward Poynting flux is generally non-zero (i.e., the BZ77 process
generally works) along open magnetic field lines threading the
ergosphere \citep[]{TT14, KO04}. Thus our findings support the existence
of the ergosphere. We note, however, that there is an alternative
suggestion that the jet sheath is launched in the inner part of the
Keplerian disk at $R \sim 10\,r_{\rm g}$ \citep[]{M16}.

\subsection{Jet Kinematics}
\label{sec:OBS-KIN}

\begin{figure*}
\centering\includegraphics[scale=0.7, angle=0]{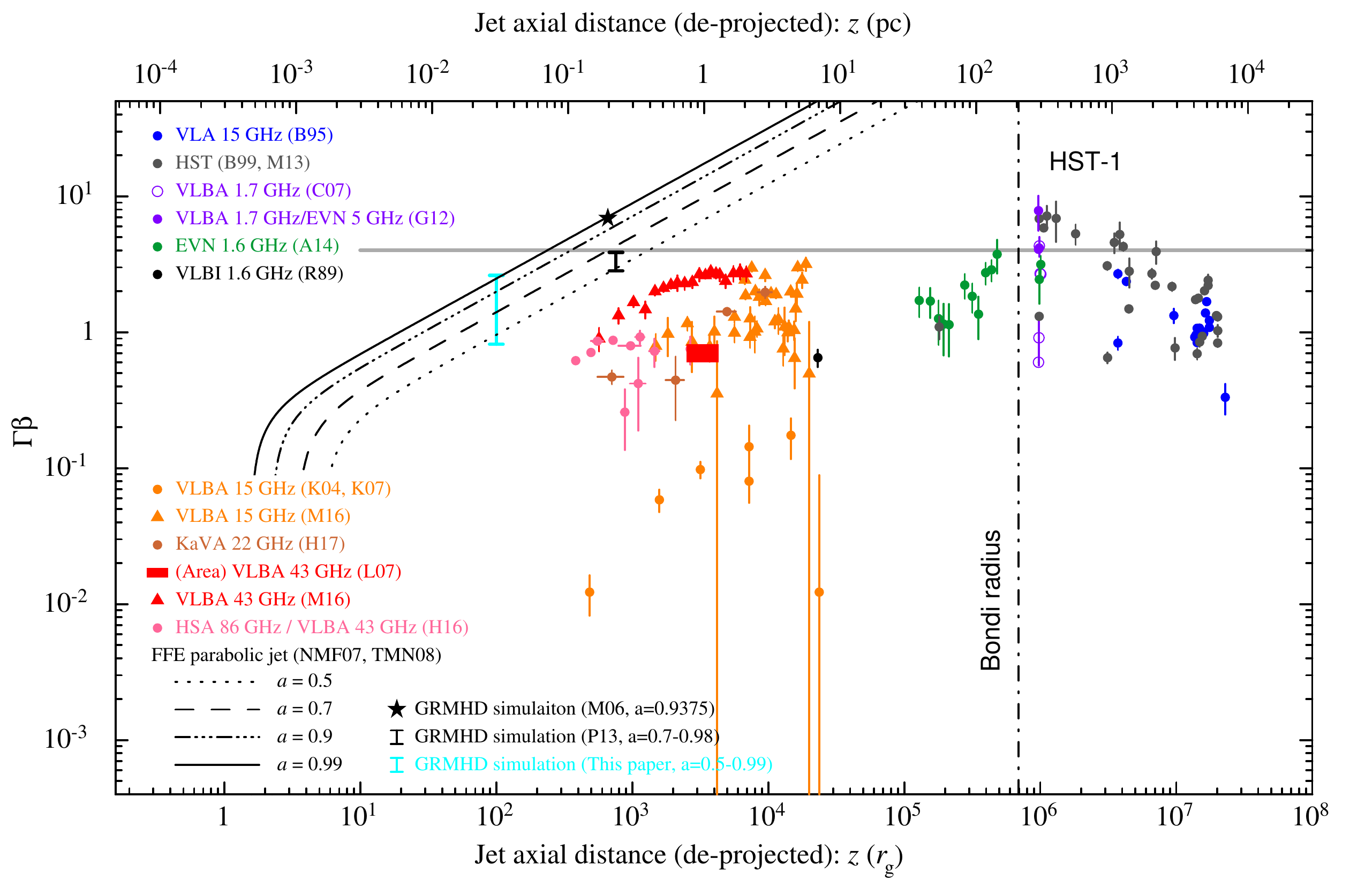}
 \caption{\label{fig:U} Distribution of $\Gamma \beta$ as a function of
 the jet axial distance $z$ (de-projected with $M=6.2 \times 10^{9}
 M_{\odot}$ and $\theta_{\rm v}=14^{\circ}$) from the SMBH in units of
 $r_{\rm g}$. The data of proper motions is taken from the literature
 \citep[][labeled as R89, B95, B99, K04, K07, L07, C07, G12, M13, A14,
 H16, M16 and H17, respectively]{R89, B95, B99, K04, K07, L07, C07, G12,
 M13, A14, H16, M16, H17}. Theoretical expectation by utilizing the FFE
 parabolic ($z \propto R^{1.8}$) jet solutions (NMF07, TMN08) is also
 displayed with varying Kerr parameters ($a=0.5$: dotted line, $a=0.7$:
 dashed line, $a=0.9$: dashed-three dotted line, and $a=0.99$: solid
 line, respectively). The vertical solid line with horizontal bars
 (cyan) indicates a range of maximum values in the jet sheath (between
 two outermost streamlines; $z \propto R^{2}$ and $z \propto R^{1.6}$),
 which are obtained in our GRMHD simulations at around $r_{\rm out}=
 100\,r_{\rm g}$ $(a=0.5$--0.99, see Figure \ref{fig:SPIN-LORENTZ}). For
 our reference, the maximum value in \citet[][labeled as M06]{M06} with
 $a=0.9375$ is marked with a filled star. Also, the vertical solid line
 with horizontal bars (black) indicates a range of maximum values in
 \citet[][labeled as P13]{P13} with $a=0.7$--0.98. The horizontal gray
 line corresponds to $\Gamma \beta$ with $\beta=\cos \theta_{\rm v}$, at
 which the Doppler beaming has a peak (see also Figure
 \ref{fig:U-BEAMING}).}
\end{figure*}

Figure \ref{fig:U} overviews the jet kinematics by compiling the data in
the literature (see the caption for references). Multi-wavelength VLBI
and optical observations reveal both sub-luminal and superluminal
features in proper motion, providing a global distribution of the jet
velocity field $V$ in M87. We display the value of $\Gamma \beta$ in
Figure \ref{fig:U} by using simple algebraic formulas with the bulk
Lorentz factor $\Gamma \equiv (1-\beta^{2})^{-1/2}$ and
$\beta=\beta_{\rm app}/(\beta_{\rm app} \cos \theta_{\rm v}+\sin
\theta_{\rm v})$, where $\beta=V/c$, and $\beta_{\rm app}$ is the
apparent speed of the moving component in units of $c$,
respectively. The value of $\Gamma \beta$ approaches $\beta$ in the
non-relativistic regime ($\Gamma \rightarrow 1$) and represents $\Gamma$
in the relativistic regime ($\beta \rightarrow 1$), thereby representing
simultaneously the full dynamic range in velocity over both regimes.

Superluminal motions ($\beta_{\rm app}>1$) have been frequently observed
at relatively large distances beyond $r_{\rm B}$. Furthermore, these
components seem to originate at the location HST-1 \citep[]{B99, C07,
G12}. On the other hand, no prominent superluminal features inside
$r_{\rm B}$ have been confirmed in VLBI observations over decades
\citep[]{R89, K04, L07}. Instead, sub-luminal features are considered as
non-bulk motions, such as growing instability patterns and/or standing
shocks \citep[e.g.][]{K07}. Thus, this discrepancy (a gap between
sub-luminal and superluminal motions along the jet axial distance) has
been commonly recognized. \citet[]{A14} discovered a series of
superluminal components upstream of HST-1 ($z/r_{\rm g} \sim
10^{5}$--$10^{6}$), providing the missing link in the jet kinematics of
M87.

Very recently, superluminal motions on the scale of $z/r_{\rm g} \simeq
10^{3}$--$10^{4}$ were finally discovered by \citet[]{M16, H17}. These
observations give a diversity to the velocity picture, and suggests the
hypothesis that the systematic bulk acceleration is taking place if the
observed proper motions indeed represent the underlying bulk flow. A
smooth acceleration from subliminal to superluminal motions upstream of
HST-1 is argued in the context of the MHD jet with an expanding
parabolic nozzle \citep[]{NA13, A14, M16, H17}, while observed proper
motions exhibit a systematic deceleration in the region downstream of
HST-1 \citep[]{B95, B99, M13} where the jet forms a conical stream.

Paired sub-/superluminal motions in optical/radio observations at HST-1
\citep[]{B99, C07} (see Figure \ref{fig:U} at around $\sim
10^{6}\,r_{\rm g}$) are modeled by the quad relativistic MHD shock
system with a coherent helical magnetic field \citep[]{NGM10,
NM14}. Taking the complex 3D kinematic features of trailing knots
downstream of HST-1 \citep[]{M13} into account, a growing current-driven
helical kink instability associated with forward/reverse MHD shocks in
the highly magnetized relativistic jet \citep[]{NM04} may be responsible
for organizing the conical jet in M87 at kpc scale.

We examine here the jet kinematics with observations far upstream of
HST-1 at $z/r_{\rm g}\simeq 10^{3}$--$10^{4}$ \citep[]{K04, K07, H16,
H17, M16}. The distribution of $\Gamma \beta$ reaches a maximum level of
$\simeq 3$ and extends to a lower value by more than two orders of
magnitude as is shown in Figure \ref{fig:U}. \citet[]{M16} interpret
that the flow consists of a slow, mildly relativistic ($\Gamma \beta
\sim 0.6$: sub-luminal) layer (the exterior of the jet sheath),
associated either with instability patterns or winds, and a fast,
relativistic ($\Gamma \beta \sim 2.3$: superluminal) layer (the jet
sheath), which is an accelerating a cold MHD jet from the Keplerian disk
(i.e., the BP mechanism). Note that $\beta_{\rm app}\simeq1$ corresponds
to $\Gamma \beta \simeq 1.46$ with $\theta_{\rm v}=14^{\circ}$ in M87.

In our numerical results, maximum values of $\Gamma \beta \simeq
0.8$--2.6 [the solid cyan vertical line in Figure \ref{fig:U}] is
obtained at around $r_{\rm out}/r_{\rm g} =100$ ($\theta \lesssim
10^{\circ}$), depending on the black hole spin. This range covers most
of the higher part of observed proper motions. For sufficiently high
spins ($a \geq 0.9$), bulk speeds of $\Gamma \beta \simeq 1.7$--2.6
could be associated with knotty structures (see, Figure
\ref{fig:SPIN-LORENTZ}). Thus, we give an additional interpretation that
superluminal motions could be interpreted as moving blobs in the
underlying flow of the jet sheath. Regarding highly sub-luminal motions,
as is shown in Figures \ref{fig:SPIN-MASSFLUX-U^r} and
\ref{fig:SPIN-LORENTZ}, we confirm a non-relativistic coronal wind
universally exists for $a=0.5$--0.99 with $\Gamma \beta \gtrsim 0.1$
\citep[see also][for $a=0$]{Y15}, which may be responsible for slow
motions immediately exterior the jet sheath. An acceleration of winds is
also of our interest, but it is unclear in our numerical results
\citep[see,][for their behaviors at $r/r_{\rm g} \gtrsim 100$]{Y15}.

Under the assumption that an observed moving component ($\beta_{\rm
app}$) represents an underlying bulk flow \citep[e.g.][]{LCH09}, we
compare observations with steady axisymmetric FFE jet solutions in
Figure \ref{fig:U}. $\Gamma \beta$ with Equation (\ref{eq:Gamma_FFE}) is
displayed with different black hole spins ($a=0.5$--0.99). Our numerical
simulations reveal that $b^{2}/\rho \simeq 0.5$-1 is sustained at the
funnel edge along the outermost BP82-type parabolic streamlines of $z
\propto R^{1.6}$. Therefore, significant acceleration through the FFE
mechanism or magnetic field conversion is not expected here.  Instead,
the inner part of the funnel (the jet sheath/limb), where high ratio of
magnetic to rest-mass energy density would be expected, is an
appropriate region to apply the FFE jet solution. Parabolic streamlines
of $z \propto R^{1.8}$ ($a=0.5$--0.99) with $\theta_{\rm fp}=\pi/3$ are
chosen as our reference solutions, taking into account that a peak in
$\Gamma$ lies asymptotically between two outermost streamlines ($z
\propto R^{2}$ and $z \propto R^{1.6}$ in Figure
\ref{fig:SPIN-LORENTZ}).

A linear acceleration of highly magnetized MHD/FFE outflows can be
expected in the moderately relativistic regime ($\Gamma \gtrsim 2$) with
$\Gamma \beta \propto R \propto z^{0.56}\ (\beta \rightarrow 1)$ as is
shown in Figure \ref{fig:U}. Similar results are obtained by
\citet[]{BN06, MN07b, PNHMWA15}. Maximum values of $\Gamma \beta$
($\simeq 0.8$--2.6) in our GRMHD simulations ($a=0.5$--0.99) are
qualitatively consistent with those of the FFE jet at $z/r_{\rm
g}=100$. We, however, consider this would be by coincidence as we cannot
find any smooth increase in $\Gamma$ well beyond $\sim 2$ within
$r/r_{\rm g}=100$ in Figure \ref{fig:SPIN-LORENTZ}. FFE jet solutions
with the parabolic shape of $z \propto R^{1.8}$ indicate $\Gamma \beta
\sim 4$--30 around the scale of $z/r_{\rm g} \sim 10^{3}$--$10^{4}$ ($a
\geq 0.5$). This velocity range corresponds to $\beta_{\rm app} \approx
4$--8 with $\theta_{\rm v}=14^{\circ}$ in M87. There is a clear
discrepancy between observed proper motions and theoretical
expectations.

In reality, AGN jets at VLBI scale may not be exactly described by the
FFE system. An agreement between the GRMHD results and FFE models is
found to be good as far as $r/r_{\rm g} \sim 10^{3}$; beyond this scale
the matter inertia becomes non-negligible ($\Gamma \gtrsim 10$) in GRMHD
simulations \citep[]{MN07a}. As a consequence, a slower evolution than
$\Gamma \beta \propto R$ may presumably take place. Nonetheless, a
departure from $\Gamma \sim 2$ at $z/r_{\rm g} \gtrsim 100$ could be
expected in a GRMHD simulation if $b^{2}/\rho$ is sufficiently large
($\gg 1$) at the jet stagnation surface. To be fair enough, $\Gamma
\beta \lesssim 7$ is achieved at $z/r_{\rm g} \simeq 700$ in
\citet[]{M06}\footnote{\citet[]{M06} uses $a=0.9375$ (the fiducial
value), but a modified floor model is adopted; factors $10^{-7}$ in both
power-law forms of $\rho_{\rm min}$ and $u_{\rm min}$ as well as a steep
gradient of $r^{-2.7}$ in both cases. This ensures a huge value of
$b^2/\rho \lesssim 10^{7}$ near the black hole in the PFD funnel.},
which is quantitatively consistent with the FFE jet with $a=0.9$--0.99
(see Figure \ref{fig:U}). On the other hand, for a moderate case of
$b^{2}/\rho \lesssim 100$, maximum values of $\Gamma \beta \simeq 3$--4
are reported at $z/r_{\rm g} \simeq 1000$ with $a=0.7$--0.98
\citep[]{P13}, indicating that a slower evolution is taking place than
in highly magnetized GRMHD/FFE outflows (see also Figure \ref{fig:U}).

As is mentioned above, the detection of faster proper motions
$\beta_{\rm app} \gtrsim 4 \ (\sim 15$ mas/yr) and a signature of their
accelerations at $z/r_{\rm g} \sim 10^{3}$--$10^{4}$, where the jet
sheath maintains a parabolic shape, will be key to confirming our GRMHD
parabolic jet hypothesis. A VLBI program with 15/22/43 GHz towards M87
with a high-cadence monitoring of less than a week (conducting each
observation every few days over a few weeks) may be feasible to find
motions faster than $\gtrsim 0.3$ mas/week.

\section{Discussions}
\label{sec:DIS}
Topical issues are discussed for applying our results to other AGN jets
and highlighting some future study on the M87 jet.

\subsection{Similarities between NGC 6251 and M87}
\label{sec:N6251}

\begin{figure}[b]
\centering\includegraphics[scale=0.36, angle=0]{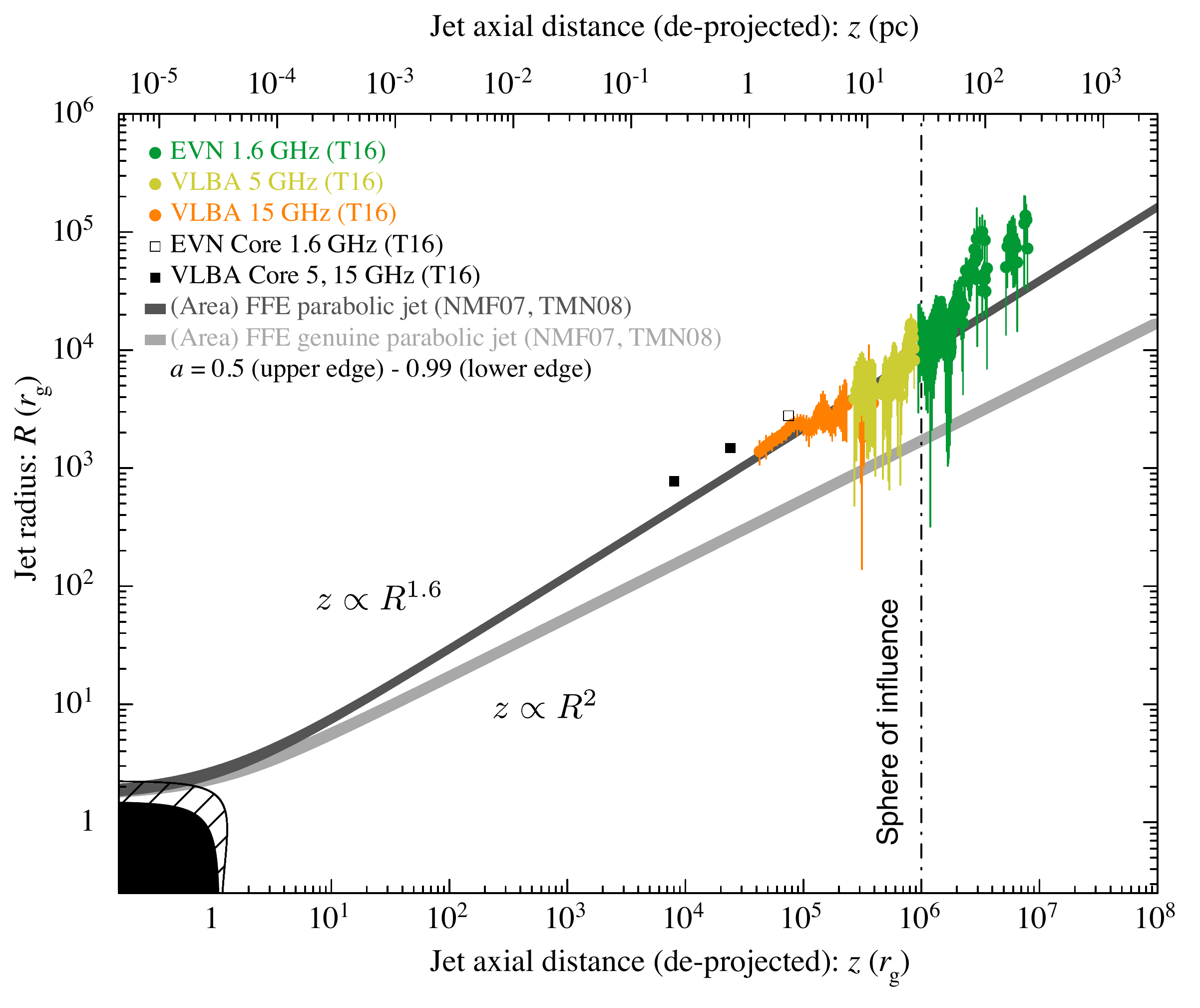}
\caption{\label{fig:R-z-N6251} Distribution of the jet radius $R$ in
NGC6251 as a function of the jet axial distance $z$ (de-projected with
$M=6 \times 10^{8} M_{\odot}$ and $\theta_{\rm v}=19^{\circ}$) from the
SMBH in units of $r_{\rm g}$ (T16). This is similar to Figure
\ref{fig:R-z}; the vertical/horizontal scales and other components,
which are shown in this figure, are identical.}
\end{figure}

\citet[][hereafter T16]{T16} analyzed multi-frequency data (VLBA, EVN,
and VLA) to investigate the jet structure in NGC 6251 and detect a
structural transition of the jet radius from a parabolic to a conical
shape at (2--4) $\times 10^{5}\, r_{\rm g}$, which is close to $r_{\rm
SOI} \simeq 10^{6}\,r_{\rm g}$ in this source. This is a remarkably
similar result to M87 (AN12); one may consider the virial equilibrium at
the center of the cooling core in the giant elliptical galaxies as a
thermodynamically stable state, which gives $r_{\rm B} \approx r_{\rm
SOI}$. Furthermore, the jet radii (in units of $2 r_{\rm g}$)
before/after the transition are quantitatively overlapped with M87 as is
shown in Figure 3 of T16. Obviously, this implies a tight correlation
between the jet sheath and the outermost BP82-type parabolic streamline
of the FFE jet solution as seen in M87 (Figure \ref{fig:R-z}). Figure
\ref{fig:R-z-N6251} confirms this at a quantitative level in NGC 6251.

T16 performed the broken power-law fitting and obtained $z \propto
R^{2.0}$ at $\lesssim 4.2 \times 10^{5}\,r_{\rm g}$ and $z \propto
R^{0.94}$ far beyond. We hereby suggest that the inner jet could be the
BP82-type parabolic geometry, which is similar to M87 (Figure
\ref{fig:R-z-N6251}), if a position offset of VLBI cores from the SMBH
$\simeq 8\times 10^{3}\,r_{\rm g}$ in T16 is taken into account. Note
that the radius of the EVN core at 1.6 GHz is almost identical to the
radii of the VLBA jet at 15 GHz, as is shown in Figure
\ref{fig:R-z-N6251} so we also confirm the VLBI core can be identified
as the innermost jet emission given at these frequencies, which is also
similar to M87 \citep[]{NA13, H13}. By comparing Figure \ref{fig:R-z}
and Figure \ref{fig:R-z-N6251}, we realize that the data points of M87
(inside $r_{\rm B}$) are distributed across more orders of magnitude
than NGC 6251 (inside $r_{\rm SOI}$). Thus, VLBI observations at higher
frequencies are needed to confirm a precise power-law index in the
parabolic stream inside the SOI.

The difference of the SMBH mass between M87 and NGC 6251 is about one
order of magnitude. If the jet radial and axial sizes are normalized in
units of $r_{\rm g}$, they are remarkably identical (see, Figures
\ref{fig:R-z} and \ref{fig:R-z-N6251} in this paper, and Figure 3 in
T16). This suggests us the structural transition may be a characteristic
of AGN jets, at least in nearby radio galaxies. It is straightforward to
seek a counterpart by studying nearby blazars with a relatively large
SMBH mass of $M \sim 10^{9}$--$10^{10}\, M_{\odot}$. Historically, the
upstream region of conical jets in blazars (i.e., inside the VLBI cores
at $\lesssim 0.1$ mas\footnote{Corresponding distances $\sim 10$ pc for
FSRQs at $\langle z \rangle \sim 1.11$ and $\sim 6$ pc for BL Lacs at
$\langle z \rangle \sim 0.37$ at an average redshift \citep[e.g.][]{D07}
with a viewing angle of $5^{\circ}$ for our reference. Thus, the region
$\lesssim 10^{5}\,r_{\rm g}$ for $M \gtrsim 10^{9}\, M_{\odot}$ has been
unexplored in many blazars.}  at mm/cm wavelengths) has been
unresolved. They are sometimes called the ``pipe-line'' from the central
energy generator to the jet, which is unknown and even said that it may
not exist \citep[see, the schematic view:][]{MG85}. Therefore, we expect
that ultra-high-angular-resolution VLBI at mm (HSA, GMVA, EHT)
wavelengths with ALMA will explore the non-conical pipe-line inside
$r_{\rm SOI}$ for bright nearby blazars.

\subsection{A Limb-brightened Feature in the M87 Jet}
\label{sec:SPINE-SHEATH}

A limb-brightened feature, one of the unanswered issues in the M87 jet,
is discussed by \citet[]{H16}; according to the jet spine-sheath
scenario by the presence of a velocity gradient transverse to the jet
\citep[e.g.][]{G05, CB13}\footnote{The radio emission seen at a large
viewing angle $\theta_{\rm v}$ is mostly coming from the slower sheath,
while the emission from the faster spine is beaming away from the line
of sight, because a Doppler factor as a function of $\beta (\equiv V/c)$
has a peak at $\beta=\cos \theta_{\rm v}$.}, viewing the M87 jet from an
angle $\theta_{\rm v}=14^{\circ}$ would not cause a limb-brightened
feature unless the jet spine was unrealistically faster than the jet
sheath, indicating alternative processes are involved \citep[see,][and
references therein]{H16}. Let us re-visit this issue based on Figure
\ref{fig:U}; a relativistic beaming effect in the M87 jet is diagnosed
with the Doppler factor $\delta = \left\{\Gamma \left(1-\beta\cos
\theta_{\rm v}\right)\right\}^{-1}$ on the scale of $z/r_{\rm g} \sim
10^{3}$--$10^{4}$, where $\beta$ and $\Gamma$ are adopted from observed
proper motions and the MHD jet theory.

\begin{figure}[t]
\centering\includegraphics[scale=0.43, angle=0]{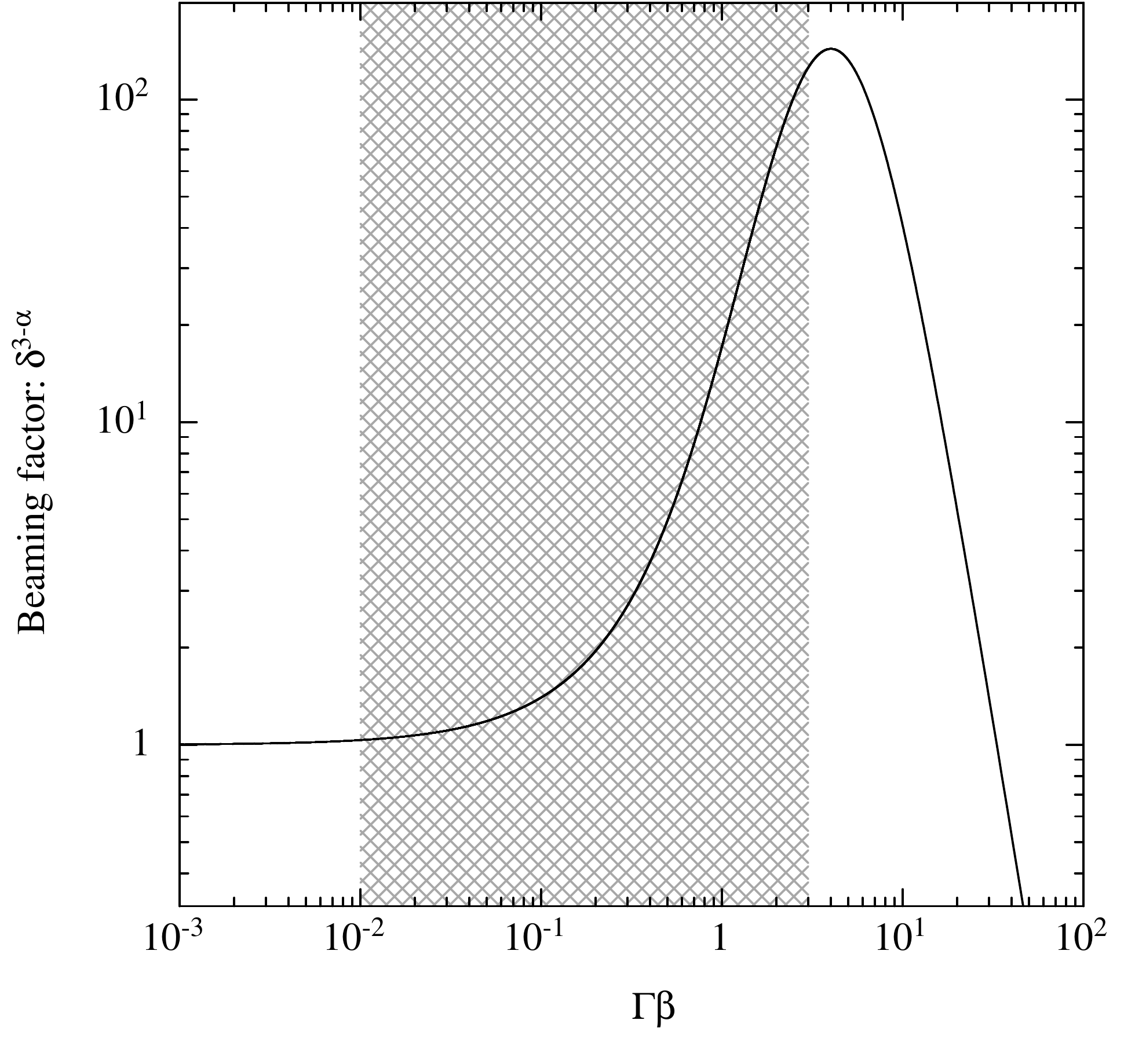}
\caption{\label{fig:U-BEAMING} Distribution of the beaming factor
$\delta^{3-\alpha}$ as a function of $\Gamma \beta$. $\theta_{\rm
v}=14^{\circ}$ and $\alpha=-0.5$ are adopted. Cross-hatched gray area
highlights the Doppler boosting with beaming factors of $\sim 1$--100 at
$10^{-2} \lesssim \Gamma \beta \lesssim 3$, corresponding to the
observations at $z/r_{\rm g} \sim 10^{3}$--$10^{4}$ in Figure
\ref{fig:U}. Note that a peak of the curve is located at $\Gamma \beta$
with $\beta=\cos \theta_{\rm v}$.}
\end{figure}

The observed synchrotron flux density $S_{\nu}$ for a relativistically
moving component is enhanced by a beaming factor $\delta^{3-\alpha}$
\citep[]{J05, S10}, where $\alpha$ is the spectral index defined as
$S_{\nu} \propto \nu^{+\alpha}$. We here adopt $\alpha=-0.5$ for the
reference and the corresponding beaming factor is displayed as a
function of the 4-velocity $\Gamma \beta$ in Figure
\ref{fig:U-BEAMING}. The beaming factor becomes less than unity in the
highly relativistic regime $\Gamma \beta \gtrsim 30$. By taking into
account the observed proper motions in HSA 86 GHz, VLBA 15/43 GHz, and
KaVA 22 GHz on the scale of $z/r_{\rm g} \sim 10^{3}$--$10^{4}$, which
corresponds to $\Gamma \beta \simeq 10^{-2}$--3 in Figure \ref{fig:U},
the beaming factor is expected to be $\sim 1$--100 at the jet sheath.
Note that a similar range of beaming factors can be expected at $\Gamma
\beta \simeq 4$--30, which is expected in the FFE jet solutions for the
parabolic geometry ($z \propto R^{1.8}$) with $a= 0.5$--0.99 at
$z/r_{\rm g} \sim 10^{3}$--$10^{4}$ in Figure \ref{fig:U}.

As Equation (\ref{eq:Gamma_FFE}) suggests, the linear acceleration of a
highly magnetized MHD/FFE outflow decreases toward the polar region of
the funnel when the power-law index of the parabola ($\epsilon$) becomes
large. A quasi-homogeneous distribution of the magnetization $\sigma$
($\approx b^{2}/\rho$) along the colatitude angle ($\theta$) at
$r/r_{\rm g} \lesssim 20$ and the structure of the jet stagnation
surface, as is revealed in our GRMHD simulations (see Figure
\ref{fig:SPIN-B-LINE-IOSS-MAG}), may provide a feasible reasoning for an
efficient bulk acceleration at the outer jet sheath, where the magnetic
nozzle effect would be expected under the progress of a differential
bunching of the poloidal magnetic flux toward the polar axis.

Therefore, the jet spine can be intrinsically less beamed than the jet
sheath as the distribution of $\Gamma$ exhibits \citep[Figure
\ref{fig:SPIN-LORENTZ}; see also][]{TMN08, TNM10, P13}. A lateral
stratification of $\Gamma$ is naturally expected so that a
limb-brightened feature may be fundamental in AGN jets if they consist
of black hole-driven GRMHD outflows. Interestingly, limb-brightened
features are also observed in the best known TeV BL Lac objects
Markarian (Mrk) 421 \citep[]{P10} and Mrk 501 \citep[]{G08, P09} on
$z/r_{\rm g} \simeq 10^{4}$--$10^{5}$ even with smaller angles of
$\theta_{\rm v} = 4^{\circ}$ \citep[][]{G04,L12}.

Based on our numerical results up to $r_{\rm out}/r_{\rm g}=100$, we
find no relevant evidence of a concentration of the poloidal magnetic
flux at the funnel edge \citep[]{G09} and/or a pileup of the material
along the funnel edge \citep[]{Z08}, which might be related to a
funnel-wall jet \citep[]{DHK03, DHKH05}, as a possible mechanism of a
limb-brightened feature. Note these may conflict with the physical
conditions necessary to accelerate MHD jets, e.g., a high ratio of
magnetic to rest-mass energy density and the magnetic nozzle effect. It
would, however, be necessary to conduct a further investigation of this
issue at the corresponding scale ($z/r_{\rm g} \gtrsim 10^{3}$).

We comment on the power-law acceleration (see the footnote 7) in the jet
sheath (possibly, a less-collimated parabolic stream than the genuine
parabolic one). As is shown in Figure \ref{fig:U}, steady axisymmetric
FFE jet solutions for streamlines of $z \propto R^{1.8}$ ($a=0.5$--0.99)
with $\theta_{\rm fp}=\pi/3$ (as the jet sheath) does {\em not} exhibit
a transition from the linear to power-law acceleration at $z/r_{\rm g} <
10^{5}$ (it takes place at $z/r_{\rm g} > 10^{10}$ for $a=0.99$). Thus,
the outer jet sheath is always faster than the inner jet spine, even if
the jet spine is launched with a sufficiently high value of $b^{2}/\rho$
at the jet stagnation surface and $\Gamma$ follows the linear
acceleration due to an efficient magnetic nozzle effect.  Both of these
factors, however, are not supported by our GRMHD simulations. Therefore,
we suggest the limb-brightened feature in M87 may be associated with the
intrinsic property of an MHD parabolic jet powered by the spinning black
hole, rather than the result of a special viewing angle as is previously
discussed in \citep[]{H16}.

Finally, as is mentioned in Section \ref{sec:OBS-MOR}, the radius of the
jet sheath starts to deviate slightly (becoming narrower) from the
outermost BP82-type streamline $z \propto R^{1.6}$ at $z/r_{\rm g}
\gtrsim 10^{4}$ (see also Figure \ref{fig:R-z}). If the jet sheath
follows the linear acceleration up to this scale, as is examined in
Figure \ref{fig:U}, the underlying flow would reach $\Gamma \beta \simeq
30$ ($a=0.9$--0.99) and result in a weaker Doppler de-boosting (Figure
\ref{fig:U-BEAMING}). Furthermore, it is interesting to note that the
emission of the parabolic jet sheath further downstream disappears at
$z/r_{\rm g} \gtrsim 4 \times 10^{5}$ \citep[]{A14}, where $\theta_{\rm
j} \simeq 0.5^{\circ}$ is obtained (Figure \ref{fig:R-z}). If the
empirical relation $\Gamma \theta_{\rm j} \sim 0.1$ -- 0.2
\citep[]{CB13} is applied, $\Gamma \simeq 11$--22 would be
expected. This is close to the velocity range, at which a Doppler
de-boosting may arise.

\subsection{VLBI Cores in M87}
\label{sec:VLBI-CORES}

\begin{figure}[h!]
 \centering\includegraphics[scale=0.45, angle=0]{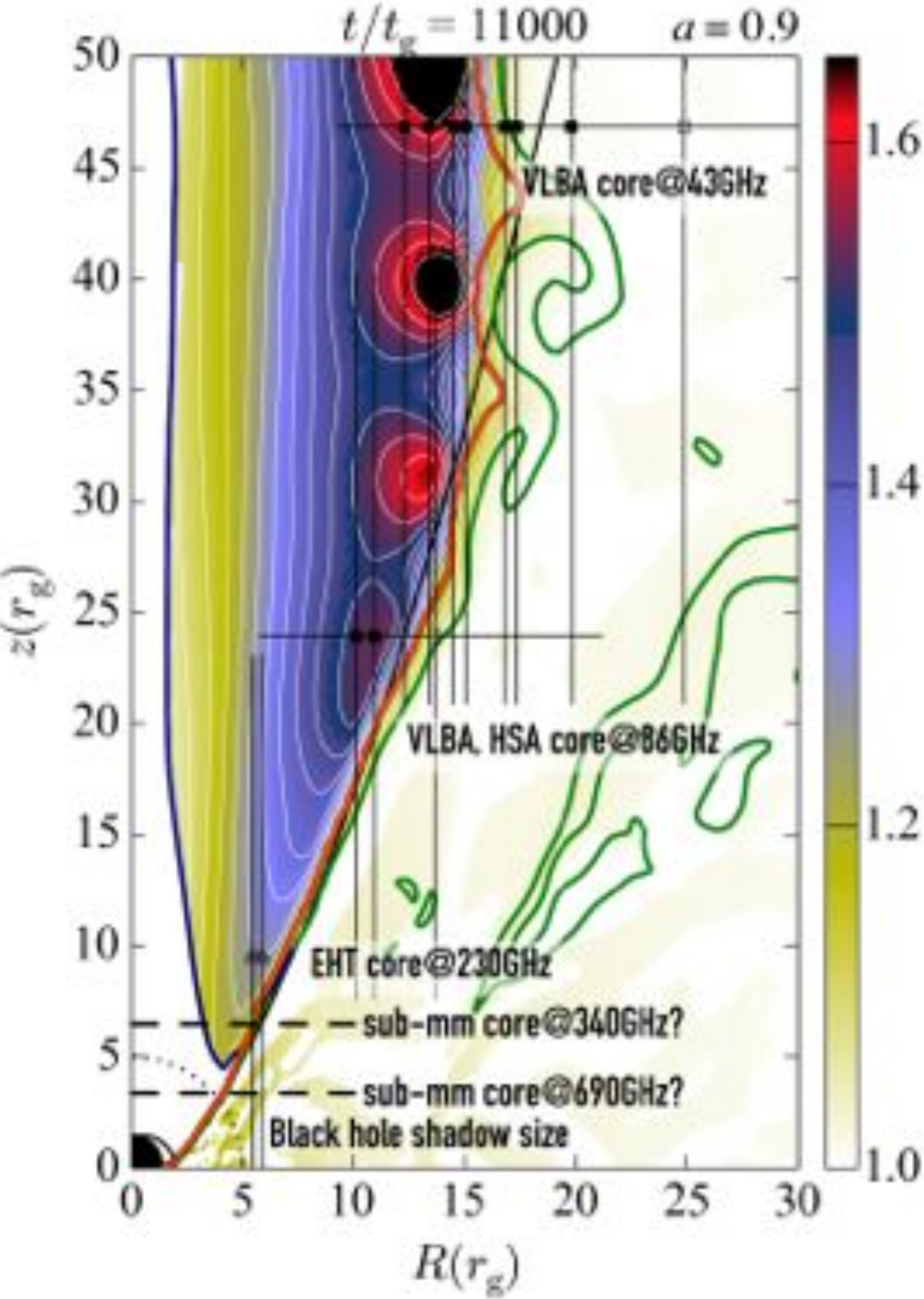}
 \caption{\label{fig:LORENTZ-VLBI-CORE} Innermost jet radii are
 displayed as the FWHM/2 of mm VLBI cores at 43, 86, and 230 GHz, by
 utilizing the VLBI core shift. Our GRMHD simulation result ($a=0.9$) in
 the quasi-steady phase ($t/t_{\rm g}=11000$) is overlaid. Expected
 positions of sub-mm VLBI cores at 345 and 690 GHz are also indicated
 with a horizontal dashed line. A color filled contour of the Lorentz
 factor $\Gamma$ (only where $u^{r}>0$) is shown as well as $\beta_{\rm
 p}=1$ (green solid lines), the jet stagnation surface $u^{r}=0$ (a navy
 solid line: only inside the PFD funnel), and $b^{2}/\rho=1$ (an orange
 solid line).  Other components are identical to those in Figure
 \ref{fig:Ergo}, but the black hole spin is adjusted. The size of the
 black hole shadow is indicated with the dotted circle with the average
 radius $\sim 5\, r_{\rm g}$ for our reference. See also Figures
 \ref{fig:FID-PR_TOT-LINES}, \ref{fig:SPIN-LORENTZ}, and \ref{fig:R-z}
 for details.}
\end{figure}

We now discuss the (sub-)mm VLBI cores in M87, which are considered the
innermost jet emission---at the given frequencies---in the vicinity of
the SMBH (see also Figure \ref{fig:R-z}). Figure
\ref{fig:LORENTZ-VLBI-CORE} shows the radius and location of VLBI cores
at mm bands (43, 86, and 230 GHz) and their expectation at sub-mm bands
(345 and 690 GHz), by an extrapolation of the VLBI core at frequencies
higher than 43 GHz \citep[]{H13, NA13} and utilizing the frequency
depending VLBI core shift \citep[]{BK79} in M87 \citep[]{H11}. Our GRMHD
simulation result for $a=0.9$ is overlaid for reference.  What we
currently know about the (sub-)mm VLBI cores of M87 from observations
are the size, the flux density, the brightness temperature \citep[]{D12,
A15}, and the energetics \citep[]{K14, K15}.

\subsubsection{(Sub-)mm VLBI Core as a Neighborhood
   of the Jet Origin}
\label{sec:VLBI-CORES1}

The synchrotron self-absorption (SSA) theory is applied in order to
examine the energy balance between electrons ($U_{e}$) and magnetic
fields ($U_{B}$) for the VLBA core at 43 GHz; it can be highly
magnetized or at most roughly in equipartition \citep[$10^{-4} \lesssim
U_{e}/U_{B} \lesssim 0.5$;][]{K14}. Furthermore, \citet[]{K15} derived
the energy balance of electrons and positrons ($U_{\pm}$) and $U_{B}$ in
the EHT core at 230 GHz as $8 \times 10^{-7} \leq U_{\pm}/U_{B} \leq 2
\times 10^{-3}$. These constraints, together with their locations
($R/r_{\rm g},\, z/r_{\rm g}$) in the funnel, may provide some hint for
how to discriminate between the (sub-)mm and mm cores as is shown in
Figure \ref{fig:LORENTZ-VLBI-CORE}.

Under the hypothesis that (sub-)mm VLBI cores consist of the optically
thick (against SSA) non-thermal synchrotron emission from the innermost
jet, mm VLBI core emission ($\leq 86$ GHz) may be dominated by the jet
sheath close to the funnel edge ($b^2/\rho \simeq 1$ and $\beta_{\rm p}
\simeq 1$). The (sub-)mm VLBI core emission ($\geq 230$ GHz) may be
dominated by the jet spine further inside the funnel edge ($b^2/\rho \gg
1$ and $\beta_{\rm p} \ll 1$), though we may not rule out a possible
contribution of the RIAF body ($b^2/\rho \ll 1$ and $\beta_{\rm p} \geq
1$) \citep[]{K15, A15}; the brightness temperature of $\sim (1-2)\times
10^{10}$ K of the EHT core is broadly consistent with the electron
temperature of $\sim 10^{9}$--$10^{10}$ K in the advection-dominated
accretion flow \citep[]{Mah97}.

Temporal variations of the FWHM size as well as the flux density of
(sub-)mm VLBI cores (spanning from months to years) may also provide
another important clue for the local behavior of the jet as is shown in
Figure \ref{fig:LORENTZ-VLBI-CORE}. It is notable that the FWHM size of
VLBI cores at 43/86 GHz is changing with variable flux density within a
factor of $\sim 2$ \citep[]{NA13, H13, H16}, while the EHT core at 230
GHz is fairly stable without a significant change of the flux density
spanning three years \citep[]{A15}. On the other hand, the light curve
of the SMA data at 230 GHz appears to exhibit a monthly scale variation
\citep[]{H14}, implying that the variability may arise on the scale
larger than $10\,r_{\rm g}$, which corresponds to the size of the jet
radius ($\gtrsim$ a few tens of $r_{\rm g}$) where blobs emerge and
propagate in our GRMHD simulations.

Thus, no significant variations of the FWHM size as well as the flux
density spanning three years \citep[]{A15} and a theoretical constraint
toward a highly magnetically dominated state \citep[]{K15} in the EHT
core give us a favor that a emission comes form the innermost jet
(inside the funnel) within $z/r_{\rm g} \lesssim 20$, whereas further
multi-epoch observations would be desired to confirm our hypothesis. As
is shown in Figure \ref{fig:LORENTZ-VLBI-CORE}, position uncertainties
of (sub-)mm VLBI cores allows us to consider either the jet and/or the
RIAF as an emission source. Thus, polarization structures at (sub-)mm
bands would be important to study the origin of the synchrotron emission
of VLBI cores; either a toroidal component (from the RIAF and/or
corona/wind) or a helical component (from the funnel). Furthermore, an
existence of sub-mm VLBI cores at 345/690 GHz (or perhaps non-existence
due to a truncation of the core emission beyond the jet stagnation
surface) may provide a further constraint.

Figure \ref{fig:LORENTZ-VLBI-CORE} also indicates the FWHM/2 size of the
EHT core $\simeq 5 \, r_{\rm g} \sim 20 \, \mu$as \citep[][]{D12, A15}
at 230 GHz is comparable to a largest extent of the stagnation surface
(i.e., a minimum extent of the funnel jet radius at the approaching
side) for the BH spin $a \sim 0.9$. This is also similar size of the
photon ring (i.e., the black hole shadow) with the average radius of
$\sim 5 r_{\rm g}$ \citep[][]{CPO13}. Therefore, the observed (sub-)mm
VLBI core structure at 230 GHz (and above) may be affected by the photon
ring and/or the gravitational lensing of surrounding emission (e.g. the
RIAF, the counter jet, and so on). Our discussion does not consider
this, while our results are not affected by this. We speculate that some
prominent feature associated to the jet base, which can be connected to
the spinning black hole with $\Omega_{\rm H}$, may be expected if the
stagnation surface is the initiation site of the particle acceleration
\citep[e.g.][]{BT15, PWY17}.

\subsubsection{Origin of Superluminal Blobs and Shock-in-jet Hypothesis}
\label{sec:VLBI-CORES2}

Distribution of $\Gamma$ in Figure \ref{fig:LORENTZ-VLBI-CORE} clearly
exhibits an existence of the cylindrical core with $\Gamma \lesssim 1.2$
inside the funnel ($R/r_{\rm g} \lesssim 5$), accompanied with a lateral
increase of $\Gamma$ along the $R$-axis (see also Figure
\ref{fig:SPIN-LORENTZ}). The funnel jet does not exhibit a significant
acceleration with remaining $\Gamma \lesssim 1.5$ at $z/r_{\rm g}
\lesssim 50$, where it does not fully enter the linear acceleration
regime ($R \Omega \gg 1$). Outside the funnel, the bound wind exists,
but $\Gamma = 1$ is sustained. However, it is notable that there is an
emergence of blobs with $\Gamma \gtrsim 1.5$ in the funnel jet (we
confirmed similar events take place when $a \ge 0.7$).

We consider that a formation of high-$\Gamma$ blobs in the underlying
low-$\Gamma$ bulk flow near the black hole may be a fundamental
phenomena in the system, giving a physical origin of superluminal
motions as seen in mm/cm VLBI observations (see Figure \ref{fig:U}).  A
blob, which may be a compressional magnetosonic wave triggered by the
axisymmetric distortion such as an $m=0$ mode instability inside the
funnel edge, could steepen into a magnetosonic shock. Therefore, moving
a shock in the jet \citep[``shock-in-jet'', e.g.][]{BK79, M80} is
presumably expected as a counterpart of enhanced synchrotron emission,
especially at the jet sheath. Thus our GRMHD simulations provide a
self-consistent process how superluminal blobs in AGNs could be
originated in the vicinity of the SMBH.

This feature has never been seen in previous simulations with a fixed
curvilinear boundary wall \citep[]{KBVK07, KVKB09, TMN08, TNM10},
implying a dynamical consequence of the external boundary during the
evolution of PFD funnel jets. It seems that a blob appears beyond $R
\simeq 10\,r_{\rm g} (> R_{\rm L})$ \citep[where $R_{\rm L}$ is the
outer light surface and $R_{\rm L} \sim 5$ for the case of $a=0.9$,
e.g.][]{M06, PNHMWA15} due to a lateral compression of the funnel-wall
as is shown in Figure \ref{fig:LORENTZ-VLBI-CORE}. Note that previous
GRMHD simulations \citep[]{M06} also experience that, beyond $r \approx
10\,r_{\rm g}$, as the jet undergoes poloidal oscillations due to
toroidal pinch instabilities; $\Gamma$ is larger in pinched regions than
non-pinched regions.

We suggest that the pressure driven (interchange) and/or the current
driven instability, such as sausage/pinch mode (the azimuthal mode
number: $m=0$) may play a dynamical role. In the highly magnetized PFD
funnel, the outer Alfv\'en surface is fairly close to the outer light
surface, where the azimuthal component of the magnetic field is
comparable to the poloidal one \citep[]{M06, PNHMWA15}. The azimuthal
component of the magnetic field becomes dominant in the super-Alfv\'enic
flow and if the ratio of the toroidal to poloidal field strength becomes
larger than $\sqrt{2}$, such an instability may take place
\citep[e.g.][]{K66, P82}. $\beta_{\rm p} \simeq 1$ is located just at
the funnel edge (see also Figures \ref{fig:FID-PR_TOT-LINES},
\ref{fig:SPIN-LORENTZ}, and \ref{fig:LORENTZ-VLBI-CORE}) and we thus
consider that the $m=0$ mode is excited at around the jet sheath at
$R/r_{\rm g} \gtrsim 10$ , but it could be suppressed at the inner jet
spine and the vicinity of the black hole where $\beta_{\rm p} \ll 1$.

One of the good examples is provided by VLBI observations with the HSA
at 86 GHz \citep[]{H16}; we remark the axisymmetric ``bottle-neck''
structure at $\sim 0.2$--0.3 mas ($\sim 230$--340 $r_{\rm g}$ in
de-projection) from the core. We further point out knot-like enhanced
intensity features as an appearance of paired 'blobs' at both northern
and southern limbs (labeled as N1/S1--N4/S4) up to $\sim 1$ mas ($\sim
10^{3} \, r_{\rm g}$) in \citet[]{H16}. Distribution of
quasi-axisymmetric blobs (northern/southern limbs) extends up to $\sim
10$ mas ($\sim 10^{4} \, r_{\rm g}$), which is revealed by the VLBA at
43 GHz \citep[]{W18} and the KaVA at 23 GHz \citep[]{N14}. Oscillatory
patterns, most likely reflecting over-collimation/-expansion of the
flow, are seen at $\lesssim 10$ mas scale of the jet sheath
\citep[]{M16}.

The very high energy (VHE; $> 100$ GeV) $\gamma$-ray flares in M87
\citep[see][for an overview]{A12} may originate in the jet base within
$\sim 100\,r_{\rm g}$, which is associated to the radio core at 43 GHz
in 2008 \citep[]{A09}, 2010 \citep[]{H12}, and 2012
\citep[]{H14}. During an enhanced VHE $\gamma$-ray state in 2012, mm
VLBI (EHT) observations of M87 at 230 GHz have been conducted. There is
little possibility of the VHE $\gamma$-ray event in a compact region of
$\lesssim 20\,r_{\rm g}$ \citep[neither obvious structural changes nor
associated flux changes;][]{A15}. Such observational evidence may
also support the shock-in-jet scenario associated with a VHE event in
M87, as some counterpart is reproduced in our GRMHD simulations. We thus
suggest that different behaviors of VLBI cores at different frequencies
(43/86 GHz and 230 GHz) can originate from the lateral extent of their
sizes ($R/r_{\rm g} \gtrsim 10$ or small), which specifies the location
of an emerging high-$\Gamma$ blob in the PFD funnel jet.

It would be favored to conduct simultaneous observations with
multi-frequency, multi-epoch mm/sub-mm VLBI study. This enables us to
examine the dynamical structure of the innermost jet in M87 as well as
proper motions by diagnosing the core size and the flux variations. For
example, an emergence of the blob with $\Gamma \sim 1.5$ near the VLBI
core at 86 GHz, as is shown in Figure \ref{fig:LORENTZ-VLBI-CORE},
corresponds to an apparent speed of $\sim 0.6$--0.7 $c$ with
$\theta_{\rm v}=14^{\circ}$. This is equivalent to a motion of $\sim
2.2$--2.6 mas yr$^{-1}$ in the proximity to M87. By considering the
distance between VLBI cores at 86 GHz and 43 GHz as $\sim 20\,r_{\rm
g}\sim 0.02$ mas in projection ($\theta_{\rm v}=14^{\circ}$), a delay of
$\sim 2$ days is expected before rising the flux density in the VLBI
core at 43 GHz, if a passing blob through VLBI cores indeed does cause
flare-ups.

\subsubsection{Comparisons with Other Models and Future Studies}
\label{sec:COMPARISONS} Based on our examination, we briefly discuss
other studies on modeling the M87 jet. A ``state-of-art'' 3D GRMHD
simulation model with radiative transfer (RT) computations is proposed
by \citet[]{MFS16}. The model considers that the radio emission comes
from the jet sheath \citep[funnel wall:e.g.][]{HK06}, in which the
plasma is constantly supplied from a less magnetized ($\beta_{\rm
p}=1$--50) accretion disk \citep[e.g.][]{MFN16}.  On the other hand,
this paper suggests the jet sheath is powered by the spinning black hole
and located inside the parabolic funnel where a highly magnetized plasma
exists ($b^{2}/\rho \gtrsim 1$ and $\beta_{\rm p} \lesssim 1$).  Our
GRMHD simulations exhibit the bulk acceleration and the superluminal
blobs are activated inside, but near the funnel edge at $\gtrsim 10 \,
r_{\rm g}$ if the black hole spin is moderately large ($a \geq 0.7$; see
Figure \ref{fig:LORENTZ-VLBI-CORE}). We therefore suggest that a proper
shape of the funnel play an important role in modeling the M87 jet
because it may provide a suitable jet sheath if the Doppler beaming and
non-thermal particle acceleration by the emerging superluminal blobs are
responsible for the limb-brightened feature. As is also discussed in
Section \ref{sec:VLBI-CORES}, we remark a highly magnetically dominated
state of the VLBA core at 43 GHz \citep[]{K14} and the EHT core at 230
GHz \citep[]{K15}, which may provide an additional constraint on these
models. We leave our direct comparison in the forthcoming paper with a
post-processing with RT computations.

\citet[]{M16} examine kinematics of the M87 jet on scales of $z/r_{\rm
g}=$ 200--2000 based on multi-epoch VLBA observations at 43 GHz and
discuss the jet acceleration\footnote{Authors conduct the
wavelet-based image segmentation and evaluation method to derive proper
motions. A wider variety of velocity fields is extracted, but the
fastest motions at each axial distance are selected to examine the jet
acceleration.} in the context of an MHD jet launched by
magneto-centrifugal mechanism from the Keplerian disk (BP82). It is
unclear whether the BP process indeed takes place at the inner accretion
flow near the ergosphere based on our GRMHD simulations (the BP process
requires a high magnetization $b^{2}/\rho \gg 1$ at the jet launching
region of the accretion flow and an existence of the coherent poloidal
magnetic field, which possibly penetrates the equatorial plane). Authors
consider an invisible/dimmer faster spine (than the slower sheath) due
to either not present in the flow (i.e., a lower synchrotron particle
energy density) or de-boosted. However, as we examined in \S
\ref{sec:SPINE-SHEATH}, such a de-beaming effect may not be expected on
that spacial scale ($z/r_{\rm g} \simeq 1000$) with $\theta_{\rm v} \sim
10^{\circ}$--$20^{\circ}$ \citep[e.g.][]{H16} even if the FFE jet model
(an upper limit in the MHD acceleration as the plasma inertia is
negligible) is adopted (unless a lower emissivity is expected).

Above two models seem to favor the accretion disk as the origin of the
jet sheath, which is contrary to our model. It may be also true that a
large-scale coherent poloidal magnetic flux, which threads the
equatorial plane, exist in the MAD-type accretion flow
\citep[e.g.][]{TNM11, TM12}. We, however, speculate that a highly
magnetized (i.e., a low mass loaded, relativistic) outflow may not be
initiated in such an environment. It is therefore necessary to examine
our hypothesis whether the jet sheath is surely originated from the
spinning black hole or not with a MAD-type configuration. Because of how
we thread the initial torus with magnetic field, larger disks will have
more available magnetic flux to accrete \citep[e.g.][]{N12}. More
accreted flux will then open the possibility of exploring how the MAD
state may affect the jet model we propose here. We leave this
exploration to future work.

Our results also need to be confirmed in 3D simulations. As observations
indicate, projected VLBI images of the M87 jet on the plane of the sky
exhibit an almost axisymmetric shape inside the Bondi radius although
some internal (non-) axisymmetric patterns exist. This may suggest that
internal modes ($m \geq 0$) of plasma instabilities are growing, while
external non-axisymmetric modes ($m \geq 1$) seem to be suppressed. This
could be the case if the highly magnetized jet is confined by the weakly
magnetized external medium \citep[e.g.][]{NLL07}. 3D GRMHD simulation
model of M87 \citep[e.g.][]{MFS16} exhibits a non-disturbed jet, which
may not be subject to external non-axisymmetric modes, at distance up to
a few hundred of $r_{\rm g}$.

Finally, we remark on the recent theoretical examination of the
limb-brightened jet feature by \citet[]{TTKNH18}. Based on a steady
axisymmetric jet model from the Keplerian accretion disk to synthesize
radio images of the M87 jet \citep[]{BrLo09}, authors examine larger
parameter spaces of locating a plasma loading and the angular frequency
$\Omega$ of the poloidal magnetic field lines. They find that
symmetrically limb-brightened jet images as is seen in M87 can be
reproduced only if the poloidal magnetic field lines of the jet
penetrate a fast-spinning black hole, while the jet with poloidal
magnetic field lines that pass through a slowly spinning black hole or
the Keplerian accretion disk (at $R/r_{\rm g} \gtrsim 10$), seems to be
disfavored\footnote{The model does not rule out the possibility of a
systematic limb-brightened jet if disk-threading poloidal magnetic field
lines are spinning fast and concentrated around the ISCO under the high
magnetization $b^{2}/\rho \gg 1$ \citep[e.g.][]{TT14}.}.

\section{Conclusions}

Our study deals with the formation of parabolic jets from the spinning
black hole by utilizing semi-analytical solutions of the steady
axisymmetric FFE jet model \citep[]{NMF07, TMN08} and the 2D public
version of the GRMHD simulation code \texttt{HARM} \citep[]{GMT03,
NGMDZ06}. Funnel jets in GRMHD simulations, which have been widely
investigated during the last decade (see Section \ref{sec:INT} for
references),
%\citep[e.g.][]{DHK03, DHKH05, H04, MG04, HK06, M06, BHK08, P13, S13}
are of our particular interest because their nature in a parabolic shape
have still been unknown. Our recent observational efforts toward M87
(see Section \ref{sec:INT} for references)
%\citep[]{AN12, A14, ANP16, NA13, H11, H13, H16, H17}
provide a case study on this context. We examined funnel jets,
especially for their shape, physical conditions at the boundary, and
their dependence on the black hole spin, by following \citet[]{MN07a,
MN07b} that provided quantitative agreements of the funnel jet interior
between the GRMHD simulations and (GR)FFE solutions. We conducted
extensive runs up to $r_{\rm out}/r_{\rm g}=100$ with various black hole
spins $a=0.5$-0.99.  Our results highlight a formation of quasi-steady
funnel jets in the less-collimated parabolic shape (than the genuine
parabolic one: $z \propto R^{2}$), which does not depend on the black
hole spin.

\begin{figure}[ht!]
 \centering\includegraphics[scale=0.45, angle=0]{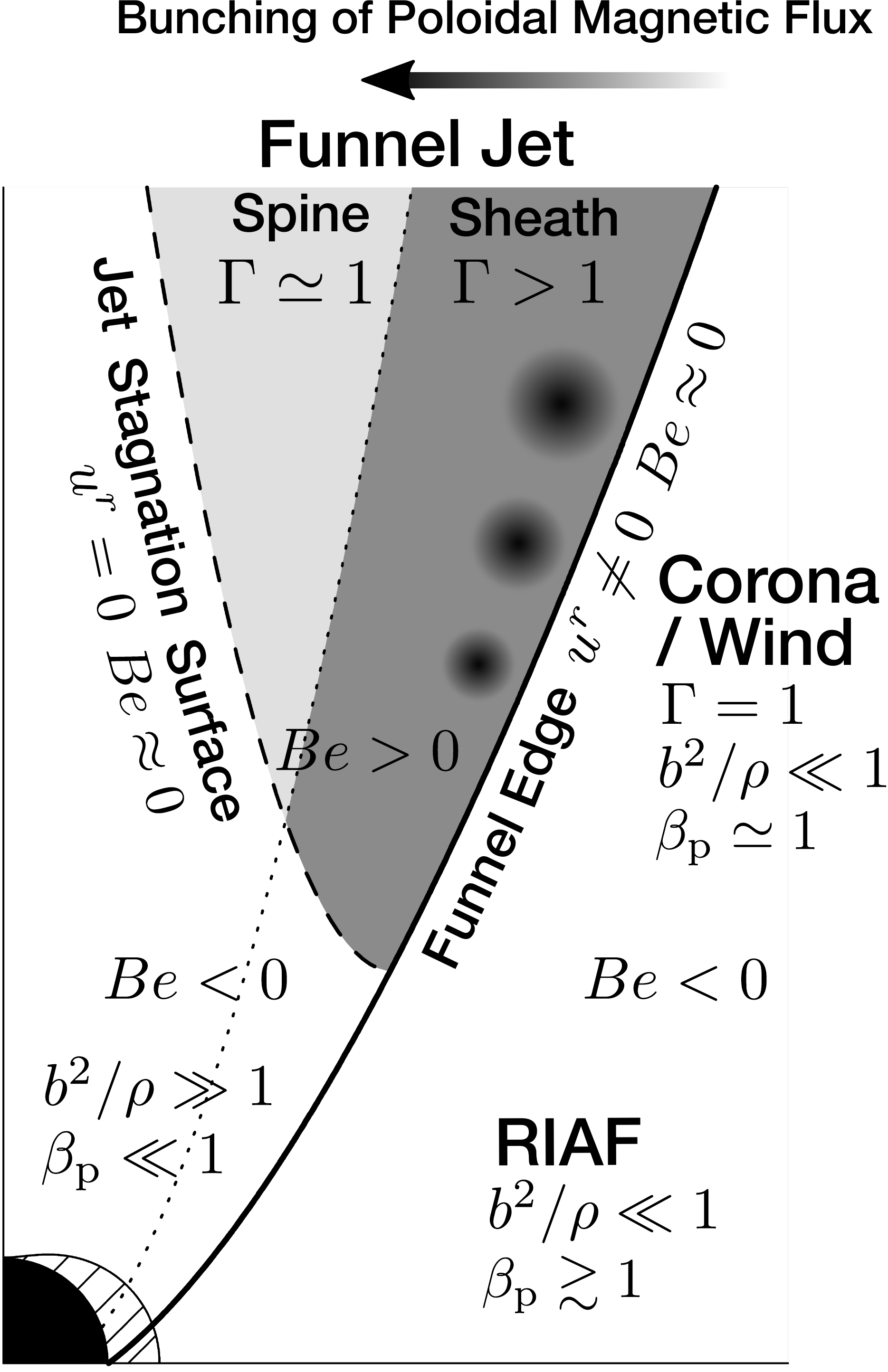}
 \caption{\label{fig:MODEL} Schematic view (arbitrary scale) of our
 parabolic GRMHD jet with a moderately high spin ($a\geq0.7$). The
 system is organized by the highly magnetized funnel (parabolic jet) and
 the weakly magnetized coronal/wind above the RIAF body. Typical values
 of the ratio of magnetic to rest-mass energy: $b^{2}/\rho$, the
 plasma-$\beta$: $\beta_{\rm p}$, and the Bernoulli parameter: $Be \,
 (\approx -u_{t}-1)$ are specified. The limb-brightened (i.e.,
 spine-sheath) structure is expressed as a context of the lateral
 stratification of the bulk Lorentz factor: $\Gamma$ (dark shaded area:
 $\Gamma > 1$ and light shaded area: $\Gamma \simeq 1$,
 respectively). The funnel edge ($b^{2}/\rho \simeq 1, \, \beta_{\rm p}
 \simeq 1, \, Be \approx 0, \, u^{r} \neq 0$) and the jet stagnation
 surface ($Be \approx 0, \, u^{r} = 0$) are shown as the thick solid and
 dashed lines, respectively. Emerging blobs are illustrated near the
 funnel edge.}
\end{figure}

Schematic view of our parabolic GRMHD jet model is displayed for a
moderately high spin ($a\geq0.7$) in Figure \ref{fig:MODEL}. The funnel
jet area is highly magnetized (PFD: $b^{2}/\rho \gg 1, \ \beta_{\rm p}
\ll 1$), while the outer area is weakly magnetized ($b^{2}/\rho \ll 1$)
that consists of the RIAF body ($\beta_{\rm p} \gtrsim 1$) and
corona/wind ($\beta_{\rm p} \simeq 1$). The funnel edge is approximately
determined by the outermost BP82-type parabolic ($z \propto r^{1.6}$)
streamline (the ordered, large-scale poloidal magnetic field line) of
the FFE solution, which is anchored to the event horizon with an almost
maximum angle $\theta_{\rm fp} \simeq \pi/2$ (a thick solid curve in
Figure \ref{fig:MODEL}), with the following equipartition of i) the
magnetic and rest-mass energy densities ($b^{2}/\rho \simeq 1$) and ii)
the gas and magnetic pressures in the fluid frame ($\beta_{\rm p} \simeq
1$). The distribution of $Be \approx 0$ forms a {\sf V}-shaped geometry
in the PFD funnel jet; $Be \approx 0$ is sustained at the funnel edge
along the outermost BP82-type parabolic streamline, while the jet
stagnation (inflow/outflow separation) surface ($u^{r}=0$: a dashed
curve in Figure \ref{fig:MODEL}) inside the funnel, the location of
which depends on the black hole spin (shifting inward with increasing
$a$) \citep[]{TNTT90, PNHMWA15}, approximately coincides with the
bound/unbound separation: $Be \lessgtr 0$. Note that $u^{r} \neq 0$ is
confirmed at the funnel edge.

At the funnel exterior ($Be < 0$), the coronal wind carries a
substantial mass flux \citep[1--3 orders magnitude higher than the
funnel jet: cf. ][]{S13, Y15}. The magneto-centrifugal mechanism (BP82)
would be unlikely to be operated; it is because of the absence of a
coherent poloidal magnetic field outside the PFD funnel jet, where the
toroidal magnetic field is dominant in the SANE state \citep[see
also][for 3D simulations]{H04} and the plasma is not highly magnetized
($b^{2}/\rho \ll 1, \ \beta_{\rm p} \simeq 1$). Thus, a process of the
relativistic MHD acceleration might not be activated (see Equation
\ref{eq:TOTAL-TO-MATTER-ENG.FLUX}) so that $\Gamma=1$ is sustained for
all cases with $a=0.5$--0.99 \citep[see also][with $a=0$]{Y15}. Note
that the situation seems to be unchanged even in the MAD state as is
shown in 3D simulations \citep[e.g.][with $a=0$--0.9]{P13}.

On the other hand, external environments (corona/wind and RIAF) provide
a sufficient pressure support for deforming the funnel jet into a
parabolic shape \citep[see also][]{GL16}; the total (gas $+$ magnetic)
pressure (i.e., $\beta_{\rm p} \simeq 1$) is dominant in the corona/wind
region, whereas, the ram pressure of the accreting gas near the black
hole ($r/r_{\rm g} \lesssim 10$) is the primal component rather than the
total pressure of the RIAF body ($\beta_{\rm p} \gtrsim 1$) even in the
SANE state. Thus, the latter may conceptually similar to the MAD
scenario \citep[][]{NIA03, I08, TNM11} although the SANE accretion flow
does not possess an arrested poloidal magnetic flux on the equatorial
plane at $R > r_{\rm H}$.

There is a lateral stratification of $\Gamma$ in the funnel outflow as a
consequence of the unique distribution of $b^{2}/\rho$ along the jet
stagnation surface (a weak dependence on the colatitude angle $\theta$
at a few $\lesssim r/r_{\rm g} \lesssim 20$) and the efficiency of the
magnetic nozzle effect (see Section \ref{sec:INT} for references).  The
poloidal magnetic flux is differentially bunched towards the central
axis by the hoop stress, causing that the magnetic nozzle effect
predominantly works at the outer layer in the funnel (sheath: $\Gamma >
1$), while it does not work efficiently at the inner layer of the funnel
(spine: $\Gamma \simeq 1$). This is rather general feature in MHD jets
from the rotating black hole \citep[e.g.][]{KBVK07, KVKB09, TMN08,
TMN09, TNM10, P13}. Thus, the spine-sheath structure can be naturally
expected in MHD jets, being responsible for the limb-brightened feature
in AGN jets if the Doppler beaming plays a role (see Figure
\ref{fig:MODEL}).

An accurate border between the inner jet spine ($\Gamma \simeq 1$) and
outer jet sheath ($\Gamma > 1$) is still undefined in our simulations up
to $r_{\rm out}/r_{\rm g}=100$ (though a conceptual border is drawn in
Figure \ref{fig:MODEL}), but we would tentatively favor to propose the
genuine parabolic inner streamline (BZ77: $z \propto R^2$). Our
preceding study \citep[]{ANAL17} gives an additional support; jet radii
estimated with VLBI cores for blazars are wider than those of BZ77-type
genuine parabolic streamlines ($a=0.5$--0.998) on $z/r_{\rm g} \simeq
10^{4}$--$10^{7}$. This is rather surprising if we consider the
classical scenario \citep[e.g.][]{G05, CB13}. A departure from $\Gamma =
1$ may {\em not} be expected at the funnel edge as far as $b^{2}/\rho
\simeq 1$ is sustained. Therefore, a peak of $\Gamma$ would be located
at some inner layer of the jet sheath. Further numerical study with
$r_{\rm out}/r_{\rm g} \geq 10^{3}$ will be presented in a forthcoming
paper.

Based on above results, we apply BP82-type outermost parabolic
streamlines of the FFE solution ($a=0.5$--0.99) to the radius of the jet
sheath in M87 derived in multi-frequency VLBI observations. A
quantitative agreement (between the FFE solution and VLBI
observations) is obtained on the scale of $z/r_{\rm g}\simeq 10$--$4
\times 10^{5}$. Same examination is applied to another nearby radio
galaxy NGC 6251 on the scale of $z/r_{\rm g}\simeq 10^{4}$--$10^{6}$; we
also obtain a similar consistency between the theory and observation
\citep[]{T16}. Furthermore, a jet structural transition (from parabolic
to conical stream) seems to be taken place at around the SOI
\citep[]{AN12, T16}, suggesting a characteristic of AGN jets, at least
in nearby radio galaxies, but possibly even in distant blazars
\citep[e.g.][]{ANAL17}.

We consider limb-brightened features in M87 as a consequence of the bulk
acceleration of MHD jets driven by the spinning black hole. The jet
sheath, which is organized by an expanding layer between the genuine
parabolic (BZ77-type: $z \propto R^{2}$) and less-collimated parabolic
streamlines (BP82-type: $z \propto R^{1.6}$), may be responsible for the
Doppler boosted emission toward us with a viewing angle $\theta_{\rm v}
= 14^{\circ}$. Our simulations also exhibit that an emergence of the
blob-like knotty feature in the underlying bulk flow (Figure
\ref{fig:MODEL}). A blob is presumably triggered by the $m=0$ mode
(pressure driven interchange and/or current-driven sausage/pinch)
distortion at the funnel edge ($b^{2}/\rho \simeq 1, \ \beta_{\rm p}
\simeq 1$). We propose that it will evolve as a superluminal knot ;
axisymmetric knotty patterns are frequently identified in mm/cm VLBI
observations \citep[e.g.][]{N14, H16, W18}.

There is a wider range (more than two orders of magnitude in units of
$\Gamma \beta$) of observed proper motions ($\beta_{\rm app} \lesssim
3$) of the jet sheath in M87 at $z/r_{\rm g} \simeq 10^{3}$--$10^{4}$
\citep[]{K04, K07, H16, H17, M16}. Velocity range fairly matches motions
of knotty structures in our simulations (at $z/r_{\rm g} \sim
100$). Therefore, we may expect a blob could be steepen into a shock;
one of the possible origins of the shock-in-jet phenomena is
reproduced. We expect further detailed examination by utilizing a joint
analysis of the VLBI core variability with the EHT, GMVA, HSA, VLBA, and
KaVA in order to confirm our hypothesis of the moving shock in the
jet. At the same time, as our examination of the beaming factor
suggests, much faster motions ($\beta_{\rm app} \approx 4$--8 with
$\theta_{\rm v}=14^{\circ}$) can be expected in the jet sheath at
$z/r_{\rm g} \simeq 10^{3}$--$10^{4}$ if the underlying flow follows the
highly magenetized MHD/FFE jet evolution. Therefore, one of the
challenges for exploring the main stream of the jet sheath would be to
conduct a high-cadence VLBI monitoring less than a week (for a faster
motion $\gtrsim 0.3$ mas/week).

Our parabolic jet model can be primarily applicable to LLAGNs and/or BL
Lacs, in which the RIAF at sub-Eddington regime $\dot{m} \lesssim
10^{-2}$ falls into the central SMBH \citep[]{NM08}. We, however,
suggest that the internal structure of a magnetically driven funnel jet
($b^{2}/\rho \gg 1$ and $\beta_{\rm p} \ll 1$) seems to be general. It
would also be expected even in radio loud quasars at a (super-)Eddington
regime \citep[no matter how large/small the radiative efficiency is in
the accretion flow, a geometrically thick disk accretion plays a role in
driving a jet;][]{T15}. Therefore, a faster jet sheath may be universal
if the MHD acceleration and collimation play a fundamental role in AGN
jets \citep[see also][for some hints of wider jet radii than $z \propto
R^{2}$ in blazars]{ANAL17}. As an immediate task, our model needs to be
examined with other sources exhibiting limb-brightened structures even
with a low viewing angle such as blazars Mrk 421 \citep[]{G06, P10} and
Mrk 501 \citep[][]{G04, G08, P09, K16}.

M.N. acknowledges Roger Blandford for careful reading of the manuscript
and helpful comments. M.N. also thanks Ruben Krasnopolsky for his help
with the Python programming. Instructive comments by the anonymous
referee helped us to improve the manuscript. TIARA summer school on
numerical astrophysics 2015 at ASIAA provided a tutorial for the
\texttt{HARM} code. K. Aasada is supported by the Ministry
of Science and Technology of Taiwan grants MOST 106-2119-M-001-027 and
MOST107-2119-M-001-017. K. Hada was supported by JSPS Grant Number
18K13592 and the Sumitomo Foundation Grant for Basic Science Research
Projects Grant Number 170201. K. Toma acknowledges JSPS Grants-in-Aid
for Scientific Research 15H05437 and JST grant “Building of Consortia
for the Development of Human Resources in Science and
Technology”. This work is partially supported by JSPS
KAKENHI grants No. JP18K03656 (M. Kino) and JP18H03721 (K. Niinuma,
M. Kino, and K. Hada). J.-C. Algaba acknowledges support from the
National Research Foundation of Korea (NRF) via grant
NRF-2015R1D1A1A01056807. K. Akiyama is financially supported by the
Jansky Fellowship of the National Radio Astronomy Observatory (NRAO) and
a grant from the National Science Foundation (NSF; AST-1614868). The
NRAO is a facility of the NSF operated under cooperative agreement by
Associated Universities, Inc.

\appendix
\section{A.  Dependence of the Funnel Jet Shape on Initial
Plasma-$\beta$ Values (Lower/Higher Limits)}
\label{sec:AP.PARAMETERS}

\begin{figure*}
 \centering\includegraphics[scale=.65, angle=0]{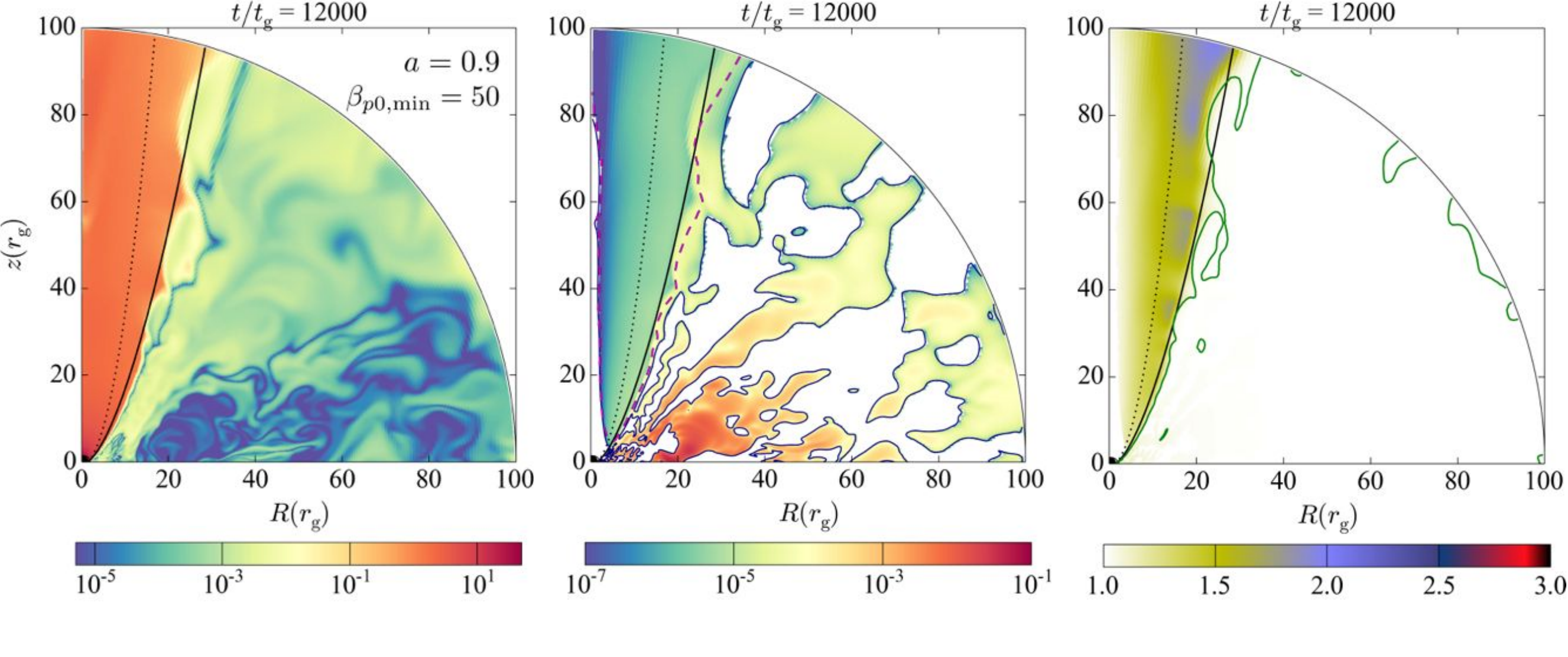}
 \centering\includegraphics[scale=.65, angle=0]{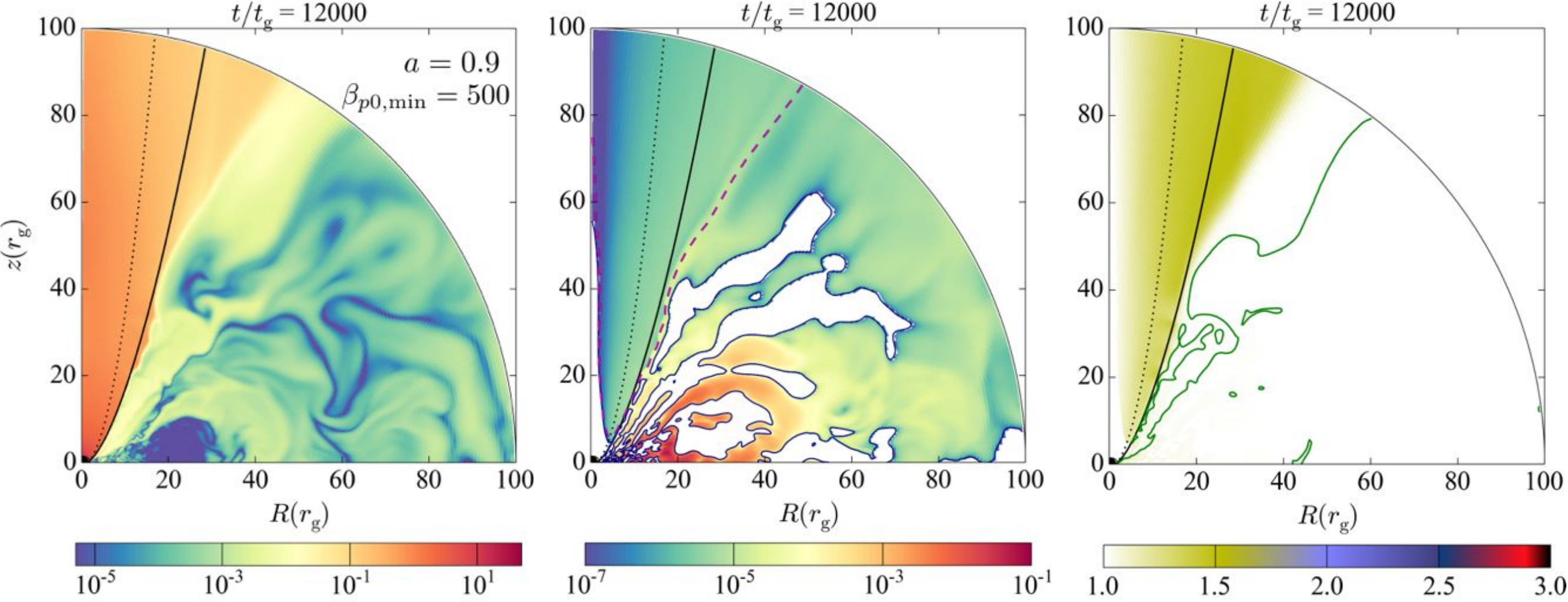}
 \caption{\label{fig:BT0-MAG-LORENTZ-MASS-FLUX}
 Final ($t/t_{\rm g}=12000$) snapshots of two different initial
 conditions (the black hope spin $a=0.9$ is fixed); $\beta_{\rm p0, 
 \, min}=50$ ({\em top}) and $\beta_{\rm p0, \, min}=500$ ({\em bottom}).
 A color-filled contour of the magnetic energy per unit particle
 $b^{2}/\rho$ ({\em left}), the magnitude of the outgoing radial mass
 flux density ({\em middle}), and the Lorentz factor $\Gamma$ ({\em
 right}). Readers can refer to Figures \ref{fig:SPIN-MAG},
 \ref{fig:SPIN-MASSFLUX-U^r}, and \ref{fig:SPIN-LORENTZ} ($a=0.9$)
 for comparison.}
\end{figure*}

This appendix provides the range of validity of parabolic funnel jets by
showing results with different parameters. We fix the dimensionless Kerr
parameter $a=0.9$, but changes $\beta_{\rm p0, \, min}$ to 50 or 500;
snapshots of physical quantities are shown in Figure
\ref{fig:BT0-MAG-LORENTZ-MASS-FLUX}. Compared with the case of
$\beta_{\rm p0, \, min}=100$, overall structures are unchanged (the
funnel edge follows the parabolic outermost streamline $z \propto
R^{1.6}$: BP82) when we start with $\beta_{\rm p0, \, min}=50$. A highly
magnetized funnel is formed and $b^{2}/\rho \simeq 1$ is sustained at
the funnel edge. The external corona/wind region is qualitatively
identical. Note that simulations with $\beta_{\rm p0, \, min}<50$
sometimes induce numerical errors around the polar axis due to the
extremely high magnetization $b^{2}/\rho \gg 1$.

On the other hand, the magnetization in the funnel is weaken if we start
with $\beta_{\rm p0, \, min}=500$ and $b^{2}/\rho \simeq 0.3$ is obtained
at the funnel edge. The magnitude of the outgoing radial radial mass
flux in the funnel region is almost similar with the case of 
$\beta_{\rm p0, \, min}=50/100$, while that in the outer coronal wind area is
different; about one order of magnitude smaller in $\beta_{\rm p0, \,
min}=500$ than the case of $\beta_{\rm p0, \, min}=50/100$. It is notable
that a departure of the funnel edge from $z \propto R^{1.6}$ is seen at
$r/r_{\rm g}>40$, following a conical expansion (see the distributions
of $Be \approx 0$ and $\beta_{\rm p}=1$ in bottom panels of Figure
\ref{fig:BT0-MAG-LORENTZ-MASS-FLUX}). The distribution of the Lorentz
factor exhibits no feature of the inhomogeneous
distribution of $\Gamma$ (blobs) at the outer layer in the funnel for
$\beta_{\rm p0, \, min}=500$.

\begin{figure}
 \centering\includegraphics[scale=.55, angle=0]{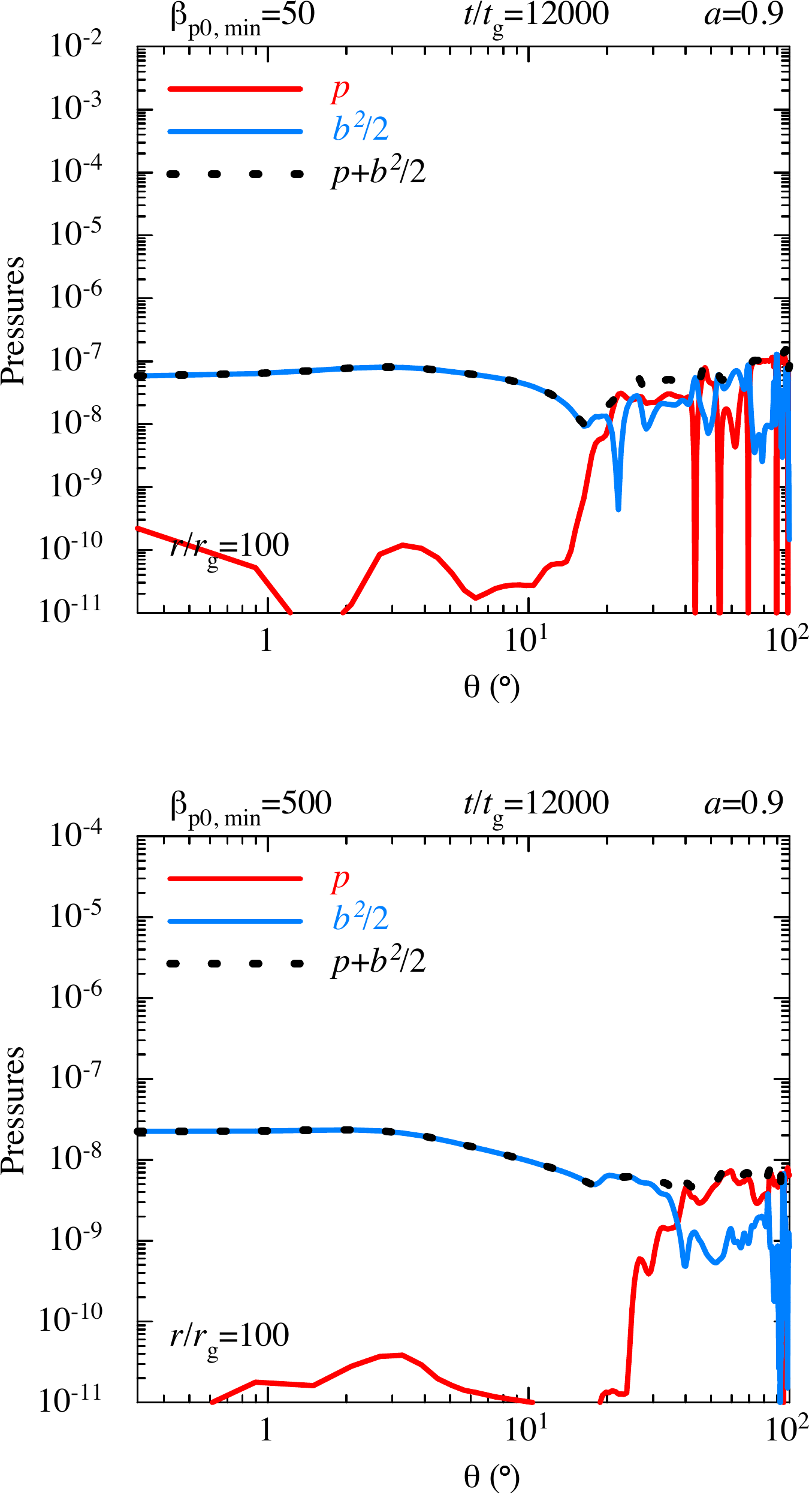}
 \caption{\label{fig:BT0-PRS}
 A $\theta$ cross-section at $r/r_{\rm g} = 100$ showing the gas
 pressure (red solid line), the magnetic pressure (blue solid line), and
 their sum (the total pressure: black dotted line) for two different
 initial conditions ($a=0.9$ is fixed) at the final stage ($t/t_{\rm
 g}=12000$); $\beta_{p0, {\rm min}}=50$ ({\em left}) and $\beta_{p0,
 {\rm min}}=500$ ({\em right}). Readers can refer to the {\em top} panel
 of Figure \ref{fig:SPIN-PRS} ($a=0.9$) for comparison.}
\end{figure}

Thus, Figure \ref{fig:BT0-MAG-LORENTZ-MASS-FLUX} exhibits an example how
the parabolic funnel jet deforms into a conical shape ($\beta_{\rm p0,
\, min}=500$). Figure \ref{fig:BT0-PRS} provides further qualitative
analysis on this issue. The total pressure balance between the funnel
region ($\beta_{\rm p} \ll 1$) and the external corona/wind region
($\beta_{\rm p} \simeq 1$) is sustained in the case of the parabolic
funnel ($\beta_{\rm p0, \, min}=50$; {\em upper} panel). On the other
hand, magnetically dominated (total) pressure in the non-parabolic
funnel ($\beta_{\rm p0, \, min}=500$; {\em lower panel}) is
over-pressured (a factor of few) against the external coronal/wind
region where $\beta_{\rm p} \simeq 1$ is not hold (the gas pressure
dominated). The prescription of a higher value of $\beta_{\rm p} \gtrsim
500$ may not be enough to provide a sufficient total pressure,
presumably due to a lack of the magnetic pressure. In summary, we
suggest that a moderately magnetized wind/corona ($\beta_{\rm p} \simeq
1$) may play a dynamical role in maintaining the funnel jet into a
parabolic shape.

\section{B. Time Evolution of the SANE}

\begin{figure}
 \centering\includegraphics[scale=.75, angle=0]{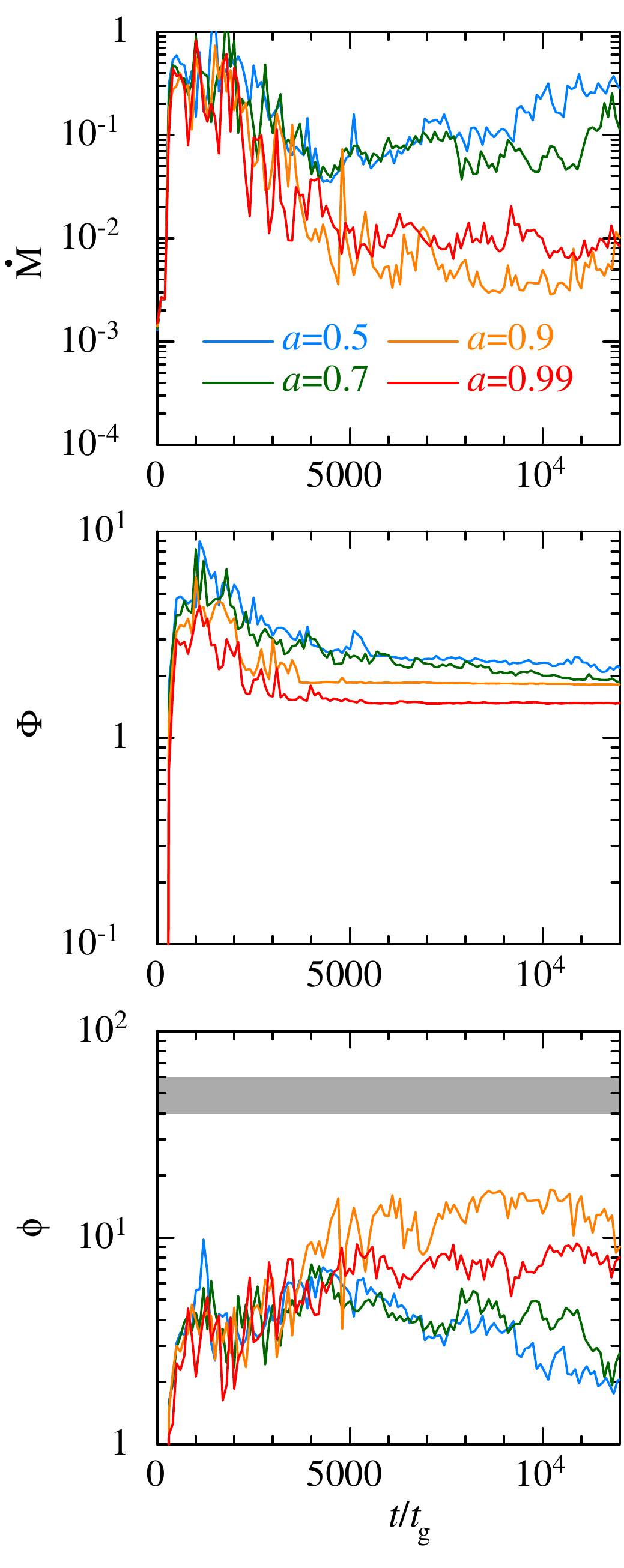}
 \caption{\label{fig:EVO-B-FLUX} Variations of $\dot{M}$
 ({\em top}), $\Phi$ ({\em middle}), and $\phi$ ({\em bottom}) as a
 function of time with varying black hole spin, corresponding to four
 different cases in Section \ref{sec:RESULTS-PARA}. $\phi\approx40$--60
 (MAD state) is indicated as gray shaded area in the {\em bottom}
 panel.}
\end{figure}

Figure \ref{fig:EVO-B-FLUX} provides the time evolution of the mass
accretion rate $\dot{M}$, the poloidal magnetic flux $\Phi$ threading
the black hole horizon, and $\phi$, the dimensionless ratio of $\Phi$ to
$\dot{M} c^{2}$ for our four systems examined in Section
\ref{sec:RESULTS-PARA}. Following \citet[]{N12}, $\dot{M}$ is defined as
\beqn
\dot{M}=2\pi \int_{0}^{\pi} \rho u^{r} \sqrt{-g}d\theta.
\eeqn
$\Phi$ is defined as
\beqn
\Phi=\frac{1}{2}\left(2\pi \int_{0}^{\pi} |B^{r}|
\sqrt{-g}d\theta\right).
\eeqn
In addition, the normalized poloidal magnetic flux $\phi$ is considered
as follows;
\beqn
\phi=\frac{\Phi}{\sqrt{\dot{M}}},
\eeqn
which characterizes the degree of the magnetization of the inner accretion
flow \citep[e.g.][]{T15}. We evaluate the above quantities at the
horizon $r_{\rm H}$.

There is almost no variation of $\Phi$ after $t/t_{\rm g} \simeq 4000$
for all cases. We confirm the values of $\Phi$ threading the black hole
horizon are quantitatively consistent in between our 2D and 3D runs
\citep[]{MFN16} with various BH spins. On the other hand, $\dot{M}$
increases in our moderate spin cases ($a=0.5$ and 0.7) at $t/t_{\rm g}
\gtrsim 4000$, while it remains sustained at lower values with time
variations in our high spin cases ($a=0.9$ and 0.99). We note that
qualitatively similar tendency ($\dot{M}$ increases at $t/t_{\rm g}
\gtrsim 3000$--4000) is confirmed in 3D runs for wider spin cases
\citep[$a=0.1$--0.98;][]{MFN16}. As a consequence, $\phi$ never reaches
a level of the MAD state in our 2D runs ($\phi \gtrsim 10$ is confirmed
in $a=0.9$ at $t/t_{\rm g} \gtrsim 4000$).

\end{document}